\documentclass[preprint]{aastex631}
\usepackage{color}
\usepackage{amsmath}
\usepackage{mathtools}
\usepackage{enumitem}
\usepackage{rotating}
\usepackage{graphicx}
\usepackage{mathrsfs}
\usepackage{booktabs}
\usepackage{float}
\usepackage{morefloats}
\usepackage{etoolbox}
\usepackage{esint}
\usepackage{xspace}
\usepackage{multirow}
\usepackage{wrapfig}

\definecolor{bcolor}{RGB}{0, 51, 153}
\definecolor{gcolor}{RGB}{10, 100, 10}
\definecolor{dblue}{RGB}{50, 50, 100}

\newcommand{\code}[1]{{\texttt{#1}}}

\newcommand{\ie}{{\it i.e.}}
\def\eg{{\it e.g.}}
\def\gsim{~\rlap{$>$}{\lower 1.0ex\hbox{$\sim$}}}
\def\lsim{~\rlap{$<$}{\lower 1.0ex\hbox{$\sim$}}}

\def\apjs{{\it Astrophys.~J.~Supp.}}

\def\apjl{{\it Astrophys.~J.~Lett.}}

\def\aap{{\it Astron.~\&~Astrophys.}}

\newcommand{\prot}{P_{\rm rot}}
\newcommand{\porb}{P_{\rm orb}}

\newcommand{\ttau}{$\tau$}
\newcommand{\tq}{$\mathcal{Q}$}

\newcommand{\dprot}{$dP_{{\rm rot}}/dt$}
\newcommand{\dporb}{$dP_{{\rm orb}}/dt$}
\newcommand{\decc}{$de/dt$}

\newcommand{\mtot}{\mu_{\rm tot}}
\newcommand{\Imom}{\mathcal{I}}

\newcommand{\N}[2]{$\mathcal{N}({#1}, {#2})$}
\newcommand{\U}[2]{$\mathcal{U}({#1}, {#2})$}

\newcommand{\msun}{M_{\odot}}

\newcommand{\Eorb}{E_{\rm orb}}

\newcommand{\Jorb}{J_{\rm orb}}
\newcommand{\Jrot}{J_{\rm rot}}
\newcommand{\Jtide}{J_{\rm tide}}
\newcommand{\Jmb}{J_{\rm mb}}

\newcommand{\Trot}{\mathcal{T}_{\rm rot}}

\newcommand{\Tmb}{\mathcal{T}_{\rm mb}}
\newcommand{\Ttide}{\mathcal{T}_{\rm tide}}

\newcommand{\ddt}[1]{\frac{d{#1}}{dt}}
\newcommand{\avg}[1]{\left\langle{#1}\right\rangle}

\newcommand{\vplanet}[0]{\texttt{VPLanet}\xspace}

\newcommand{\kepler}{{\it Kepler}\xspace}
\newcommand{\tess}{{\it TESS}\xspace}
\newcommand{\ktwo}{{\it K2}\xspace}
\newcommand{\gaia}{{\it Gaia}\xspace}

\shorttitle{tidal dissipation in binary stars}
\shortauthors{birky et al.}

\begin{document}

\title{\sc Prospects of Constraining Equilibrium Tides in Low-Mass Binary Stars}

\correspondingauthor{Jessica Birky}
\email{jbirky@uw.edu}

\author[0000-0002-7961-6881]{Jessica Birky}
\affiliation{Department of Astronomy, University of Washington, 3910 15th Avenue NE, Seattle, WA 98195, USA}
\affiliation{DiRAC Institute, University of Washington, 3910 15th Avenue NE, Seattle, WA 98195, USA} 
\author[0000-0001-6487-5445]{Rory Barnes}
\affiliation{Department of Astronomy, University of Washington, 3910 15th Avenue NE, Seattle, WA 98195, USA} 
\author[0000-0002-0637-835X]{James R. A. Davenport}
\affiliation{Department of Astronomy, University of Washington, 3910 15th Avenue NE, Seattle, WA 98195, USA} 
\affiliation{DiRAC Institute, University of Washington, 3910 15th Avenue NE, Seattle, WA 98195, USA}


\begin{abstract}
The dynamical evolution of short-period low-mass binary stars (with mass $M < 1.5M_{\odot}$, from formation to the late main-sequence, and with orbital periods less than $\sim$10 days) is strongly influenced by tidal dissipation.
This process drives orbital and rotational evolution that ultimately results in circularized orbits and rotational frequencies synchronized with the orbital frequency. Despite the fundamental role of tidal dissipation in binary evolution, constraining its magnitude of (typically parameterized by the tidal quality factor $\mathcal{Q}$) has remained discrepant by orders of magnitude in the existing literature. 
Recent observational constraints from time-series photometry (e.g., Kepler, K2, TESS), as well as advances in theoretical models to incorporate a more realistic gravitational response within stellar interiors, are invigorating new optimism for resolving this long-standing problem. 
To investigate the prospects and limitations of constraining tidal $\mathcal{Q}$, we use global sensitivity analysis and simulation based inference to examine how the initial conditions and tidal $\mathcal{Q}$ influence the observable orbital and rotational states. 
Our results show that even under the simplest and most tractable models of tides, the path towards inferring $\mathcal{Q}$ from individual systems is severely hampered by inherent degeneracies between tidal $\mathcal{Q}$ and the initial conditions, even when considering the strongest possible constraints (i.e., binaries with precise masses, ages, orbital periods, eccentricities, and rotation periods). 
Finally as an alternative, we discuss how population synthesis approaches may be a more promising path forward for validating tidal theories. \\
\end{abstract}

\section{Introduction} \label{sec:intro}

Binary stars make up nearly half of the galactic population \citep{DuquennoyMayor91} and form a foundation for most of modern astrophysics. 
The dynamics of systems are continuously evolving through a combination of nonlinear physical processes in which angular momentum can be exchanged between the orbit and rotations of the individual stars. As various studies have modeled so far \citep{witte_orbital_2002,repetto_coupled_2014,penev_poet_2014,bolmont_effect_2016,song_close_2018,fleming_rotation_2019,zanazzi_tidal_2021}, the dominant processes include tidal dissipation, stellar evolution, and magnetic braking. 
Tides cause orbital energy to be dissipated in the form of heat and cause angular momentum to exchange between the orbit and rotations, ultimately altering observable properties (orbital period, rotation period, eccentricity, and obliquity) over long timescales (Myr – Gyr). Additionally stellar evolution plays a role in dynamical evolution, as the radius and mass concentration alter the moment of inertia of the individual stars, thus influencing rotational angular momentum. Magnetic braking also plays an important role as stars lose angular momentum due to magnetized stellar winds and mass loss \citep{skumanich_time_1972,barnes_rotational_2003,matt_mass-dependence_2015,breimann_statistical_2021}. 
In other words, the influence of tides drives binary orbits towards a state of energy minimization \citep{counselman_outcomes_1973,hut_stability_1980}, causing systems to converge to long-term quasi-equilibrium states in which orbital periods become synchronized with rotation periods, and eccentricity decays towards circularization, which is broadly consistent with observations \citep{meibom_observational_2006,meibom_spin-down_2015}.

The influence of stellar tides extends to many systems beyond just binary stars.
Tidal interactions have been observed to cause profound dynamical effects across many scales, from exoplanet systems to Galactic populations. These effects can include tidal locking \citep{barnes_tidal_2017}, tidal heating \citep{barnes2013,leconte_is_2010,levrard_tidal_2007,jackson_tidal_2008}, spin-orbit alignment \citep{heller_tidal_2011,albrecht_stellar_2022}, capture into spin-orbit resonances \citep{goldreich_spin-orbit_1966,colombo_rotation_1966,correia_mercurys_2004}, as well as tidal destruction of exoplanets \citep{jackson_observational_2009,hamer_hot_2019}.
Additionally, on population and Galactic scales, tidal interactions play a role in the dynamics of triple and higher-order multistar systems \citep{ginat_analytical_2021,hamers_multiple_2021}, influencing the rate of capture, escape, or merging in multibody stellar encounters, as well as chaos in stellar clusters \citep[\eg,][]{mardling_tidal_2001}. 

Theories predicting stellar deformation can be broadly characterized as \emph{equilibrium tide} \citep{hut_tidal_1981}, which assume a potential raises a hydrostatic tidal deformation on a star, or \emph{dynamical tide} \citep{zahn_dynamical_1975}, which assume the gravitational influence of a companion drives hydrodynamic or fluid motions. 
Most analytic models of tides apply a ``lag-and-add'' approach \citep{greenberg_frequency_2009}, in which the gravitational response of a star is considered to be the summation of elongated bulge components.  
Among the most simplified models of equilibrium tide which have been widely applied to stars and planets, the lag components are considered to be constant in phase or constant in time. \citep{macdonald_tidal_1964,kaula_theory_1966,goldreich_q_1966,greenberg_outcomes_1974}.
More complex nonlinear mechanisms of tides have also been proposed for stars of different structures. For low-mass stars with radiative cores and convective envelopes, these models consider processes such as turbulent viscosity that dissipates the equilibrium tide in convective layers \citep{zahn_tidal_2008,vidal_turbulent_2020}, or 
dissipation due to internal gravity waves in radiative zones due to the dynamical tide
\citep{zahn_dynamical_1975,lai_dynamical_1997,terquem_tidal_1998,WitteSavonije99,witte_orbital_2002,ogilvie_tides_2013,FullerLai12,barker_tidal_2009,barker_tidal_2020,barker_2022,burkart_dynamical_2014,zanazzi_tidal_2021}.

Current observational opportunities through large-scale stellar surveys offer significant promise for rigorously testing hypotheses of tides for different stellar types. 
From time domain surveys such as \kepler\ \citep{borucki_kepler_2010}, \ktwo\ \citep{howell_k2_2014}, and \tess\ \citep[Transiting Exoplanet Survey Satellite;][]{Ricker15}, eclipsing binaries are ideal observational laboratories for constraining stellar tides, as they enable direct measurements of fundamental stellar properties (masses, densities, temperatures, etc.), and undergo observable changes in orbital/rotational dynamics under the long-term influence of tides. Time-series photometry of eclipsing binaries provides constraints on orbital periods, rotation periods (from star-spot rotation), as well as approximate eccentricities \citep{matson_fundamental_2016} and mass estimates \citep{windemuth_modeling_2019}.

Already \kepler\ and \tess\ have revolutionized discoveries of tidal synchronization \citep{lurie_tidal_2017,hobson_ritz_2025} by discovering an intriguing sub-population of FGKM-type binaries in subsynchronous spin-orbit ratios in which the star has a slower rotation period than the orbital period. \cite{fleming_rotation_2019} demonstrated that equilibrium tide models can reproduce 1:1 synchronized binaries and generate a wide distribution of subsynchronously rotating binaries. However their model falls short of reproducing the tight overdensity of observed subsynchronous rotators discover in \cite{lurie_tidal_2017} in which $\sim15\%$ of the \kepler\ eclipsing binary population was found to have rotation periods $\sim13\%$ slower than their orbital period. \cite{lurie_tidal_2017} suggested that this population of subsynchronous binaries could be the result of differential rotation in which a latitudinal shear along a star could produce starspots rotating slower than the equator at high latitudes. Futhermore, \cite{lurie_tidal_2017}, \cite{hobson_ritz_2025}, and \cite{jermyn_differential_2020} suggest that low-mass eclipsing binaries with measurable star-spot modulation are promising targets for investigating differential rotation on the surface of convective stars.

In addition to eclipsing binaries, spectroscopic follow-up has enabled precise masses and eccentricities of many binaries from radial velocity constraints. Multi-epoch surveys, such as the Apache Point Observatory Galactic Evolution Experiment \citep[APOGEE;][]{majewski_apache_2015}, have enabled the discovery and orbital constraints of thousands of spectroscopic binaries \citep{price-whelan_binary_2018,el-badry_discovery_2018,price-whelan_close_2020,kounkel_double-lined_2021}. In some cases with detailed follow-up, it is additionally possible to constrain obliquities from individual binary systems through the Rossiter-McLauglin effect \citep{mazeh_observational_2008,albrecht_stellar_2022}, or infer the distribution of obliquities from a population of binaries hierarchically \citep{morton_obliquities_2014}. Although ages are among the most challenging parameters to determine for stellar systems \citep{soderblom_ages_2010}, \ktwo and \tess have enabled Galactic coverage of many stellar populations, including binaries in open clusters, which are the gold standard when it comes to constraining orbital properties as a function of age \citep[\eg,][]{southworth_eclipsing_2006,david_k2_2015,david_new_2016,gillen_new_2017,torres_eclipsing_2018}. 

Despite the drastic improvements in observational capabilities, it remains an open question as how to best approach constraining tidal models against data. 
Early attempts to constrain tidal models considered a metric known as the circularization cut-off period \citep{MayorMermilliod84,Mathieu_mazeh_1988,mathieu_tidal_2004}, which determines the orbital period at which most binaries in a population have circularized (where $e \approx 0$ for systems with $\porb < P_{\rm cut}$). The idea behind this method relies on the assumption that binaries at close orbital separation will experience stronger tides, and should thus circularize faster --- a trend which has been confirmed in observations of open clusters \citep{Mathieu_mazeh_1988,mathieu_tidal_2004}.

Later work by \cite{meibom_robust_2005} proposed a more robust metric known as the tidal circularization period. Instead of determining $P_{\rm cut}$ from the observations at which $e > 0$, they fit a functional form to observed $\porb$ and $e$, assuming that populations of binaries could originate from a distribution of initial orbital periods and eccentricities. This circularization period method did a better job at accounting for some uncertainty in initial conditions; however, it did not consider all prior uncertainties (e.g. rotation period, stellar evolution) or incorporate the constraints from stellar rotation periods. Recent studies have called these additional model factors into question, such as \cite{bashi_features_2023}, which finds evidence from \gaia spectroscopic binaries that the circularization period depends more on stellar effective temperature than open cluster age. Additionally a study by \citet{mirouh_detailed_2023} used population synthesis to examine the impacts of tides and initial orbital period distributions on the circularization and synchronization of binaries as a function of age (using data of systems from 8 open clusters). Their findings concluded that constraining tidal efficiency based on the circularization of binary populations is difficult or impossible due to the inefficiency of tides on the main-sequence, as well as strong dependencies on the initial orbital configuration. However, their work also showed that tidal synchronization better captures the age-dependent effects of tidal dissipation, and thus the fraction of synchronous binaries at a given age would be a more promising constraint on tides.

More generally, there are debates in the literature as to whether individual or population approaches to inferring tidal \tq's are more appropriate. Some approaches have focused on using forward modeling approaches to infer unknown parameters including the initial orbital states and the tidal dissipation rate (or equivalently the tidal \tq) value that best reproduces observations of individual stars or planets \citep{goldreich_q_1966}. 
Studies such as \cite{barker_tidal_2020} advocate for a system-by-system comparison of tidal \tq\ values, given the potential dependence \tq\ may have with stellar parameters or age. On the other hand, studies such as \cite{cameron_hierarchical_2018} advocate for a population based approach using a broad sample of binaries to constrain \tq. The population approach has the advantage of accounting for uncertainties in the initial condition as a population parameter; however, it requires making an assumption about how \tq\ depends on stellar parameters (in this case, \cite{cameron_hierarchical_2018} assume a functional dependence on stellar effective temperature). 
Resulting estimates of tidal \tq\ for solar-like stars from the literature have spanned orders of magnitude with estimates of $\mathcal{Q}$ ranging from $\sim10^5$ to $10^8$ \citep{meibom_observational_2006,jackson_observational_2009,hansen_calibration_2010,PatelPenev22,penev_comprehensive_2022}, bringing into question the underlying cause of the discrepancy. 
Given that constraints on stellar tides have far-reaching consequences on dynamics from exoplanetary to galactic scales, with decades of research investment, it is critical that we rigorously examine all uncertainties (both model and observational) to understand the prospects and limitations of constraining the tidal dissipation mechanisms. 

From a theoretical perspective, there may be a variety of reasons why latent parameters may not be feasible to constrain: there may be infinite solutions, there may be finite solutions, or there may be no solutions. If solutions exist, it may be that the latent parameters do not (strongly) influence manifest parameters, or there may be significant degeneracies between latent parameters and initial conditions. As we argue in this paper, constraining tidal dissipation has been an underestimated challenge, even when considering the simplest formulations of tidal dissipation against simulated data.

In this paper, we assess the degree to which improved observations would resolve the discrepancy in the inferred values of tidal \tq\ based on individual systems.
For simplicity we will consider a commonly used and easily tractable model of the equilibrium tide, however the statistical methodology presented in Section \ref{sec:methods} may also be applicable to more complex tidal dissipation models, including dynamical tide formulations. 
In this study, our aim is to answer the questions: 
\emph{How do we constrain theories of tidal dissipation?} 
In particular, to what degree is model inference inhibited by unidentifiability or degeneracies inherent in the model formulation? To what degree do observational uncertainities limit model inference? Finally, what needs to be done next and how should we focus our efforts (\eg, such as observational follow up)?

In this paper we characterize the prospects and limitations of constraining equilibrium tides in low-mass stars using global sensitivity analysis \citep{sobol_global_2001,saltelli_variance_2010}, which has been widely used in applications from biological systems to finance \citep{saltelli_why_2017}. Although it is less widely applied to astrophysical simulations, as we present in this paper, sensitivity analysis is an effective way to systematically characterize how sensitive model outputs are to model inputs. Sensitivity analysis tells us which input parameters of the model are most influential to first order, however, it does not give us as much information about the correlation between model inputs. To further investigate the dominant effects of this non-linear model, we perform simulated Bayesian inference \citep[using Active Learning for Acelerated Bayesian Inference, or \code{alabi};][]{birky_alabi} to test how observational uncertainties influence the quality of posterior constraints as well as examine the degeneracies between input parameters.

This paper is organized as follows. Section \ref{sec:formalism} defines the terminology used throughout the analysis, as well as describes the model assumptions and numerical implementations used for all of our simulations. Section \ref{sec:methods} describes the challenges that arise when comparing tidal models to binary observations and describes the methodology for sensitivity analysis. Section \ref{sec:results} explains the results of the sensitivity analysis applied to the numerical simulations described in Section \ref{sec:formalism}. Section \ref{sec:discussion} synthesizes the main findings and discusses some of the main limitations of this analysis. Finally, Section \ref{sec:conclusion} discusses future observational and numerical work that is motivated from our results.  \\

\section{Model} \label{sec:formalism}

The rotational and orbital evolution of low-mass stars ($M_* <1.5 \msun$) is influenced by a combination of physical processes, namely tidal dissipation (Section~\ref{sec:tides}), stellar evolution (Section~\ref{sec:stellar}), and magnetic braking (Section~\ref{sec:stellar}).
Here we provide an overview of the main assumptions used in our model.
In Section~\ref{sec:states} we summarize the key model variables for the coupled system of ordinary differential equations and numerically solve the equations using the \vplanet implementation (\citealt{barnes_vplanet_2020}; see also \citealt{fleming_lack_2018,fleming_rotation_2019}).
The goal of this model set up is to analyze and establish the limitations of comparing equilibrium tidal models to datasets available in surveys (\eg\ \kepler, \ktwo, \tess, APOGEE) and the literature. 

\subsection{Energy Dissipation and Torque due to Tides} \label{sec:tides} 

To consider the rate of energy dissipated over an orbit, it is common to introduce a quantity known as the ``tidal quality factor''. This quantity is proportional to the maximum energy stored in tidal deformation divided by the average energy dissipated over an orbit. With $E_0$ defined as the maximum energy and $E$ being the orbit energy as a function of time $t$, the tidal \tq\ is defined as
\begin{equation} \label{eqn:q_def}
    \mathcal{Q}^{-1} 
    \equiv
        \frac{\textrm{energy dissipated over an orbit}}{\textrm{maximum energy stored in deformation}}
    \equiv 
        \frac{1}{2\pi E_0} 
        \oint \left(-\frac{dE}{dt} \right) dt.
\end{equation}
A lower tidal Q value indicates that the star is more efficient at dissipating tidal energy. This parameterization of tidal \tq\ can be translated into the secular evolution of the orbit by assuming that each Fourier component of the potential induces a tidal response on a perturbed star. In other words, each component of the potential drives the distortion of material at a particular frequency and phase lag, known as $\sigma$ and $\varepsilon$ respectively. 
This effective forcing frequency depends on both the mean motion $n$ and the rotation frequency $\omega$, where $\sigma = kn - m\omega$ and $k$ and $m$ are integers.  The phase lag $\varepsilon$ is equal to twice the geometric lag angle of the bulge orientation. The tidal \tq\ is related to the phase lag  \citep[Equation 105,][]{efroimsky_tidal_2009}:
\begin{equation}
    \mathcal{Q}^{-1} 
        = \frac{\tan\varepsilon}{1 - \left(\frac{\pi}{2} - \varepsilon \right) \tan\varepsilon}
        = \tan \varepsilon + \mathcal{O}(e^2).
\end{equation}
In general, the relationship between the phase lag and the effective forcing frequency $\varepsilon(\sigma)$ depends on the internal structure of the body. However, for small phase lags (assuming a weak friction approximation), $\tan\varepsilon(\sigma) \approx \varepsilon(\sigma)$, in which the response of a tidal bulge under the forcing frequency is analogous to a damped driven harmonic oscillator \citep{greenberg_frequency_2009}.
The relation between $\varepsilon$ and the tidal \tq\ can be approximately linearized by assuming either a constant phase lag $\varepsilon(\sigma) \approx 1/\mathcal{Q}$, where \tq\ is held constant, or a constant time lag that assumes $\varepsilon(\sigma) \approx n\tau$, where the time lag, \ttau\ is held constant. 
These linear parameterizations of the equilibrium tide are hereafter referred to as the constant phase lag \citep[CPL; ][]{goldreich_q_1966,ferraz-mello_tidal_2008} and the constant time lag \citep[CTL;][]{hut_tidal_1981,leconte_is_2010} models.
The derivations for the CTL and CPL models are well reviewed in literature, but we provide a summary of the assumptions and equations used in the Appendix Section \ref{sec:appendix}. 

\subsection{Stellar Evolution and Magnetic Braking} \label{sec:stellar}

In general, the relationship between external torques and the angular momentum evolution of a single star is written as
\begin{equation}
    \Trot  = \ddt{\Jrot}  = \Imom \ddt{\omega} + \omega \ddt{\Imom},
\end{equation}
with $\mathcal{I}$ being the moment of inertia of the body, and $\Trot$ being the net rotational torque due to external influences (in the case of stellar evolution, the angular momentum lost due to stellar winds, $\Trot = \Tmb$). \vplanet\ includes several implementations of magnetic braking models \citep{skumanich_time_1972,repetto_coupled_2014,reiners_radius-dependent_2012,matt_mass-dependence_2015,breimann_statistical_2021} that simulate the rotational torque due to stellar winds.
In this study we adopt the \cite{matt_mass-dependence_2015,matt_erratum_2019} model for $\Tmb$ that considers a semi-analytic saturated magnetic braking law, and uses stellar evolution models from \cite{baraffe_new_2015} to determine the change in moment of inertia, $d\Imom/dt$.

Magnetohydrodynamic simulations of solar-like stellar winds \citep{matt_magnetic_2012} show that the rotational torque on a star can generally be written in semianalytic form.
Making physically motivated assumptions, \cite{matt_mass-dependence_2015} introduce a scaling relation for the surface magnetic field $B_*$ and the stellar wind mass loss $\dot{M_w}$ as a function of stellar mass and Rossby number, in which the torque due to stellar winds can be written as:
\begin{equation} \label{eqn:torque_mb}
\Tmb \, = \, \mathcal{T}_0 \,
    \begin{cases}
        \chi^2 \left( \frac{\omega}{\omega_{\odot}} \right)  & \text{if saturated,} \ R_o \leq R_{o,\odot}/\chi \\
        \left( \frac{\tau_{cz}}{\tau_{cz \odot}} \right)^2 \left( \frac{\omega}{\omega_{\odot}} \right)^3 &  \text{if unsaturated,} \ R_o > R_{o,\odot}/\chi
    \end{cases}.
\end{equation} 
The Rossby number is the ratio of the rotation period to the convective turnover timescale, $R_o = \prot/\tau_{cz}$. The timescale $\tau_{cz}$ is computed using a relation from \cite{cranmer_testing_2011}. The transition between saturated and unsaturated regimes is specified by $\chi \equiv R_{o,\odot}/R_{o,sat}$, where we adopt a value of $\chi = 10$ as used in \cite{matt_mass-dependence_2015}. The proportionality constant is 
\begin{equation} \label{eqn:matt0}
	\mathcal{T}_0 = 6.3 \times 10^{30} \ \mathrm{erg} \ \left( \frac{R}{R_{\odot}} \right)^{3.1} \left( \frac{M}{M_{\odot}} \right)^{0.5}.
\end{equation}

In the case of a binary star system, the rotational torque acting on an individual star is influenced by both the magnetic braking torque as well as the tidal torque. We make the assumption that the torque processes due to magnetic braking and tides are linearly independent, such that the net rotational torque is the sum of the two components $\Trot = \Tmb + \Ttide$. We note that a non-linear coupling between $\Tmb$ and $\Ttide$ (e.g., considering the influence of tides on magnetic field structure of a star) would require significantly more theoretical work and magnetohydrodynamic simulations to accurately model, hence we consider only linear coupling following previous studies \citep{witte_orbital_2002,repetto_coupled_2014,penev_poet_2014,bolmont_effect_2016,song_close_2018,fleming_rotation_2019,zanazzi_tidal_2021}.

\subsection{Model State Space} \label{sec:states} 

Adopting terminology from dynamical systems analysis \citep{willems_modelling_2000}, the characterization of each of our model free parameters is listed in Table~\ref{tab:parameters}. Model variables that govern the dynamical evolution of a system can be characterized as \emph{state variables} (time-variable parameters of the coupled ordinary differential equations), \emph{manifest variables} (time-fixed parameters that are observable), or \emph{latent variables} (time-fixed parameters that are not directly observable). 

\begin{center}
\begin{longtable}{rcl} 
\caption{\normalsize Definition of model parameters, where subscript 1 is the primary and subscript 2 is the secondary for the two stars in the binary system. 
Note that we use the variables ($\omega$, $a$, $e$) in our model parameterization (Section \ref{sec:appendix}) but the state of the system can be equivalently parameterized by ($\prot$, $\porb$, $e$), with $\prot$ being the rotation period and $\porb$ being the orbital period, which are the values which we use to compare to observations.
\label{tab:parameters}
} \\
\hline
	\multicolumn{3}{c}{\textit{State Variables}} \\
\hline
    $a$				&& 		semi-major axis of the orbit \\
	$e$ 			&& 		eccentricity of the orbit \\
	$\omega_1$, $\omega_2$		&& 		rotation frequency of each star \\
	$\psi_1$, $\psi_2$			&& 		obliquity of each star \\
	$R_1$, $R_2$    			&&		average radius of each star  \\
	$r_{g1}$, $r_{g2}$ 			&& 		radius of gyration of each star \\ 
\hline
	\multicolumn{3}{c}{\textit{Manifest Variables}} \\
\hline
	$M_1$, $M_2$ 			&&		mass of each star \\	
\hline
	\multicolumn{3}{c}{\textit{Latent Variables}} \\
\hline
        $\mathcal{Q}_1$, $\mathcal{Q}_2$				&& 		tidal quality factor \\ 
	$\tau_1$, $\tau_2$			&& 		tidal time lag \\
	$k_2$			&& 		Love number of second degree \\
	$\tau_{cz}$		&&		convective turnover timescale \\
\hline
\end{longtable} 
\end{center}


\subsection{Numerical Simulations} \label{sec:simulations}

\begin{figure}[ht]
    \centering
    \includegraphics[width=.95\textwidth]{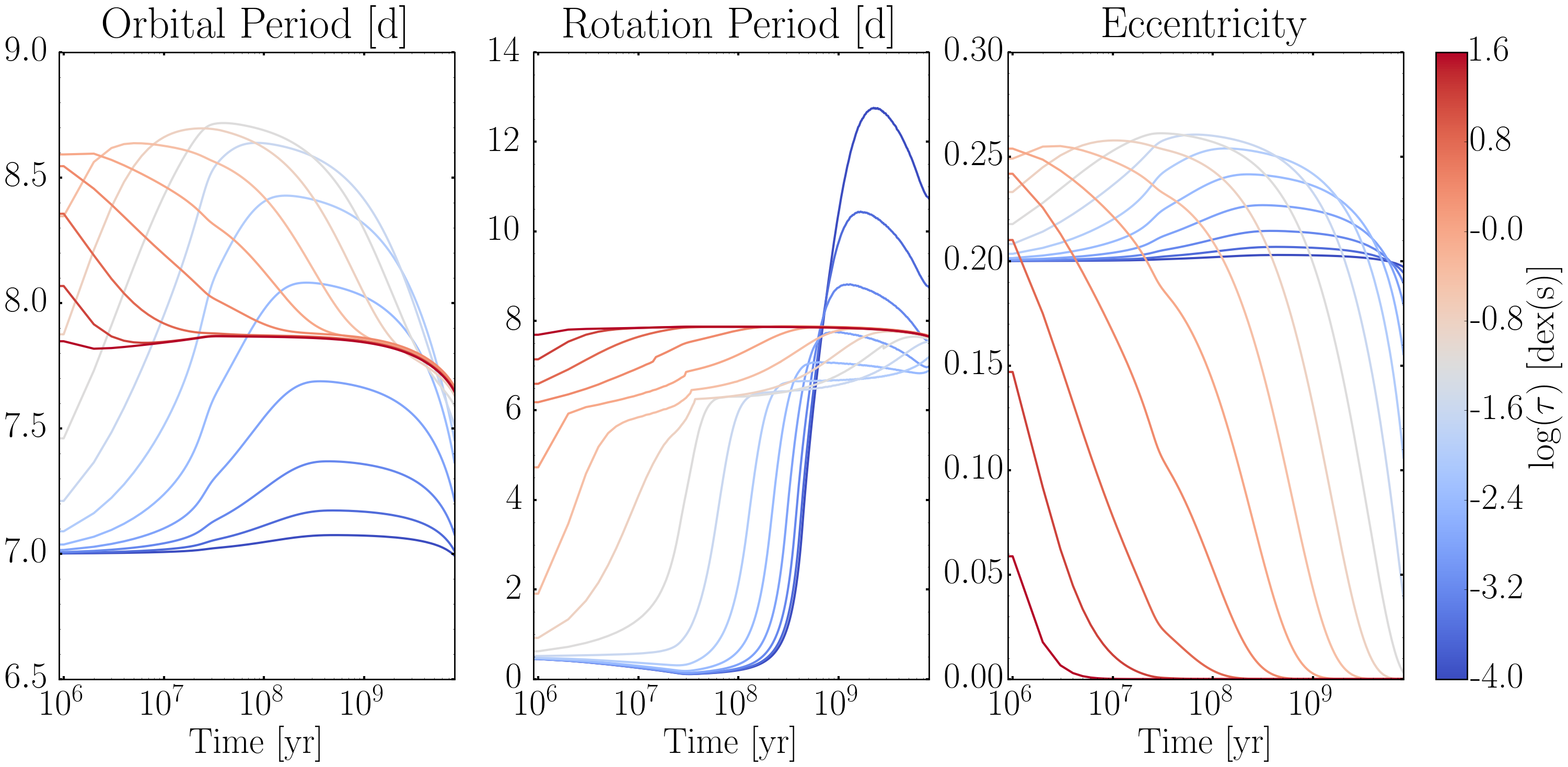} 
    \vspace{0.5em}
    \caption{\normalsize Simulations of equilibrium tide (CTL) coupled with stellar evolution. Panels show the evolution of orbital period (left), rotational period (center), and eccentricity (right) for tidal \ttau\ strengths in the range $-4.0 < \log(\tau) < 1.6$. Tidal \ttau\ influences the timescale of synchronization and circularization, where \emph{higher} tidal \ttau\ results in more rapid evolution. Each track originates from the same initial conditions for a $1\msun-1\msun$ mass binary, and we vary only tidal \ttau\ to illustrate the effect of tides on the timescale of orbital evolution.}
    \label{fig:evolution_ctl}
\end{figure}

\begin{figure}[h]
    \centering
    \includegraphics[width=.95\textwidth]{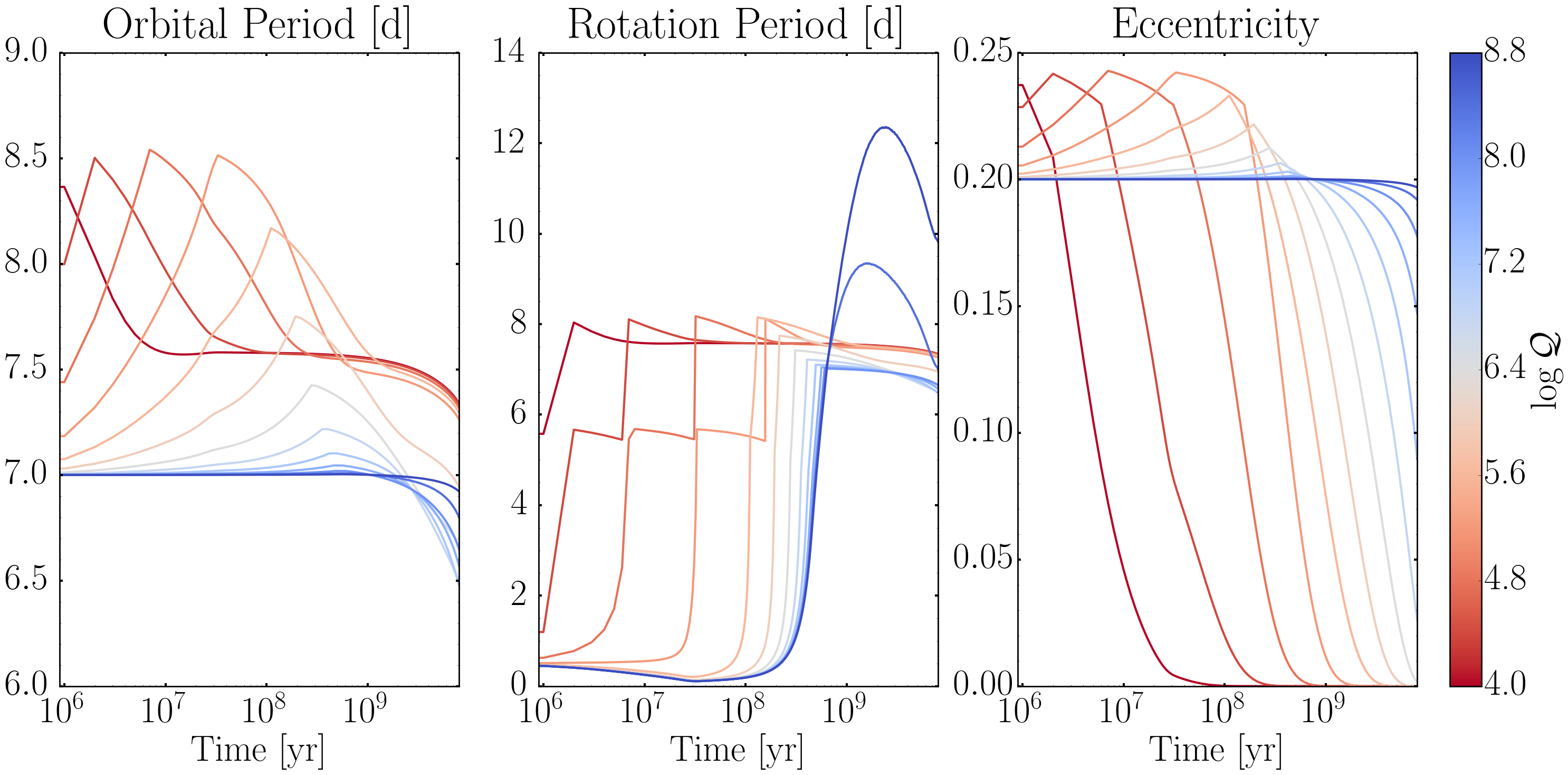} 
    \vspace{0.5em}
    \caption{\normalsize Same as Figure \ref{fig:evolution_ctl}, but with the CPL model. Panels show tidal \tq\ strengths in the range $4.0 < \log(\mathcal{Q}) < 9.0$. Tidal \tq\ influences the timescale of synchronization and circularization, where \emph{lower} tidal \tq\ results in more rapid evolution.}
    \label{fig:evolution_cpl}
\end{figure}

\begin{figure}[h]
\begin{center}
    \includegraphics[width=0.49\linewidth]{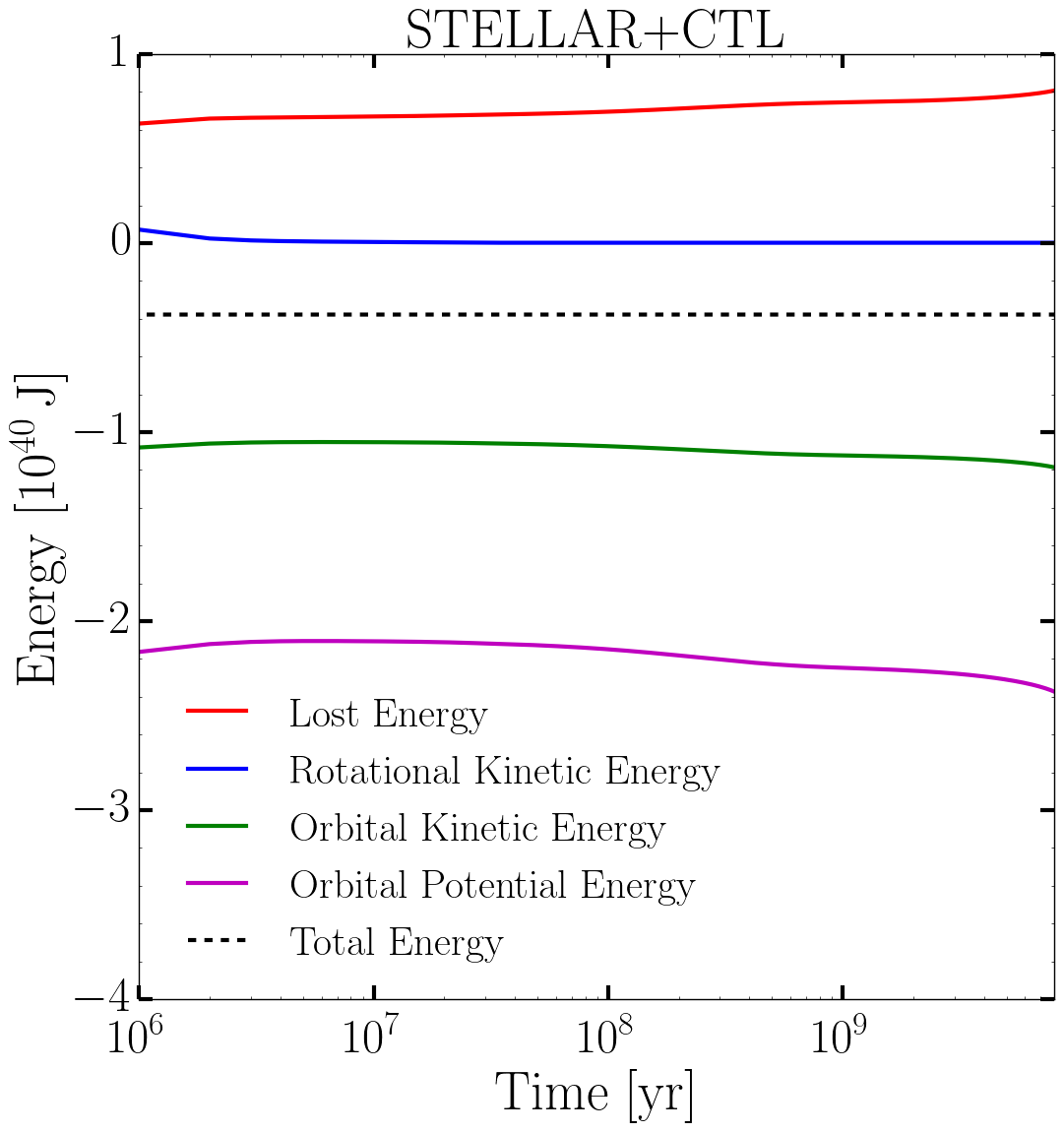} 
    \includegraphics[width=0.49\linewidth]{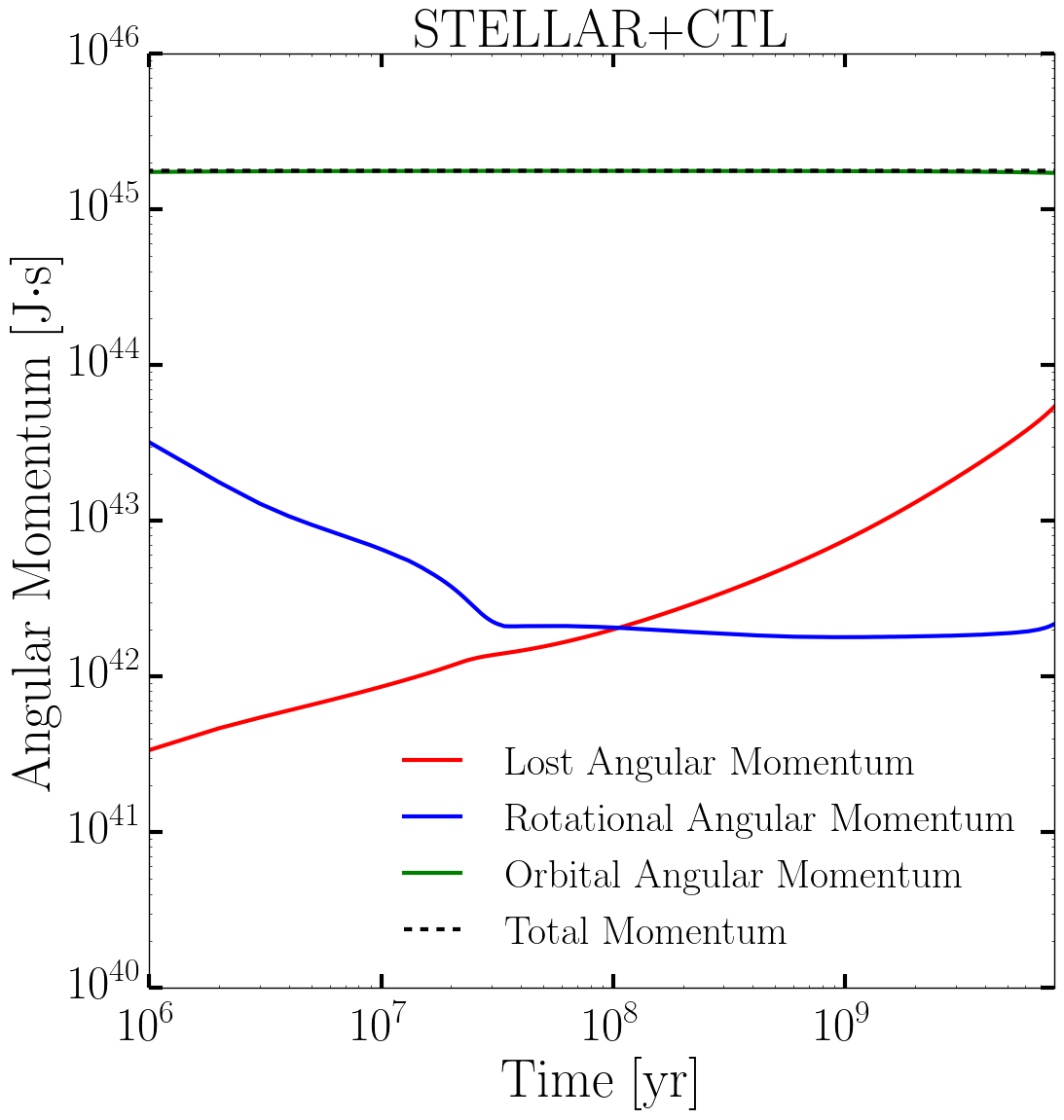} 
    \vspace{0.5em}
    \caption{\normalsize Exchange in energy (left panel) and angular momentum (right panel) between the rotation and orbit of stars in the binary system, as predicted by the CTL model \citep{hut_tidal_1981} when coupled with stellar evolution \citep{baraffe_new_2015}, and magnetic braking \citep{matt_mass-dependence_2015}. Lost energy (red line, left panel) is due to tidal heating, and lost angular momentum is due to stellar winds (red line, right panel). The black dashed line shows that energy and angular momentum remained conserved over time, showing that the coupling of our equations is behaving as expected.}
\label{fig:conservation_ctl}
\end{center}
\end{figure}

\begin{figure}[h]
\begin{center}
    \includegraphics[width=0.49\linewidth]{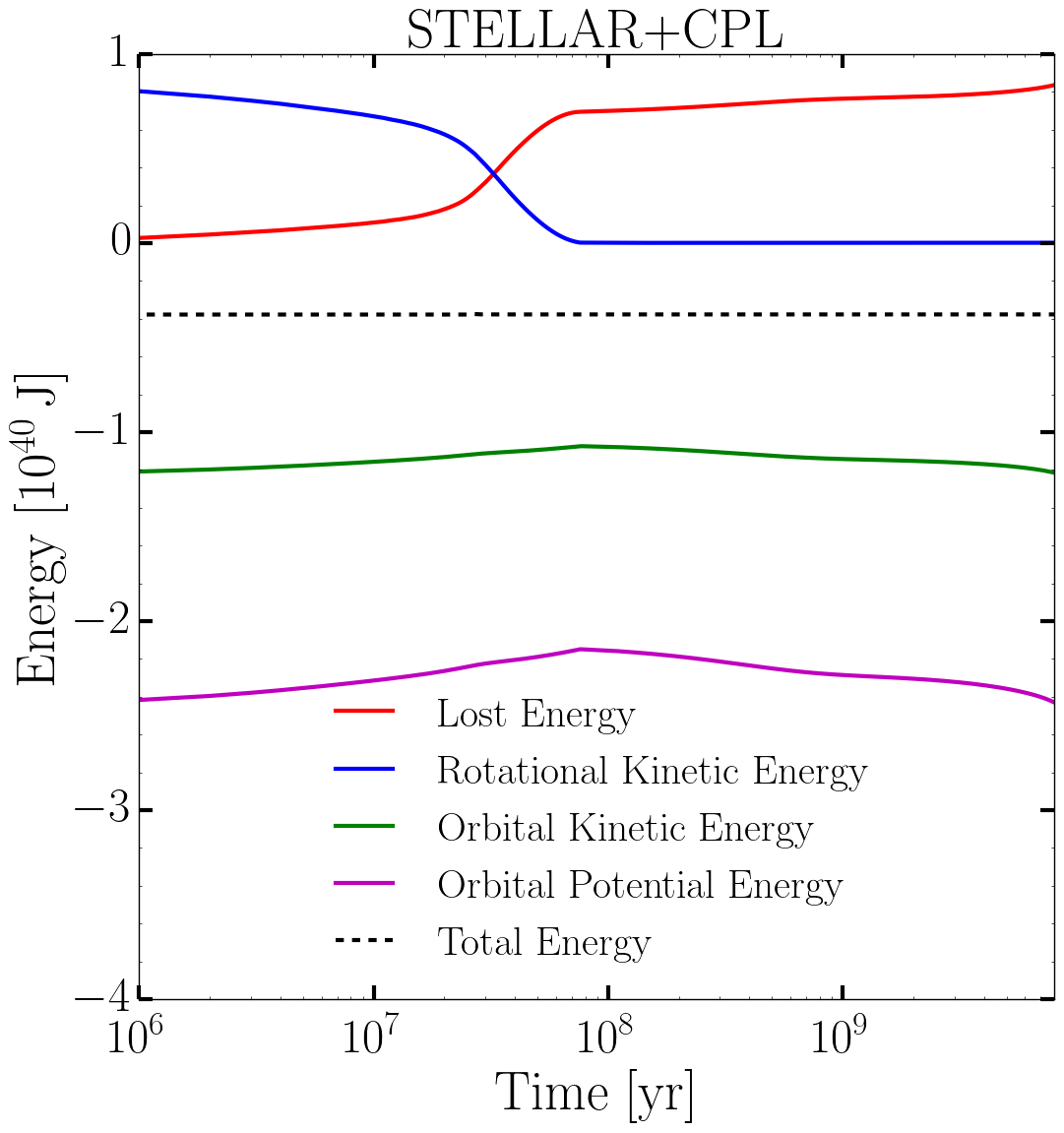} 
    \includegraphics[width=0.49\linewidth]{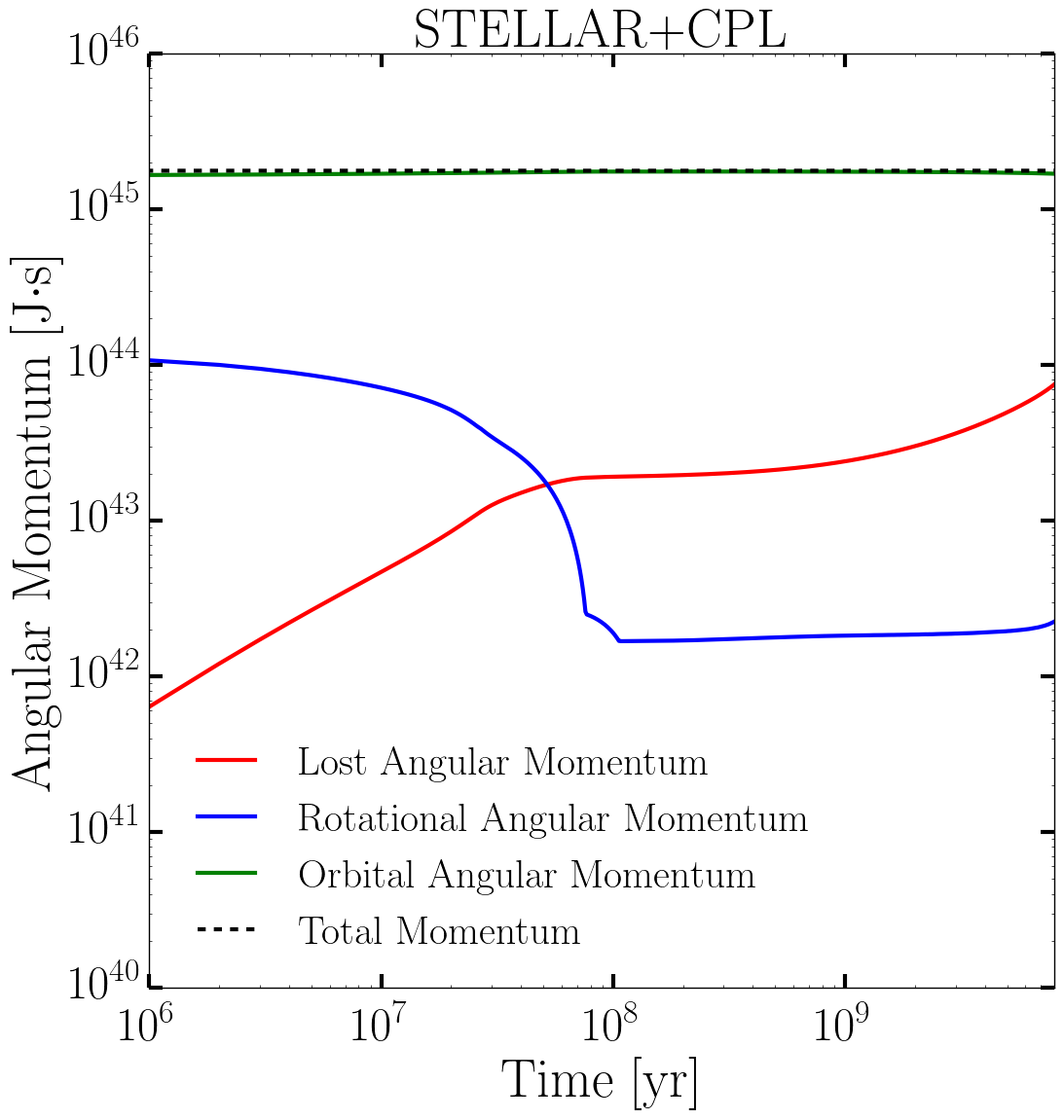} 
    \vspace{0.5em}
    \caption{\normalsize Same as Figure \ref{fig:conservation_ctl}, but for the CPL model.}
\label{fig:conservation_cpl}
\end{center}
\end{figure}

We use the \vplanet package \citep{Barnes20} to evolve the system of differential equations using fourth order Runge-Kutta with adaptive time stepping. Our implementation follows \cite{fleming_lack_2018} and \cite{fleming_rotation_2019} in which we numerically solve for each state variable (Table \ref{tab:parameters}). A demonstration of evolution trajectories is shown in Figure \ref{fig:evolution_ctl} for the CTL model and Figure \ref{fig:evolution_cpl} for the CPL model. 
The state variables are coupled under the assumption of conservation of energy and angular momentum.
Figures \ref{fig:conservation_ctl} and \ref{fig:conservation_cpl} illustrate how energy and angular momentum are exchanged between orbit and rotation over the course of 8~Gyr evolution for a $1\msun-1\msun$ binary system according to both CTL and CPL implementations.  \\

\clearpage

\section{Methods} \label{sec:methods}

\subsection{Sensitivity Analysis}  \label{subsec:sensitivity_methods}

Interpreting high-dimensional, nonlinear dynamical models is a non-trivial task. Understanding
the potential for inference (e.g., identifying systems with the strongest constraints and
inspecting degeneracies) requires a framework to assess how sensitive the model outputs are to
each input.
Overcoming these challenges requires developing a methodology for systematically analyzing high-dimensional parameter spaces to establish which input model parameters dominate the final state, and to identify which observable uncertainties are the most important to reduce (thus, informing observational priorities). 
To make this assessment, we perform sensitivity analyses. We also consider the best-case limits for inferring latent parameters of the models (\tq\ or \ttau) by applying sensitivity analysis to a simulated likelihood with optimistic uncertainties based on observational studies of the most precisely constrained systems.  

Variance-based global sensitivity analysis is a statistical technique used to identify and quantify the impact of input variables on the variance of a system's output. It is often used to assess the robustness and uncertainty of a model or simulation, as well as to identify the most influential inputs. Sensitivity analysis has been widely applied in many areas of quantitative model analysis, but has seen relatively few applications on astrophysical models \citep[\eg,][]{oleskiewicz_sensitivity_2019}. 

The most well-known variance-based method is the Sobol sensitivity indices \citep{sobol_global_2001,saltelli_variance_2010}, which decompose the output variance into main effects and interaction effects. This method involves sampling the input space using a pseudo-random design and evaluating the output for each set of inputs. The resulting data are then used to calculate sensitivity indices for each input variable, which represent the contribution of that variable to the total variance of the output.

There are several types of Sobol sensitivity indices: first-order indices that measure the contribution of each individual input variable to the output variance, total-order indices that measure the contribution of each input variable, along with all possible interactions with other variables, to the output variance, and higher-order indices that measure the contribution of specific interactions between input variables to the output variance.

The procedure for computing Sobol sensitivity indices is as follows. Any model that can be described in the form $Y=f(\bf{X})$, where $\bf{X} \in \mathbb{R}^d$ is an input vector, and $d$ is the number of input variables. If $Y$ is a chosen univariate output of the model, $Y$ can be decomposed in the form:
 \begin{equation}
	Y = f_0 + \sum_{i=1}^d f_i (X_i) + \sum_{i<j}^d f_{ij} (X_i, X_j) + \dots + f_{1,\dots,d} (X_1,\dots,X_d),
\end{equation}
and taking its variance of $Y$ yields:
\begin{equation}
    \mathrm{Var}(Y) = \sum_{i=1}^d V_i + \sum_{i<j}^d V_{ij} + \dots + V_{12\dots d}
\end{equation}
where the variance components can be computed as:
\begin{equation}
    V_i 	= {\rm Var}_{X_i} (E_{X\sim i} [Y | X_i]),
\end{equation}
\begin{equation}
    V_{ij}	= {\rm Var}_{X_{ij}} (E_{X\sim ij} [Y | X_i, X_j]) - V_i - V_j.
\end{equation}
Here the notation $X_{\sim i}$ is used to denote the set of all inputs except $X_i$. The first-order sensitivity indices, $S_i$, are defined as the fraction $V_i$ attributed to model parameter $i$, divided by the total variance $\rm{Var}(Y)$: 
\begin{equation} \label{eqn:sens_index1}
	S_i = \frac{V_i}{\rm{Var}(Y)}.
\end{equation}

We follow the procedure of \citet{herman_salib_2017} for computing variances. First, a set of $N$ random samples for $\bf{X}$ is drawn from a defined input domain. 
In practice, this sampling is performed using a pseudo-random approach, such as Sobol sampling \citep{sobol_global_2001}. The idea behind Sobol sampling is to create a low-discrepancy sequence of points that is more evenly spaced than a random sampling method and can efficiently span a large parameter space given a finite set of points.
For each parameter $i$, we calculate the variance of the output due to that parameter, $V_i$, via computing the estimator:
\begin{equation}
	\rm{Var}_{X_i} (E_{X_{\sim i}} [Y | X_i])
		\approx \frac{1}{N} \sum_{j=1}^N f(\mathbf{B})_j \left(f(\mathbf{A}^i_B)_j - f(\mathbf{A})_j \right),
\end{equation}
where $\mathbf{A}$ and $\mathbf{B}$ are each $N\times d$ matrices of input parameters sampled using the Sobol (or other pseudorandom) method. In our analysis, we use the \code{SALib} Python implementation \citep{herman_salib_2017} to compute the Sobol sensitivity indices.

\subsection{Simulated Likelihoods} \label{subsec:likelihood}

In addition to the sensitivity of each output parameter, we also consider how the relative uncertainties weight each output. To make this assessment, we consider a simulated likelihood function:
\begin{equation} \label{eqn:lnlike}
    \ln \mathcal{L} = \sum_{j=1}^n \left( \frac{y_j - \hat{y}_j}{\hat{\sigma}_j } \right)^2,
\end{equation}
where $y_j = \{ \prot, \porb, e \}$ are the outputs of a given simulation compared to a fiducial simulation with mean values $\hat{y}_j$ and uncertainties $\hat{\sigma}_j$. In other words, $y_j$ represents the model samples and ($\hat{y}_j$, $\hat{\sigma}_j$) represent the simulated data. 

We further simulate posterior probability estimates according to Bayes' Theorem,
\begin{equation} \label{eqn:bayes}
    \ln \mathcal{P} \propto \ln \mathcal{L} + \ln \Pi,
\end{equation}
where $\ln\mathcal{P}$ is the sampled posterior distribution sampled according to an uninformative prior, $\ln \Pi$, and model likelihood fit is $\ln \mathcal{L}$ (Eqn. \ref{eqn:lnlike}). Fiducial values for the simulated likelihood, as well as prior ranges, are given in Table \ref{tab:ranges}.  

\begin{table}
\begin{tabular}{|c|c|c|c|c|c|}
    \hline
    Input   & Unit & Min range & Max range & Fiducial & Prior  \\
    \hline
    $M_1$, $M_2$           & $\msun$   &   0.1     &   1.1     &   1.0     & 1.0 \\
    $\psi_{1i}$, $\psi_{2i}$        &   deg     &   0       &    30     &   10      & 0 \\
    $P_{\rm rot1,i}$, $P_{\rm rot2,i}$ &   days      &   0.1     &    10.0   &   0.5     & $\mathcal{U}(0.1, 10.0)$ \\
    $P_{\rm orb,i}$ &   days    &   0.1     &    12.0   &   7.0     & $\mathcal{U}(0.1, 12.0)$ \\
    $e_i$             &           &    0.0    &   0.5     &   0.3     & $\mathcal{U}(0.0, 0.5)$ \\
    $\log_{10}(\mathcal{Q}_1)$, $\log_{10}(\mathcal{Q}_1)$  &           &   4.0     &  12.0     &   6.0     & $\mathcal{U}(4.0, 12.0)$ \\
    $\log_{10}(\tau_1)$, $\log_{10}(\tau_2)$  & log\,(s)  & $-4.0$    & 1.0       &   $-1.0$  & $\mathcal{U}(-4.0, 1.0)$ \\
    \hline 
\end{tabular}
\vspace{1em}
\caption{\normalsize Ranges of input parameters considered in the sensitivity analysis simulations. Input parameters chosen for the fiducial simulation to compute the likelihood (Equation \ref{eqn:lnlike}). Subscript 1 and 2 denote the primary and secondary of the system, while subscript $i$ denotes an initial value of a time-varying parameter.}
\label{tab:ranges}
\end{table}

We consider the limits of inference under optimistic uncertainties, adopting the values for $\hat{\sigma}_1$ and $\hat{\sigma}_2$ listed in Table \ref{tab:constraints}.
The uncertainties in Table \ref{tab:constraints} are chosen to represent the (optimistic) order of magnitude to which each parameter has been constrained in the literature using data from eclipsing binaries and radial velocities \citep{david_k2_2015,david_new_2016,gillen_new_2017,torres_eclipsing_2018}. 
Particularly in time series photometry with long baselines such as \kepler, orbital periods can be highly precise ($\sim 1$ sec). 
Rotation periods, on the other hand, are inherently more difficult to constrain, as they depend on the complex star-spot processes that make them observable \citep[\eg\ spot lattitude, differential rotation, and emergence/decay timescale;][]{aigrain_testing_2015}. 
However studies such as \citealt{david_new_2016} and \citealt{gillen_new_2017} show that rotation periods for short-period binaries may be measured to uncertainties of $\sim0.01 - 0.05$\,d, by using a Gaussian process with a periodic kernel \citep[see 5.4 of][]{rasmussen_gaussian_2006}. While obliquities can in some cases be constrained by radial velocities \citep{triaud_rossiter-mclaughlin_2018,Hatzes2019,albrecht_spin_2007}, given that the results of our sensitivity analysis (Section \ref{subsec:sensitivity_results}, Figures \ref{fig:sensitivity_ctl_full}--\ref{fig:sensitivity_cpl_full}) suggest that obliquity is not a dominant factor in the final orbital state, for simplification in these posterior simulations we hold obliquity fixed at $\psi = 0$. \\

\begin{table}
\begin{tabular}{|c|c|c|c|}
\hline
 Parameter  & Description  & Observation Method & Uncertainty  \\
\hline
 $P_{\rm rot1,f}$, $P_{\rm rot2,f}$ & final rotation period [days]	& LC autocorrelation function & 0.01	\\ 
 $P_{\rm orb,f}$	& final orbital period [days]		& LC lomb scargle		& $10^{-5}$ \\ 
 $e_{\rm f}$		& final eccentricity 				& LC eclipse + RVs 		& 	$10^{-3}$ \\ 
\hline
\end{tabular}
\vspace{2em}
\caption{\normalsize Model parameters that are observable from lightcurve (LC) photometry and/or spectroscopic radial velocity (RV) datasets. The subscript $f$ denotes that these are the final values of each state variable, evaluated at a given age. The ``Uncertainty'' column reports typical uncertainties that are obtainable by certain types of observation, listed in the corresponding column of the ``Observation Method'' column. The uncertainties obtained from a method obviously depend on the quality of a particular dataset (instrument, observing conditions, quantity of observations, etc.); however, as argued in Section~\ref{subsec:sensitivity_results}, the relative orders of magnitude for observable precision is important for considering any likelihood-based inference.}
\label{tab:constraints}
\end{table}

\subsection{Posterior Sampling with Active Learning} \label{subsec:active_learning}

We further examine the structure of the simulated posterior by sampling the points from (Eqn. \ref{eqn:bayes}) using the priors of Table \ref{tab:ranges}. Systematically sampling points in a $d$-dimensional space at a density high enough to resolve high-posterior modes is a challenging problem especially as $d$ becomes large and forward model evaluations are computationally expensive. In the case of our problem, we sample $d=5$ input parameters (initial $P_{\rm rot,1}$, $P_{\rm rot,2}$, $e$, $P_{\rm orb}$, and \tq). The computation time for our model can take on an order of a few seconds (integrated to $\sim10s$ of Myrs) to $\sim30$ seconds (integrated to $\sim10$ Gyr).

In order to efficiently sample and interpret the structure of our 5-dimensional posterior, we adopt the methodology presented in \cite{kandasamy_query_2017} in which we train a Gaussian process (GP) surrogate model to replicate the true posterior function and apply active learning to iteratively sample points concentrated at high probability.
This methodology is also highly efficient for performing Markov chain Monte Carlo sampling with computationally expensive models \citep{fleming_approxposterior_2018,fleming_xuv_2020,birky_improved_2021}, however in this analysis we just use it to visualize high-probability modes and degeneracies in the posterior. We perform this sampling using the package \code{alabi} \citep[Active Learning for Accelerated Bayesian Inference;][]{birky_alabi}, an open-source Python implementation for training GP surrogate models to sample posteriors for computationally expensive forward models.

We train a Gaussian process defined by a mean and covariance function.
The covariance function $k(x, x')$ measures the degree to which the value of the function at one point $x$ is correlated with the value at another point $x'$. The covariance function is modeled according to a kernel function, a positive semidefinite function of input points $x$ and $x'$, that determines the smoothness and complexity of the GP. For an in-depth review, see \cite{rasmussen_gaussian_2006}.

To emulate the posterior, we use a square exponential kernel,
\begin{equation} \label{eqn:kernel}
    k(x, x') = \exp \left( - \frac{(x - x')^2}{2l^2} \right),
\end{equation}
where $x$ and $x'$ are two given input points, and $l$ is a length scale hyperparameter. We assume that each input dimension has its own scale value. The mean and length scale hyperparameters of the Gaussian process are numerically optimized by the GP's likelihood of the training data using \code{scipy.optimize} \citep{jones_scipy_2001}.

In the first step, we draw a training sample of $N=1000$ evenly sampled training points using the \cite{sobol_global_2001} pseudorandom algorithm. Next, using the procedure of \cite{kandasamy_query_2017} we apply the Bayesian Active Learning for Posterior Estimation (BAPE) algorithm to iteratively sample high posterior regions using the GP surrogate model for $N=2000$ more samples. We demonstrate the resulting posterior samples in the form of a corner plot in the following section. 

\clearpage
\section{Results} \label{sec:results}

\subsection{Model Sensitivity Analysis} \label{subsec:sensitivity_results}

We apply the global sensitivity analysis method described in Section \ref{subsec:sensitivity_methods} to simulations that combine stellar evolution, magnetic braking, and equilibrium tide. We use a quasi-monte carlo sampling scheme \citep{saltelli_variance_2010} to evenly sample $N = 4096$ simulations spanning the input parameter space. The ranges of input parameters we consider are given in Table \ref{tab:ranges}.

\subsubsection{Sensitivity of the orbital period, rotation period, and eccentricity}

\begin{sidewaysfigure}[ht!]
    \centering
    \begin{turn}{180}
    \begin{minipage}{\linewidth}
    \includegraphics[width=\columnwidth]{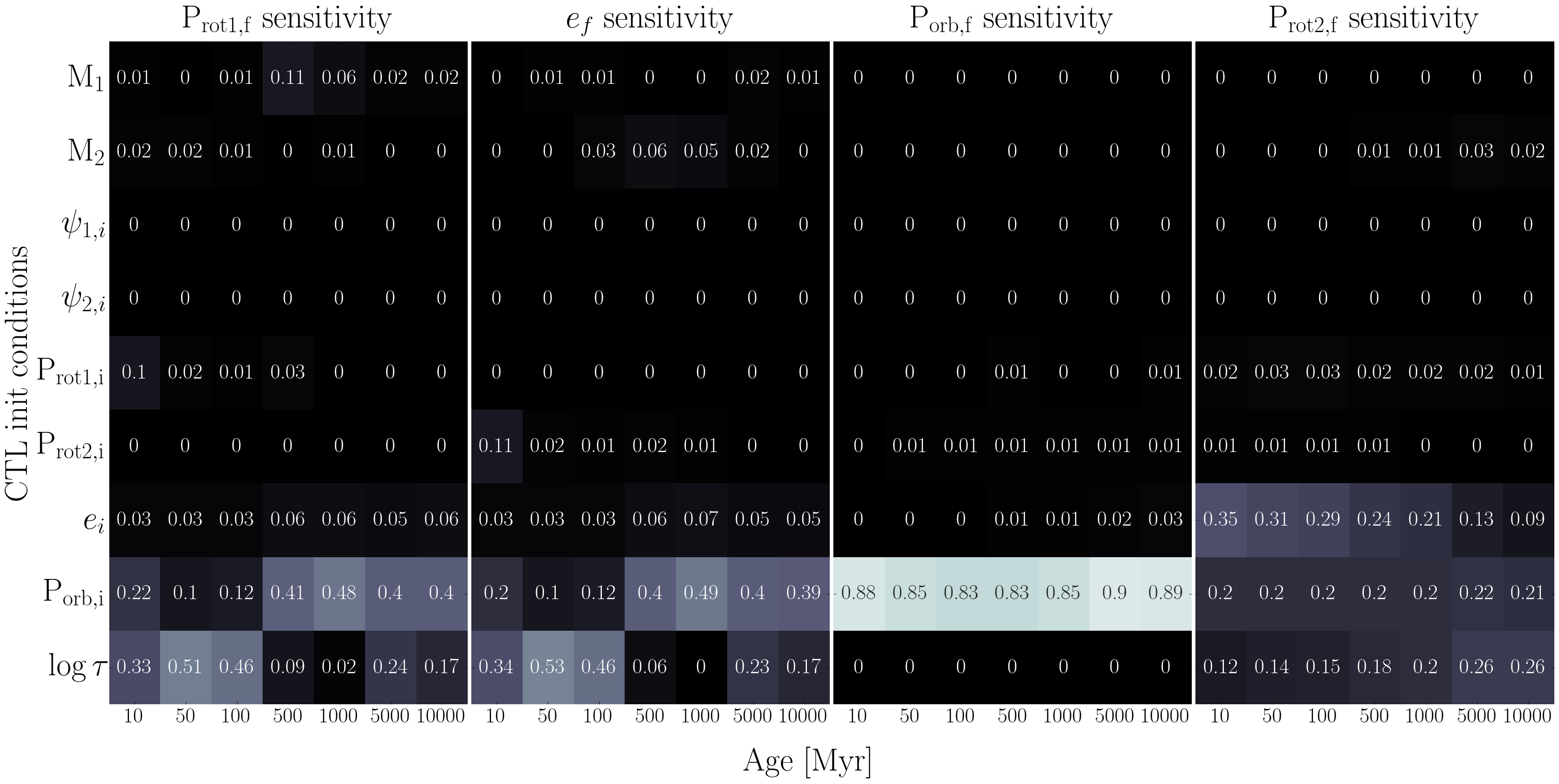}
    \caption{\normalsize Global sensitivity analysis performed on a set of 4096 simulations with varying initial conditions (y-axis) and age (x-axis). Each panel quantifies the sensitivity of an output parameter (final $\prot$, $\porb$, $e$) in terms of the amount of variance each input parameter contributes to the distribution of final parameters.}
    \label{fig:sensitivity_ctl_full}
    \end{minipage}
    \end{turn}
\end{sidewaysfigure}

\begin{sidewaysfigure}[ht!]
    \centering
    \includegraphics[width=\columnwidth]{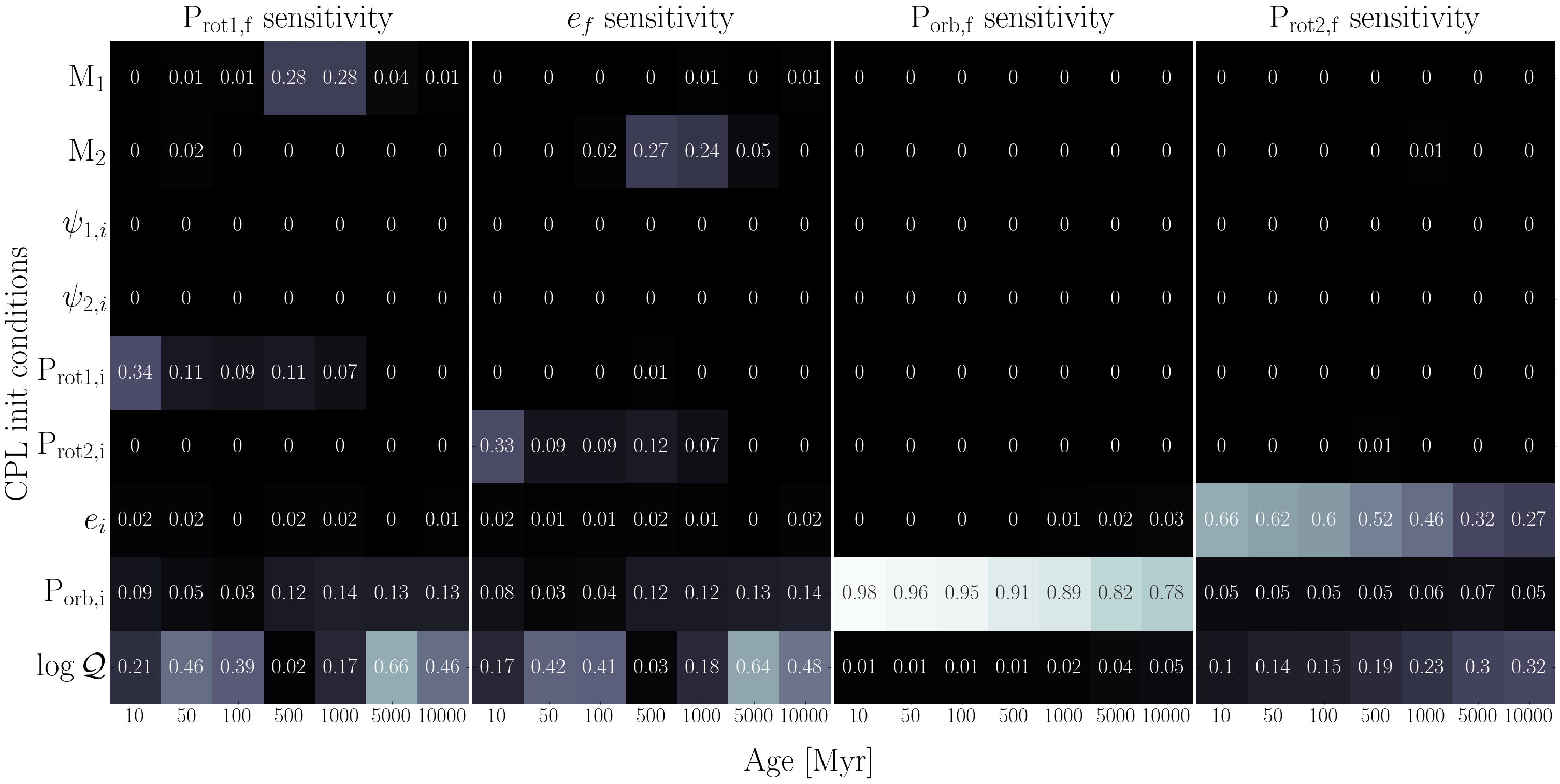}
    \caption{\normalsize Global sensitivity analysis performed on a set of 4096 simulations with varying initial conditions (y-axis) and age of evolution (x-axis). Each panel quantifies the sensitivity an output parameter (final $\prot$, $\porb$, $e$) in terms of the amount of variance each input parameter contributes to the distribution of final parameters.}
    \label{fig:sensitivity_cpl_full}
\end{sidewaysfigure}

Figures \ref{fig:sensitivity_ctl_full} and \ref{fig:sensitivity_cpl_full} visualize the results of the sensitivity analysis applied to the full set of model parameters. Figure \ref{fig:sensitivity_ctl_full} shows the sensitivity analysis applied to the CTL model, and the following Figure \ref{fig:sensitivity_cpl_full} shows the  the CPL model. Within each of the figures, there are three grids displaying the sensitivity for three different model output parameters: final primary rotation period (left grid), final orbital period (center grid), and final orbital eccentricity (right grid). The x-axis for each subpanel shows the model input parameters that were sampled over the ranges listed in Table \ref{tab:ranges}, and the y-axis of each subpanel shows the age to which the set of $N$ simulations evolved (ranging from 10 Myr to 10 Gyr). Each value in the grid (denoted by color and labeled with numerical value) is the first-order sensitivity index (see Section \ref{subsec:sensitivity_methods}) corresponding to a given input parameter at a specific age. Thus, we can interpret which output parameters are most sensitive to which input parameters as a function of age. 

In Figure \ref{fig:sensitivity_ctl_full} our sensitivity analysis shows that the final $\porb$ and $e$  values are dominated by initial orbital conditions for systems across all ages (as shown by the rows with highest variance). 
Picking out the regions of age/initial condition parameter space with the highest variance due to \tq\ (or \ttau), elucidates the observables (final $\prot$, $\porb$, $e$) that are most sensitive to the tidal dissipation.
The analysis in Figures \ref{fig:sensitivity_ctl_full} and \ref{fig:sensitivity_cpl_full} suggest similar results for both the CTL and CPL models. In particular, the primary $\prot$ of young systems $\sim50 - 100$\,Myr, or the eccentricity or secondary $\prot$ of old systems $\sim5 - 10$\,Gyr hold the most promise for constraining tidal \tq\ (or \ttau), in other words, whether or not tides were strong enough to circularize and synchronize most initial conditions by that age.
The results shown in Figures \ref{fig:sensitivity_ctl_full} and \ref{fig:sensitivity_cpl_full} suggest that the final rotation periods tend to be more sensitive to tidal \tq\ than the final orbital period or eccentricity. However, given that rotation periods are difficult to measure precisely from lightcurves ($\sigma\sim0.01$\,d optimistically, Section \ref{subsec:likelihood}), rotation periods generally have less constraining power than orbital periods or eccentricities.

\subsubsection{Effects of observational uncertainties}
\begin{figure}[ht!]
\begin{center}
    \includegraphics[width=0.65\linewidth]{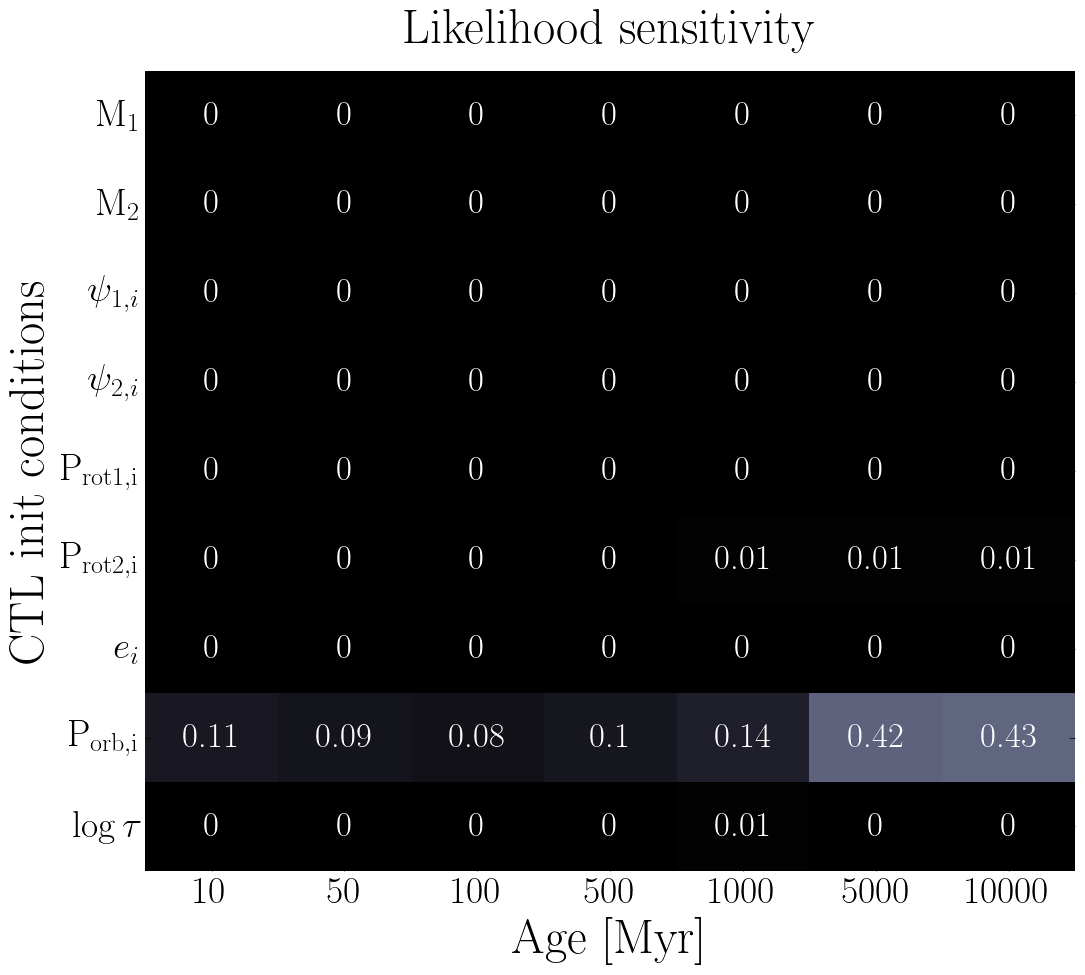}
    \includegraphics[width=0.65\linewidth]{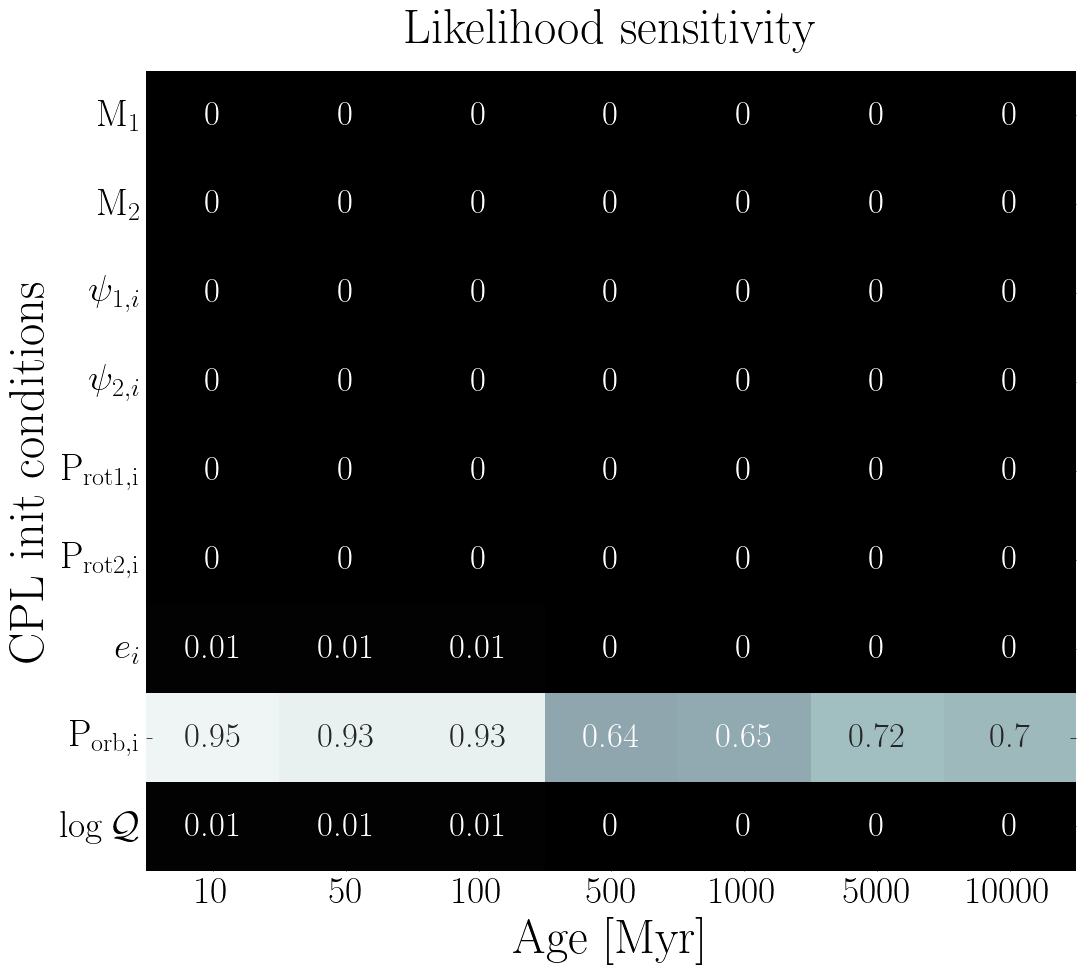}
\vspace{0.5em}
\caption{\normalsize Likelihood sensitivity for the CTL (top) and CPL (bottom) model. The results for both models indicate that model goodness-of-fit (as quantified by a Gaussian likelihood, with uncertainties described in Section \ref{subsec:likelihood}) is most sensitive to the initial orbital period.}
\label{fig:sensitivity_lnlike}
\end{center}
\end{figure}

In addition to looking at the sensitivities of individual parameters, we can also look at the combined constraint of using measured $\prot$, $\porb$, $e$ and their uncertainties.
Figure \ref{fig:sensitivity_lnlike} shows how the combined likelihood (computed at fiducial parameters; see Table \ref{tab:constraints}) is sensitive to each input parameter. The likelihood sensitivity tests how sensitive the model goodness-of-fit would be when combining the fit from all of the model observables and their uncertainties. Figure \ref{fig:sensitivity_lnlike} shows that when uncertainties are taken into account, the initial $\porb$ completely dominates the model goodness of fit across all ages. This implies that inference of tidal \ttau\ or \tq\ is very dependent (and may be systematically biased) by the prior chosen for the initial orbital period, which we explore further in Section \ref{subsec:posterior}. 

\clearpage

\begin{figure}[ht!]
\begin{center}
    \includegraphics[width=0.6\linewidth]{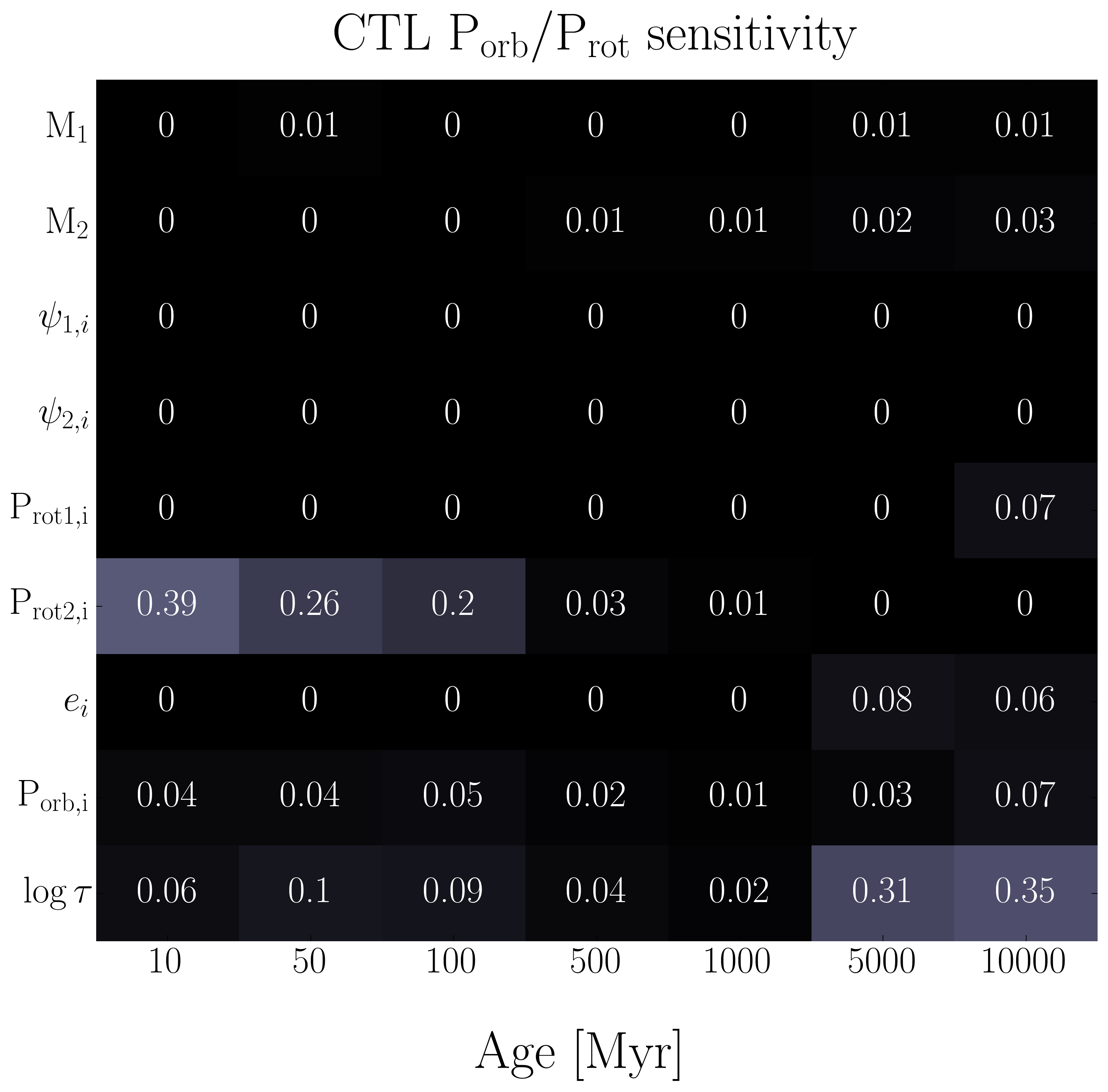}
    \includegraphics[width=0.6\linewidth]{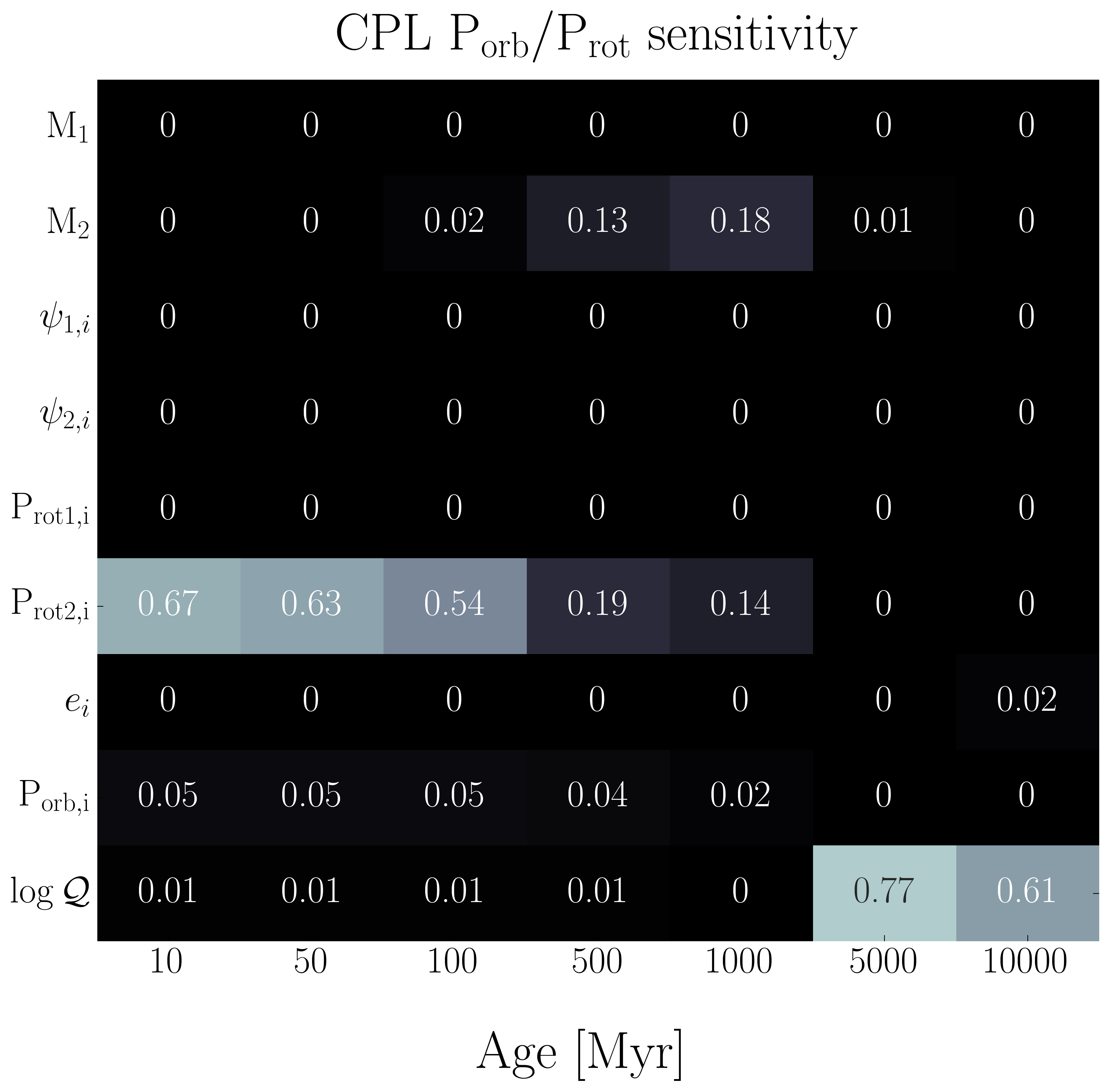}
\vspace{0.5em}
\caption{\normalsize Sensitivity of the period ratio (orbital period / primary rotation period) for the CTL (top) and CPL (bottom) model. At ages of 5-10 Gyrs, \ttau\ and \tq\ dominate the final period ratio $\rm P_{orb} / P_{rot1}$.}
\label{fig:sensitivity_ratio}
\end{center}
\end{figure}
\clearpage

\subsubsection{Sensitivity of the period ratio}

In summary, the final values of individual orbital parameters (Figures \ref{fig:sensitivity_ctl_full}--\ref{fig:sensitivity_cpl_full}) and likelihood constructed from these parameters (Figure \ref{fig:sensitivity_lnlike}) are more dominated by the initial conditions than tidal \ttau\ or \tq. However, if we look at the sensitivity of the ratio $\rm P_{orb}/P_{rot}$ as a function of evolution (Figure \ref{fig:sensitivity_ratio}), we find that the sensitivity is less dependent on initial conditions at older ages. At ages of 5--10 Gyrs, the variance in the orbital period to rotation period ratio is dominated by \ttau\ and \tq\ for the CTL and CPL models respectively. 

\begin{figure}[ht!]
\begin{center}
    \includegraphics[width=0.55\linewidth]{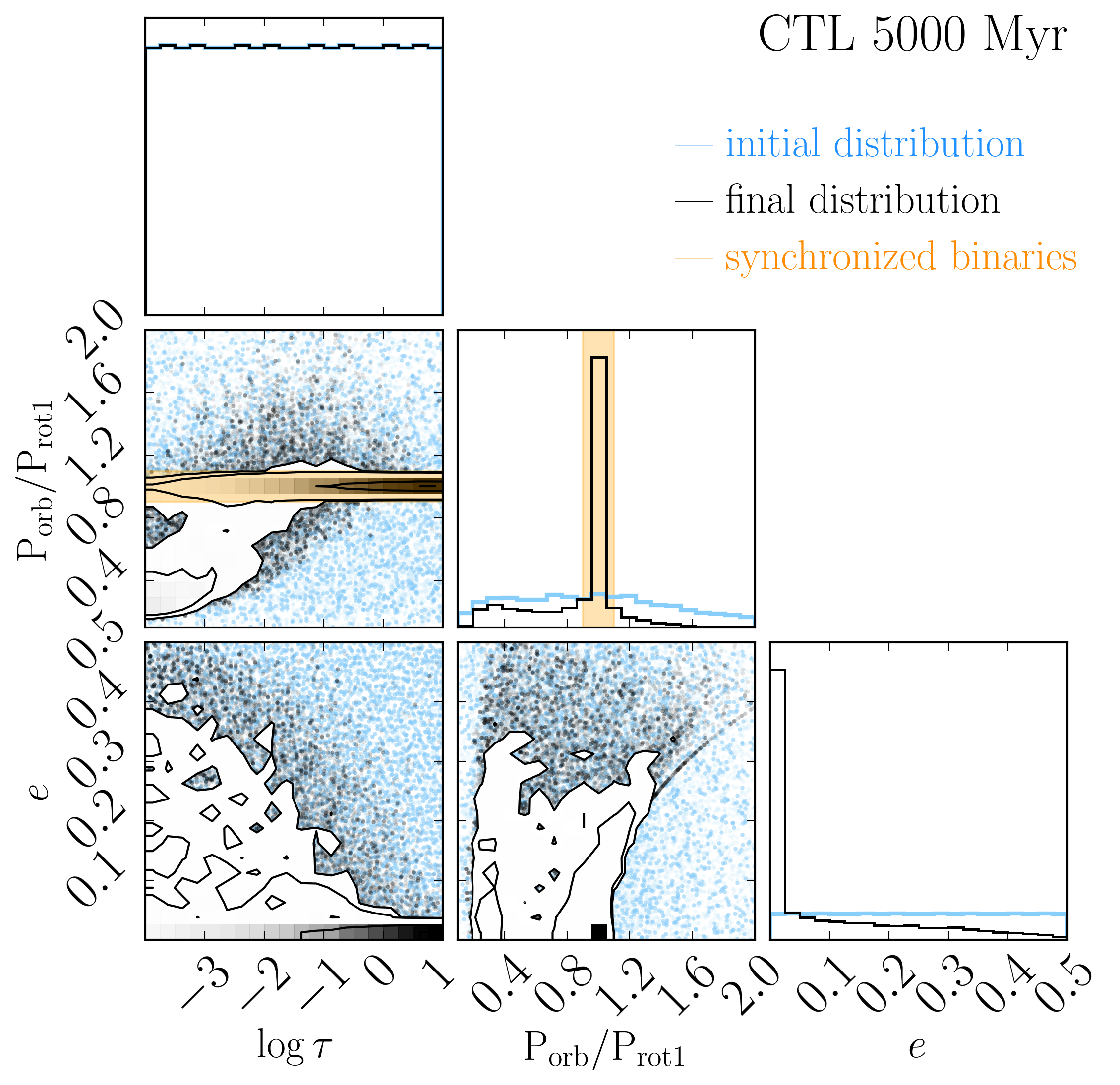}
    \includegraphics[width=0.55\linewidth]{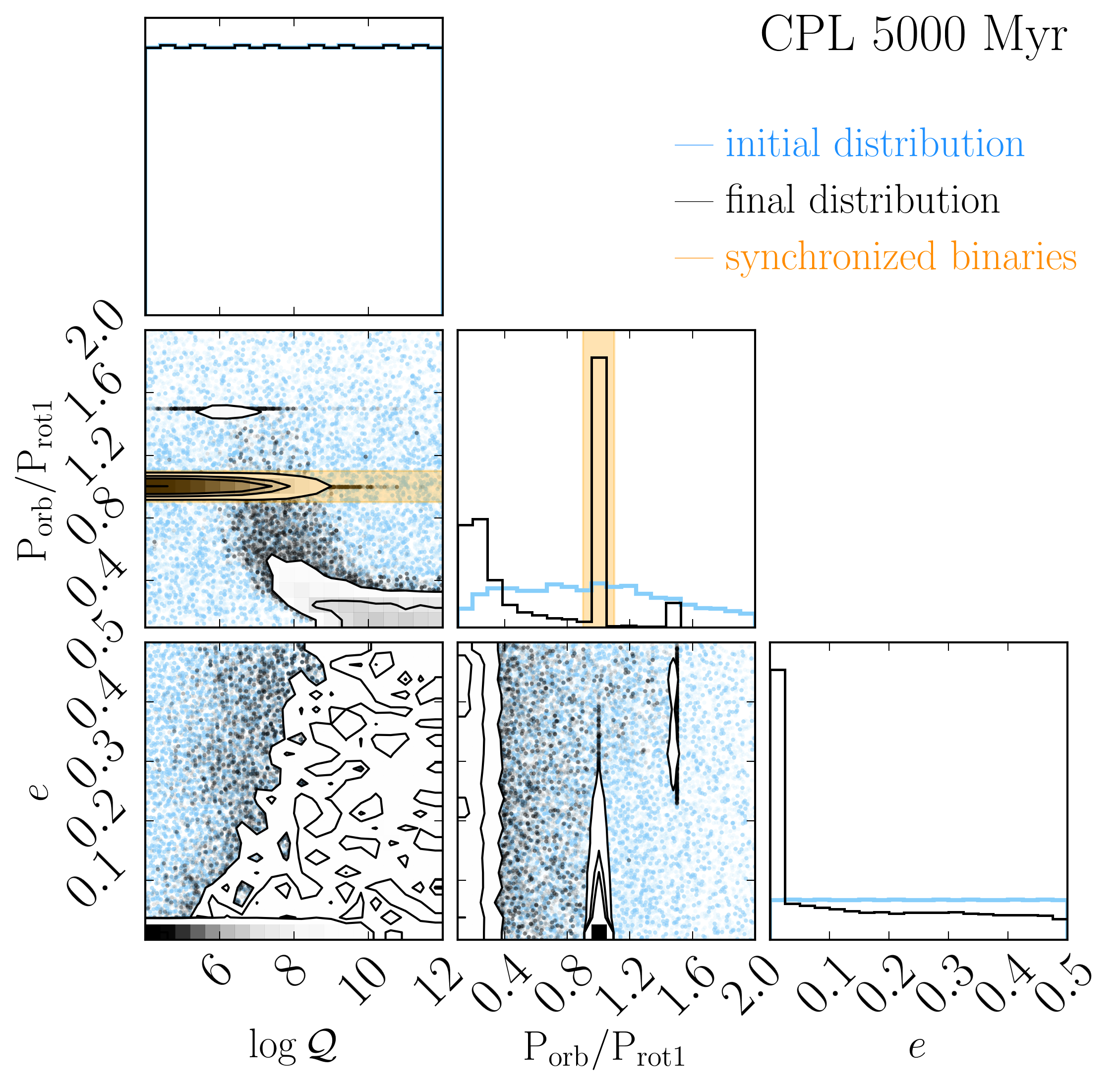}
\caption{\normalsize Distribution of orbital parameters for the CTL model (top) and CPL model (bottom) evolved to an age of 5 Gyr, taken from the same simulations as Figures \ref{fig:sensitivity_ctl_full}--\ref{fig:sensitivity_ratio}. The blue scatterpoints show the initial distribution of parameters (sampled uniformly in eccentricity, orbital period, rotation period, and tidal \ttau\ or \tq), the black contours show the final distribution of parameters after 5 Gyrs of evolution, and the orange range highlights synchronized binaries. At 5 Gyr, there is a clear bimodal separation between systems that are synchronized ($\rm P_{orb}/P_{rot1} \approx 1$), and those that are subsynchronous ($\rm P_{orb}/P_{rot1} < 1$).}
\label{fig:corner_ratio}
\end{center}
\end{figure}

Figure \ref{fig:corner_ratio} examines the $\rm P_{orb}/P_{rot}$ dependence on tidal parameters further. The blue scatterpoints show the initial distribution of parameters (with uniformly sampled initial conditions) and the black contours show the final distribution of parameters evolved to 5 Gyrs (where the period ratio is most dependent on tidal parameters, based on Figure \ref{fig:sensitivity_ratio}). In Figure \ref{fig:corner_ratio}, the 5 Gyr simulations converge towards a bimodal distribution: one cluster with synchronized systems (highlighted in orange), and another cluster of subsynchronous, eccentric binaries. 
For the CTL model, there is a fairly clear separation where simulations with $\log\tau > 0$ are nearly all tidally locked. Similarly for the CPL model, most systems with $\log\mathcal{Q} < 6$ are tidally locked. For the range of weaker tides ($\log\tau < 0$, or $\log\mathcal{Q} > 6$) there is a higher variance in period ratio where some systems are synchronized, but a larger fraction of systems are subsynchronous.

\begin{figure}[ht!]
\begin{center}
    \includegraphics[width=\linewidth]{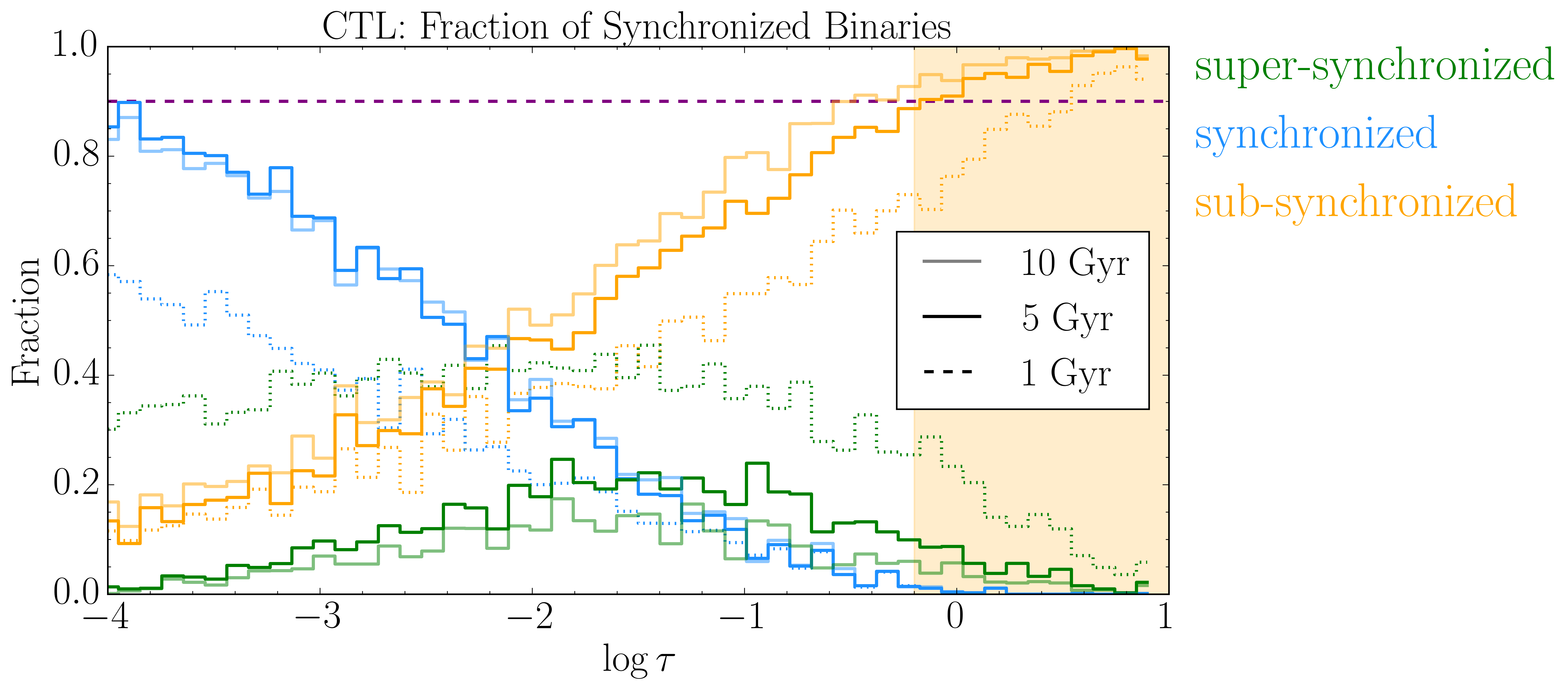}
\caption{\normalsize Fraction of binaries that are synchronized ($\rm P_{orb}/P_{rot1} \approx 1$, orange), subsynchronous ($\rm P_{orb}/P_{rot1} < 1$, blue) and supersynchronous ($\rm P_{orb}/P_{rot1} > 1$, green) according to the CTL model. The dotted lines show the systems evolved to an age of 1 Gyr, the dark solid lines show 5 Gyr, and the light solid lines show 10 Gyr. For each bin of $\log\tau$, we compute relative fractions of synchronous, subsynchronous, and supersynchronous in each sample for a given age. 
The shaded orange region shows hypothetical the lower bound on $\log\mathcal{Q}$ given that a majority of sources ($>90\%$) are observed subsynchronous after 5 Gyrs.}
\vspace{-15pt}
\label{fig:ctl_sync_frac}
\end{center}
\end{figure}

\begin{figure}[ht!]
\begin{center}
    \includegraphics[width=\linewidth]{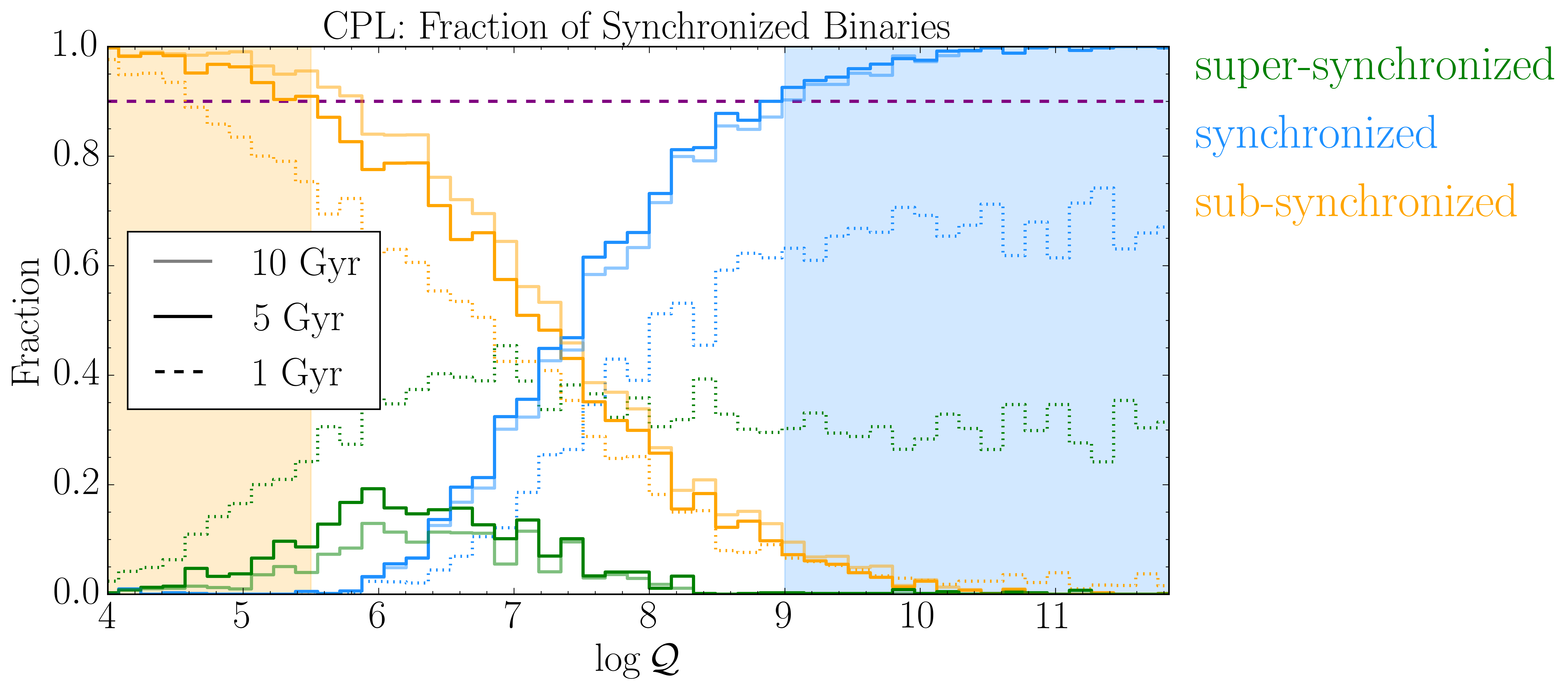}
\caption{\normalsize Same as Figure \ref{fig:ctl_sync_frac}, but for the CPL model. The shaded orange region shows hypothetical the lower bound on $\log\mathcal{Q}$ given that a majority of sources ($>90\%$) are observed subsynchronous after 5 Gyrs. The shaded blue region shows hypothetical the upper bound $\log\mathcal{Q}$ given that a majority of sources ($>90\%$) are subsynchronous after 5 Gyrs.}
\label{fig:cpl_sync_frac}
\end{center}
\end{figure}

Figures \ref{fig:ctl_sync_frac} and \ref{fig:cpl_sync_frac} investigate whether the bimodality between the subsynchronous and synchronous populations is distinguishable as a function of $\log\tau$ and $\log\mathcal{Q}$ respectively. Figure \ref{fig:ctl_sync_frac} shows the histograms of the sample categorized into three different populations: synchronized ($\rm P_{orb}/P_{rot1} \approx 1$, orange), subsynchronous ($\rm P_{orb}/P_{rot1} < 1$, blue) and supersynchronous ($\rm P_{orb}/P_{rot1} > 1$, green), each a three different ages: 1 Gyr, 5 Gyrs, and 10 Gyrs. 

First we can note the imporance of age on the synchronization fraction. The distributions evolved for 5 Gyrs closely overlap with distributions evolved for 10 Gyrs, but diverge with the distributions evolved for 1 Gyr, showing that most systems will have reached their final synchronization state within 5 Gyrs. This means that if we have a population of binaries that are roughly older than 5 Gyrs, we can place limits on $\log\tau$ based on the fraction of systems synchronized, without having to have precise age estimates for each of the individual systems, which is promising, as precise ages are observationally challenging for old systems \cite{soderblom_ages_2010}. 

Next we can look at the importance of $\log\tau$ on the synchronous fraction. Most importantly, we want to understand the limits where $\log\tau$ dominates over the effects of the initial conditions. As shown in Figure \ref{fig:ctl_sync_frac}, after 5 Gyrs, nearly all orbits ($>90\%$ marked by the purple dashed line) with $\log\tau \gtrsim 0$ are expected to become synchronized (shaded orange region), regardless of initial orbital configuration. On the opposite side (not shown on the bounds of the x-axis of the plot), it appears that $\log\tau \lesssim -4$ is roughly the lower limit where we expect the majority ($>90\%$) of systems to be synchronized. Similarly for the CPL model in Figure \ref{fig:cpl_sync_frac}, $\log\mathcal{Q} \lesssim 5.5$ would be the lower limit given a majority fraction of the population are subsynchronized, and $\log\mathcal{Q} \gtrsim 9$ would be the upper limit given a majority are synchronized. This observationally implies that we could measure the fraction of synchronous and subsynchronous systems among populations of old ($>5$ Gyr) systems: if a strong majority of systems are synchronized, then we could determine a lower bound on $\log\tau$ (or upper bound on $\log\mathcal{Q}$), and if a strong majority are subsynchronized, then we could determine an upper bound on $\log\tau$ (or lower bound on $\log\mathcal{Q}$).

We note that an upper or lower bound constraint would still leave several orders of magnitude of uncertainty in the tidal parameters, and would be roughly consistent with the range of constraints currently presented in literature \citep{meibom_observational_2006,jackson_observational_2009,hansen_calibration_2010,PatelPenev22,penev_comprehensive_2022}. 
Gaining more precise constraints (within the middle regions $-4 \lesssim \log\tau \lesssim 0$, or $5.5 \lesssim \mathcal{Q} \lesssim 9$) becomes more difficult, however. Figures \ref{fig:ctl_sync_frac} and \ref{fig:cpl_sync_frac} suggest that the relative fractions of synchronous and subsynchronous systems vary consistently as a function of tidal parameter, and that these fractions could be compared to the fractions measured from observations to constrain the tidal parameters more precisely. These fractions are the result of using uniform priors for the initial conditions (Table \ref{tab:ranges}). Further work would need to explore how different (possibly more realistic) distributions of initial conditions might change the synchronization fraction as a function of tidal parameter, and very careful observational work would need to be done obtain an unbiased sample with measured orbital periods and rotational periods.

\clearpage

\subsection{Simulated Posteriors} \label{subsec:posterior}

Section \ref{subsec:sensitivity_results} shows that, to first order, constraining tidal \tq\ or \ttau\ is not very promising: the final orbital states are more sensitive to the initial orbital and rotational states than they are to the effects of the tidal dissipation efficiency. However, looking at Figures \ref{fig:sensitivity_ctl_full} and \ref{fig:sensitivity_cpl_full}, we do see that tidal \tq\ or \ttau\ accounts for a significant amount of variance in the final rotation periods and eccentricities (up to $\sim0.5$ at some ages). For the sake of completeness, we further examine the covariance structure based on simulated posteriors and attempt to rigorously test whether informative constraints on \tq\ or \ttau\ could be teased out from higher-order effects in the correlations between variables.

\begin{figure}[ht!]
\begin{center}
    \includegraphics[width=\linewidth]{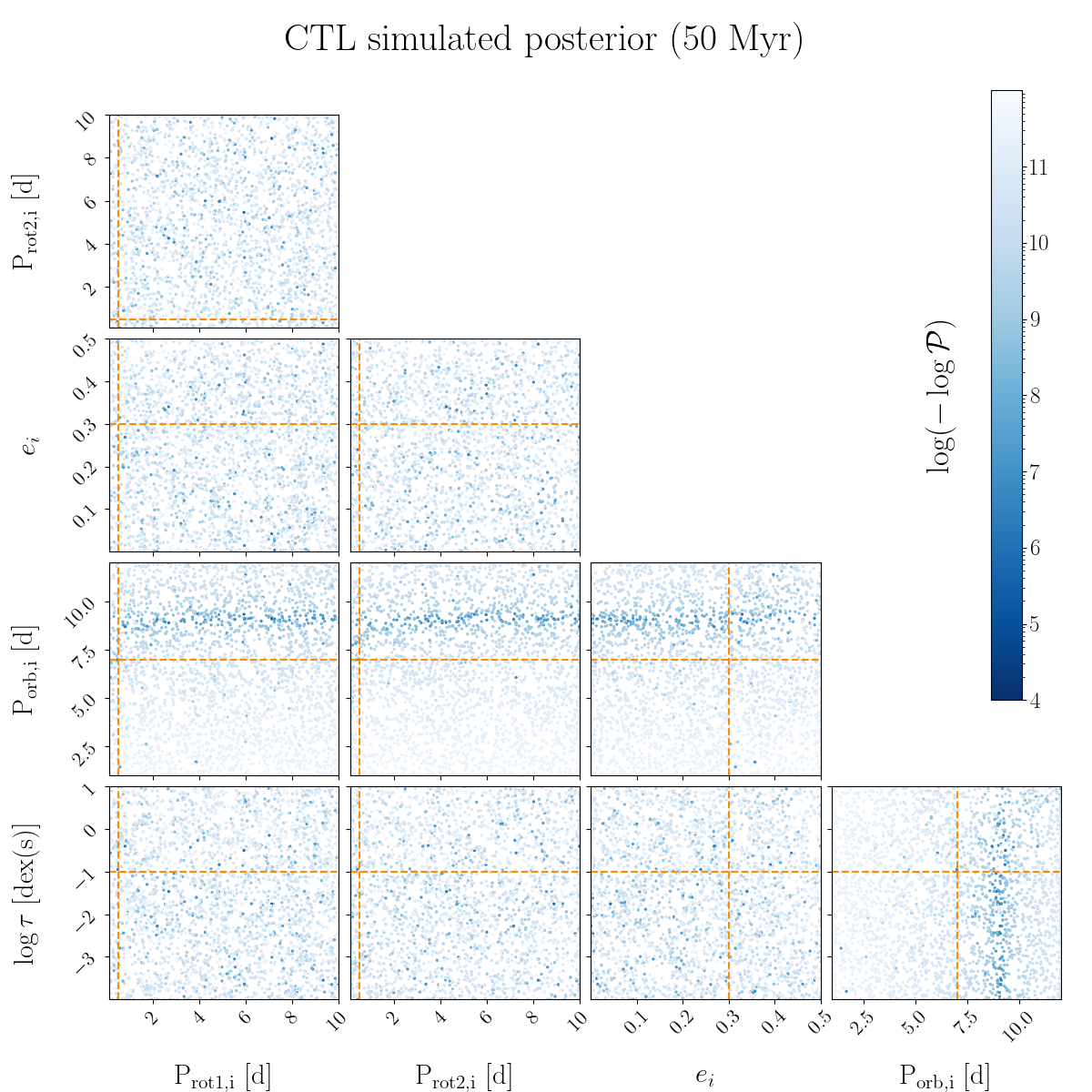}
\vspace{0.5em}
\caption{\normalsize Samples from the simulated posterior for the CTL model at an early age of 50 Myr (blue points). The posterior function is sampled using a Gaussian process and active learning \citep{kandasamy_query_2017} using the package \code{alabi} \citep{birky_alabi} in order to visualize the structure of high posterior modes/degeneracies. Points are colored according to their posterior values. Since the posterior spans an order of magnitude range, we present these values as $\hat{P} = \log(-\log\mathcal{P})$, where lower values of $\hat{P}$ (darker blue) represent higher posterior probability. Orange dashed lines mark the true values of the initial parameters.
}
\label{fig:posterior_ctl_5myr}
\end{center}
\end{figure}

\begin{figure}[ht!]
\begin{center}
    \includegraphics[width=\linewidth]{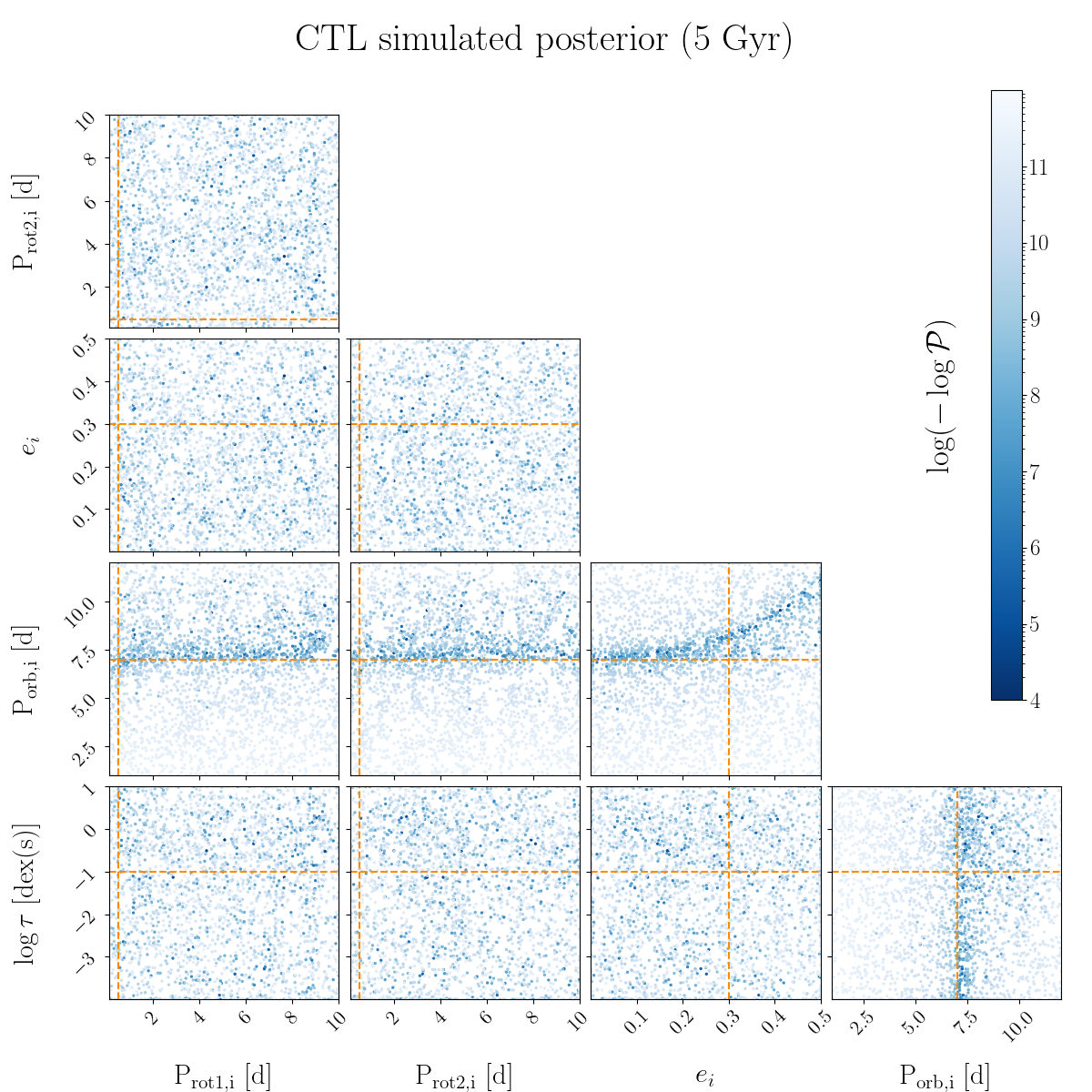}
\vspace{0.5em}
\caption{\normalsize Same as Figure \ref{fig:posterior_ctl_5myr}, but for 5 Gyr.
}
\label{fig:posterior_ctl_5gyr}
\end{center}
\end{figure}

\begin{figure}[ht!]
\begin{center}
    \includegraphics[width=\linewidth]{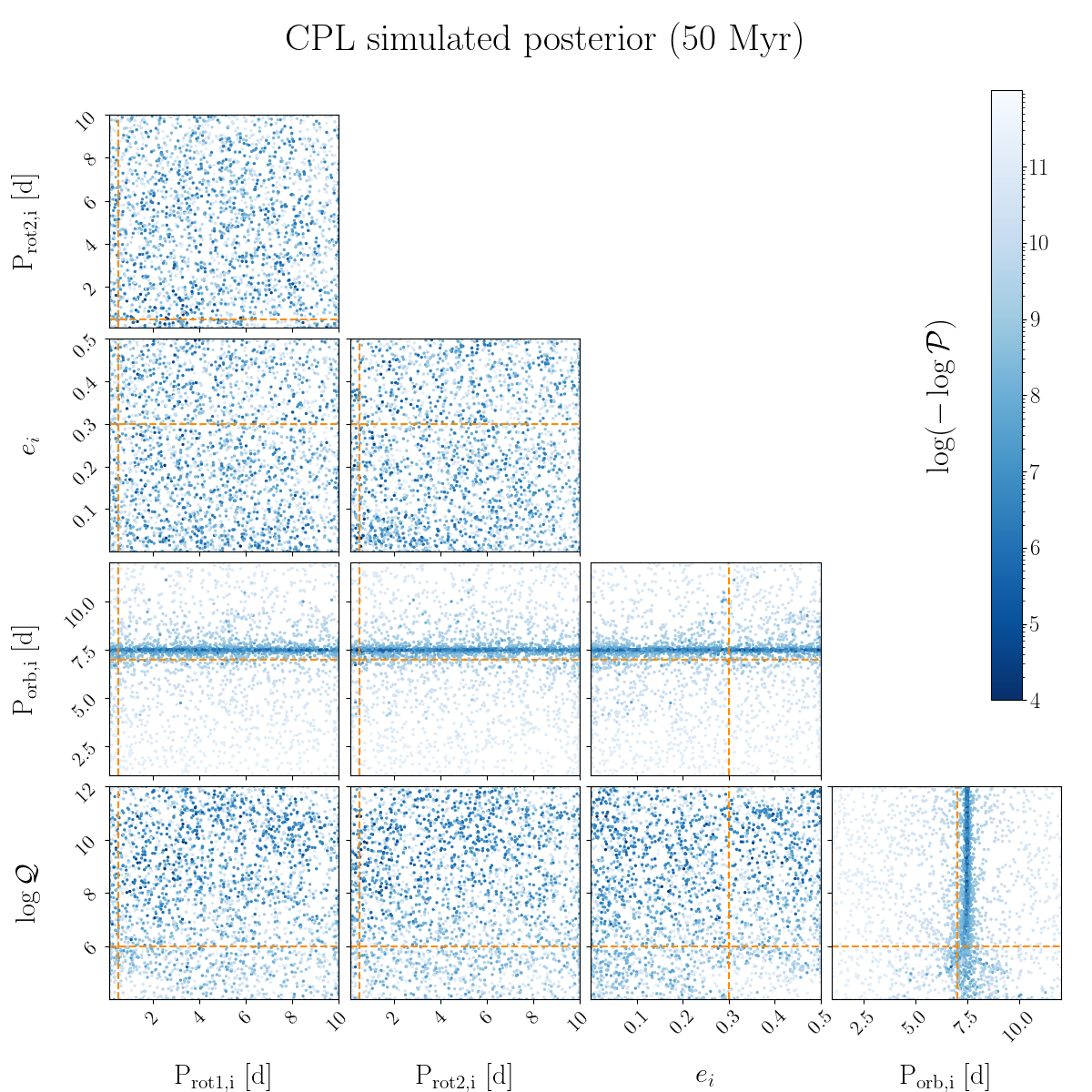}
\vspace{0.5em}
\caption{\normalsize Same as Figure \ref{fig:posterior_ctl_5myr}, but for the CPL model.
}
\label{fig:posterior_cpl_5myr}
\end{center}
\end{figure}

\begin{figure}[ht!]
\begin{center}
    \includegraphics[width=\linewidth]{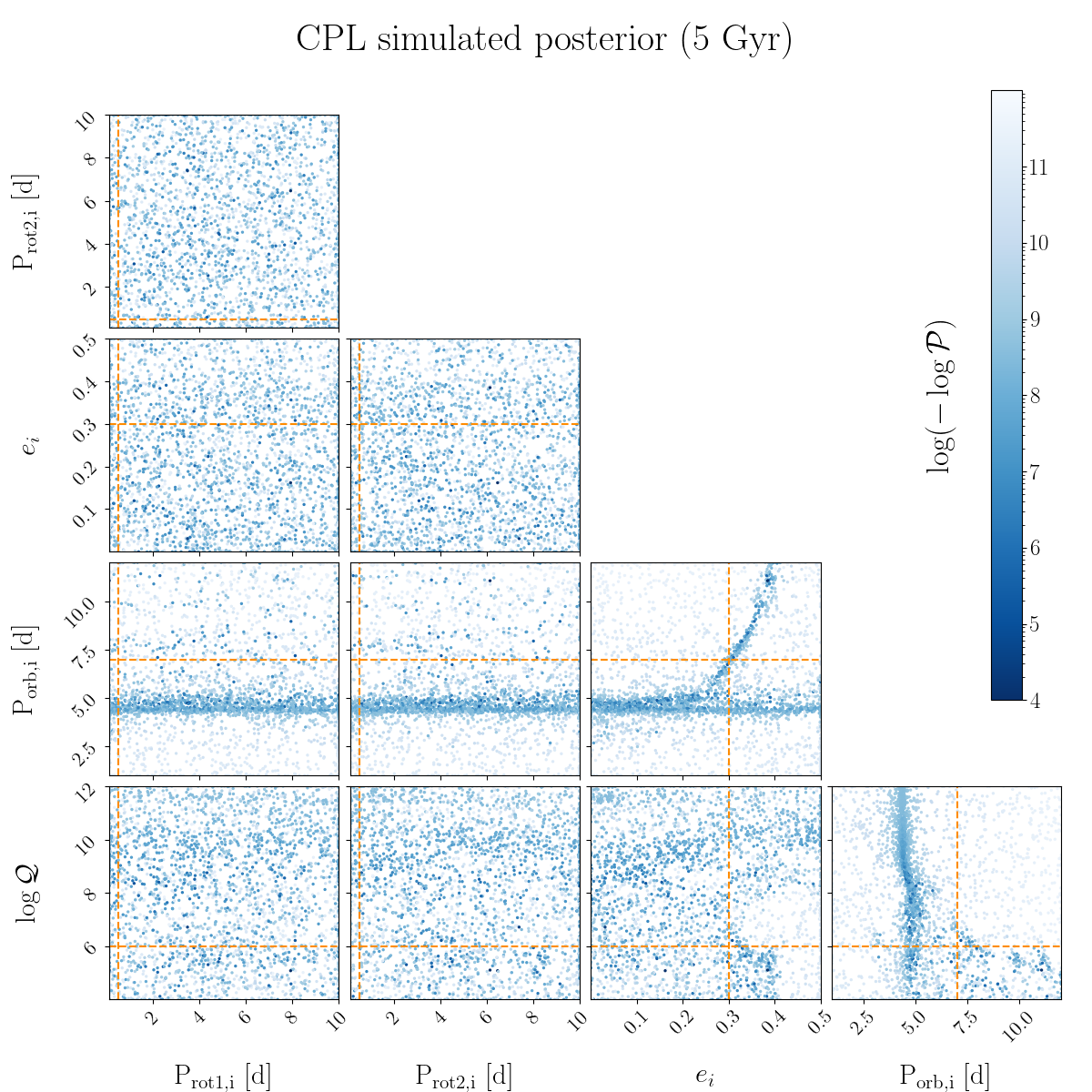}
\vspace{0.5em}
\caption{\normalsize Same as Figure \ref{fig:posterior_cpl_5myr}, but for 5 Gyr.
}
\label{fig:posterior_cpl_5gyr}
\end{center}
\end{figure}

We generate 5-dimensional posteriors according to the prescription described in Section \ref{subsec:likelihood}. In this test we allow for uninformed priors for five parameters (the initial states $P_{\rm rot1,i}$, $P_{\rm rot2,i}$, $e_i$, $P_{\rm orb,i}$, and \tq) given that the initial states of a system are generally unknown. To isolate the effects of the other input parameters of the model, we fix the masses, initial obliquities, and ages to the true value. We performed a total of four simulated posterior tests encompassing both tidal models at young and old ages. These tests include: CTL at 50 Myr (Figure \ref{fig:posterior_ctl_5myr}), CTL at 5 Gyr (Figure \ref{fig:posterior_ctl_5gyr}), CPL at 50 Myr (Figure \ref{fig:posterior_cpl_5myr}), and CTL at 5 Gyr (Figure \ref{fig:posterior_cpl_5gyr}). We visualize the posteriors in the form of a corner plot, in which samples are colored according to their posterior probability, where darker blue represents higher probability values.

In simulations at young ages, the results are qualitatively similar for both CTL and CPL (Figures \ref{fig:posterior_ctl_5myr} and \ref{fig:posterior_cpl_5myr} respectively). Here, we see that the high posterior regions most strongly depend on the initial orbital period, which is consistent with the sensitivity analysis results. We also see that the high posterior regions are nearly flat in the previous range of \tq\ (or \ttau). The reason is that at only a few Myrs of evolution, we expect hardly any evolution, even for stronger tides. Given that orbital periods are generally measured to the best level of uncertainty (Table \ref{tab:constraints}), we find that sets of values that match around $P_{\rm orb,i} \approx P_{\rm orb,f}$ with weak values of \tq\ generally tend to maximize the posterior. Our results suggest that we may expect young systems to provide a lower bound on \tq\ (or an upper bound on \ttau). However even with optimistic uncertainties, this degeneracy between $P_{\rm orb,i}$ and \tq\ is poorly traced out. 

\begin{figure}[ht!]
\begin{center}
    \includegraphics[width=\linewidth]{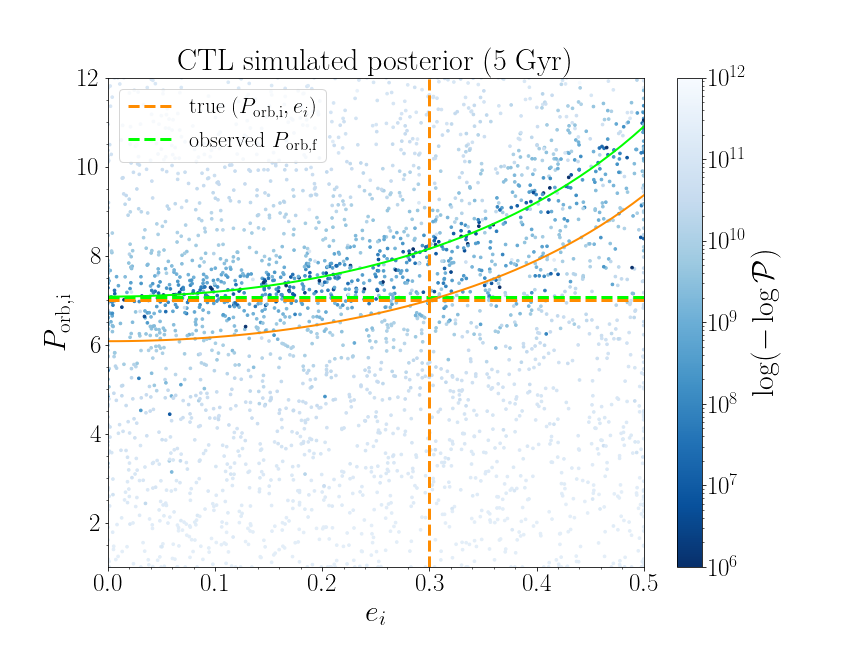}
\caption{\normalsize Samples from the simulated posterior for the CTL model at a late age of 5 Gyr (from Figure \ref{fig:posterior_ctl_5gyr}) zoomed in on the marginal distribution of initial orbital period and eccentricity. The dashed orange line shows the true initial orbital period that would be unknown, and the green dashed line shows the final orbital period that would be observed. The solid lines show the curve of $P_{\rm orb,i}$ and $e_i$ for a constant value of orbital angular momentum. The orange solid curve shows the initial orbital angular momentum ($J_{\rm orb,i} = 1.80 \times 10^{45} \, \mathrm{kg \cdot m^{-2}}$), and the green solid line shows the final orbital angular momentum ($J_{\rm orb,f} = 1.61 \times 10^{45} \, \mathrm{kg \cdot m^{-2}}$). Instead of converging to the true initial orbital period, the posterior is biased towards picking solutions with $P_{\rm orb,i} \approx P_{\rm orb,f}$ (as seen by the overlapping horizontal dashed lines). Furthermore, we see that the high posterior samples lie along the curve of final orbital angular momentum, meaning that the posterior is more informative of the final conditions than it is of the initial conditions that we are trying to infer.
}
\label{fig:posterior_ctl_5gyr_zoom}
\end{center}
\end{figure}

The overfitting for initial orbital period is a trend that occurs not only for young ages, but persists for inference at old ages as well. The results for the CTL and CPL model for a system of 5 Gyr age are shown in Figures \ref{fig:posterior_ctl_5gyr} and \ref{fig:posterior_cpl_5gyr}, respectively. At higher ages we see two degeneracy features emerge between the initial orbital period and initial eccentricity for both CTL and CPL models. To show this structure in more detail, Figure \ref{fig:posterior_ctl_5gyr_zoom} shows a zoom-in for the marginal posterior of ($P_{\rm orb,i}$, $e_i$) for the CPL model. Here, we see that two main degeneracy features appear in the posterior points sampled. One of the features is similar to the young age posterior, which is flat across eccentricity with $P_{\rm orb,i} \approx P_{\rm orb,f}$, and corresponds to weak tidal dissipation (high \tq\ or low \ttau). The other main feature shows a curved $P_{\rm orb,i}-e_i$ degeneracy, which corresponds to stronger tidal dissipation (low \tq\ or high \ttau). The curved degeneracy intersects with the true initial orbital period and eccentricity. However, the true values are statistically indistinguishable from other samples along the degeneracy.

\clearpage


\clearpage

\clearpage

\clearpage

\clearpage

\clearpage

\begin{figure}[ht!]
\begin{center}
    \includegraphics[width=\linewidth]{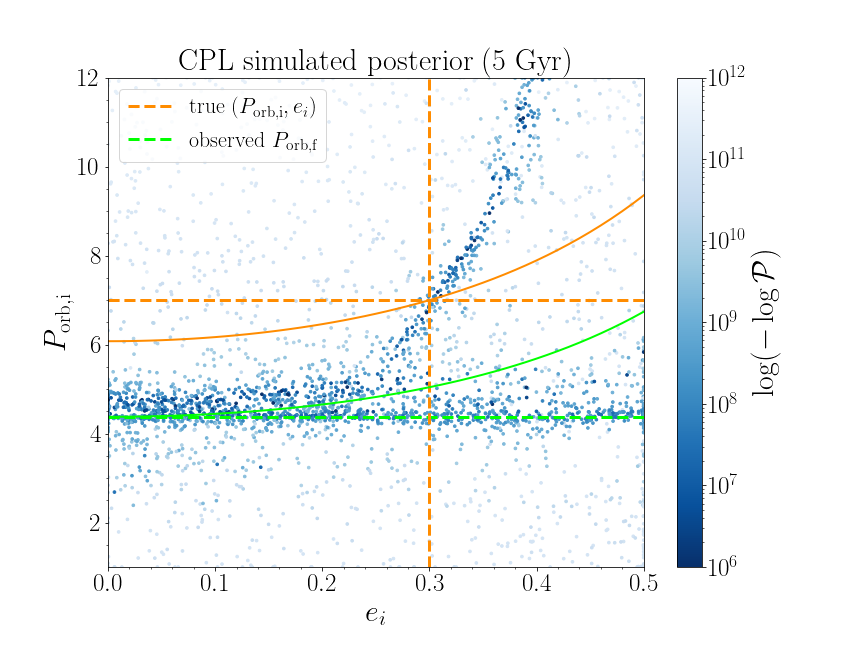}
\caption{\normalsize Same as Figure \ref{fig:posterior_ctl_5gyr_zoom}, but for the CPL model. Two degeneracy features emerge from the blue sampled posterior points: those flat across eccentricity with $P_{\rm orb,i} \approx P_{\rm orb,f}$ (with negligible values of tidal dissipation), and those with curved $P_{\rm orb,i}-e_i$ degeneracy (with stronger values of tidal dissipation).}
\label{fig:posterior_cpl_5gyr_zoom}
\end{center}
\end{figure}

\clearpage

\subsection{Degeneracies in the most simplified simulated model} \label{subsec:1d_likelihood}

Finally to test the limitations of inference, we perform a likelihood recovery test by considering the most optimistic inference case possible: 1 free parameter (\tq\ or \ttau), using simulated data with ideal uncertainties, see Table \ref{tab:constraints}. This test is set up as follows: we run a simulation with fiducial initial values to compute a set of final values. This procedure is very similar to Section \ref{subsec:posterior}, but instead of five free parameters, we only vary one. The values of the fixed parameters are given in Table \ref{tab:fiducial1d}.

\begin{table}[ht!]
\begin{tabular}{|r|c|c|c|c|c|}
\hline
Parameter & $M_1$, $M_2$ & $\psi_{1i}$, $\psi_{2i}$ & $P_{\rm rot1,i}$, $P_{\rm rot2,i}$ & $P_{\rm orb,i}$ & $e_i$ \\
\hline
Unit & $M_{\odot}$ & deg & days & days & \\
\hline
Fiducial value & 1.0 & 0 & 0.5 & 7.0 & 0.15 \\
\hline
\end{tabular}
\vspace{1em}
\caption{\normalsize Values of input parameters used in the 1-dimensional inference test. The subscript $i$ an initial value of a time-varying parameter.}
\label{tab:fiducial1d}
\end{table}

\begin{figure}[ht!]
\begin{center}
    \includegraphics[width=.49\linewidth]{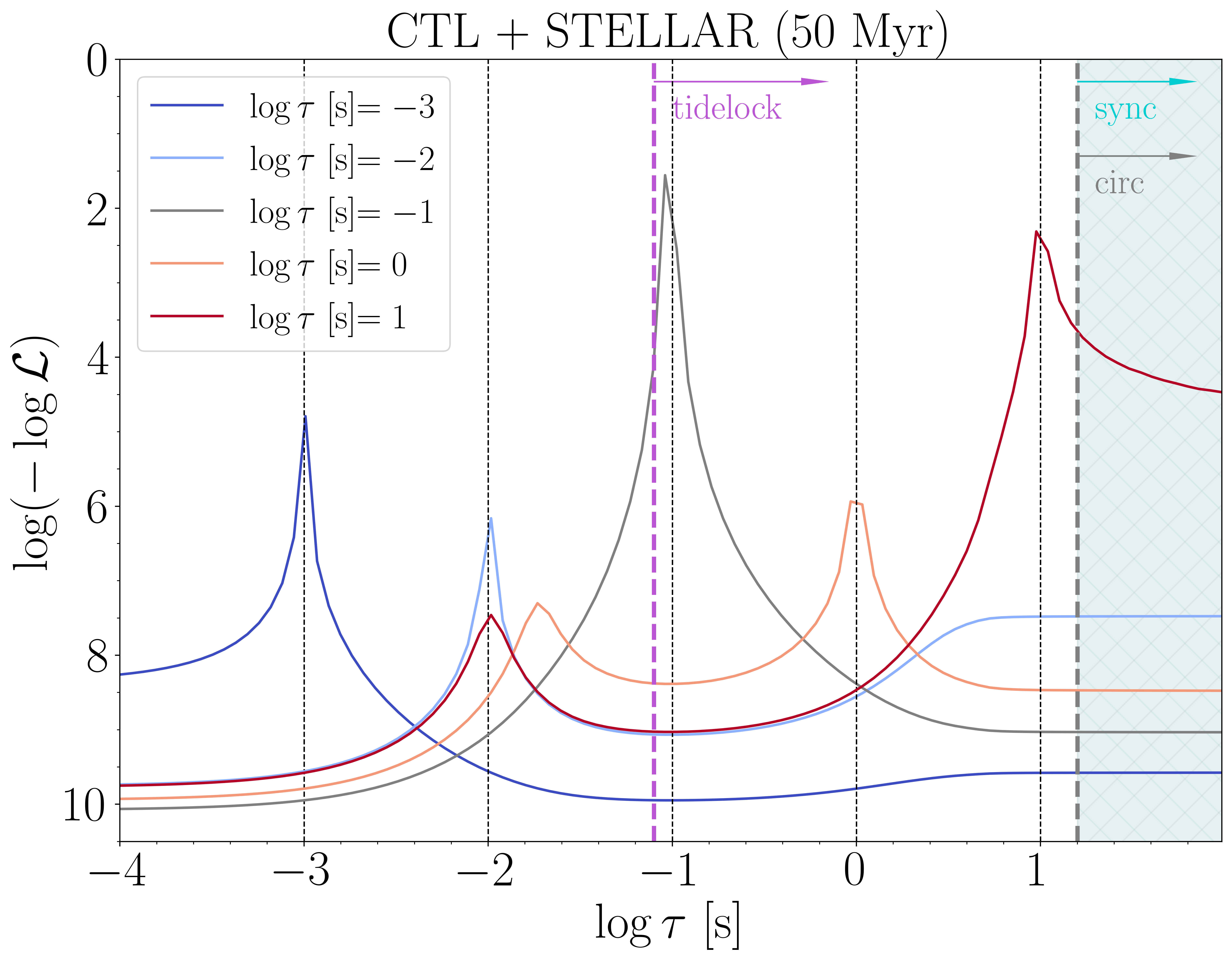} 
    \includegraphics[width=.49\linewidth]{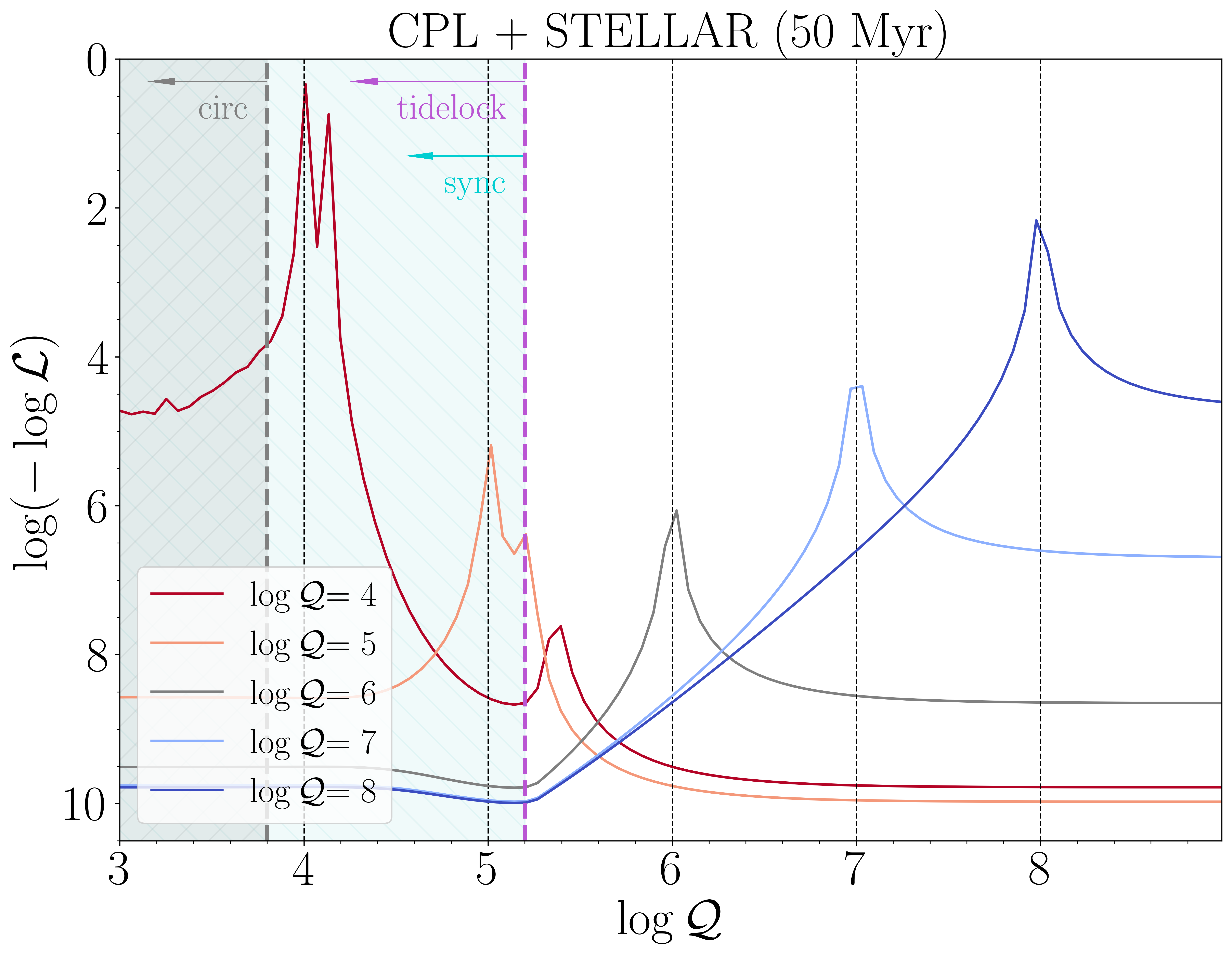} 
    \vspace{0.5em}
    \caption{\normalsize Simulated 1-dimensional likelihood for CTL model (left) and CPL model (right) evaluated at an age of 50 Myrs. Each of the colored lines shows the likelihood value as a function of varying \ttau\ or \tq\ for different ``true'' values ($-3, -2, -1, 0, +1$) for \ttau, and ($4, 5, 6, 7, 8$) for \tq. For both plots, blue lines represent weaker tides and red lines represent stronger tides. The y-axis shows the likelihood value when \ttau\ is allowed to vary and is compared to the fiducial simulation (with the values of Table \ref{tab:fiducial1d}, and uncertainties of Table \ref{tab:constraints}). On the y-axis we compress the magnitude range for visualization by showing the likelihood as $\log(-\log\mathcal{L})$, where lower values of $\log(-\log\mathcal{L})$ (higher on the y-axis) represent higher likelihood values (similarly to Figures \ref{fig:posterior_ctl_5myr}--\ref{fig:posterior_cpl_5gyr}). The magenta dashed line and arrows indicate the range of tidal values that are strong enough to cause tidal locking by an age of 50 Myr. Similarly, the cyan arrows and shaded regions show the range of tidal values that are strong enough to cause synchronization, and the grey arrows and shaded regions show the range of tidal values that are strong enough to cause circularization. } 
    \label{fig:degeneracy1d}
\end{center}
\end{figure}

\begin{figure}[ht!]
\begin{center}
    \includegraphics[width=.49\linewidth]{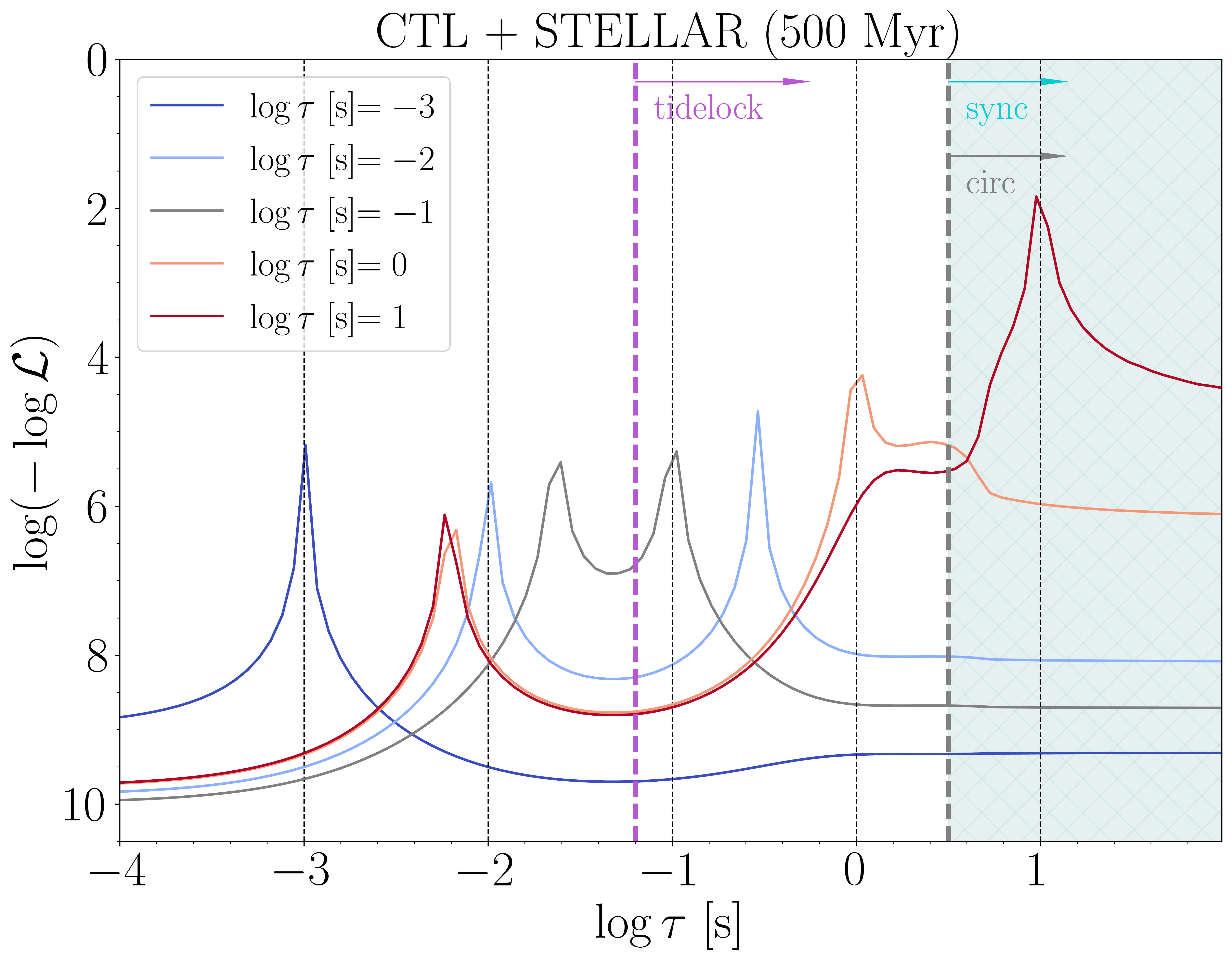} 
    \includegraphics[width=.49\linewidth]{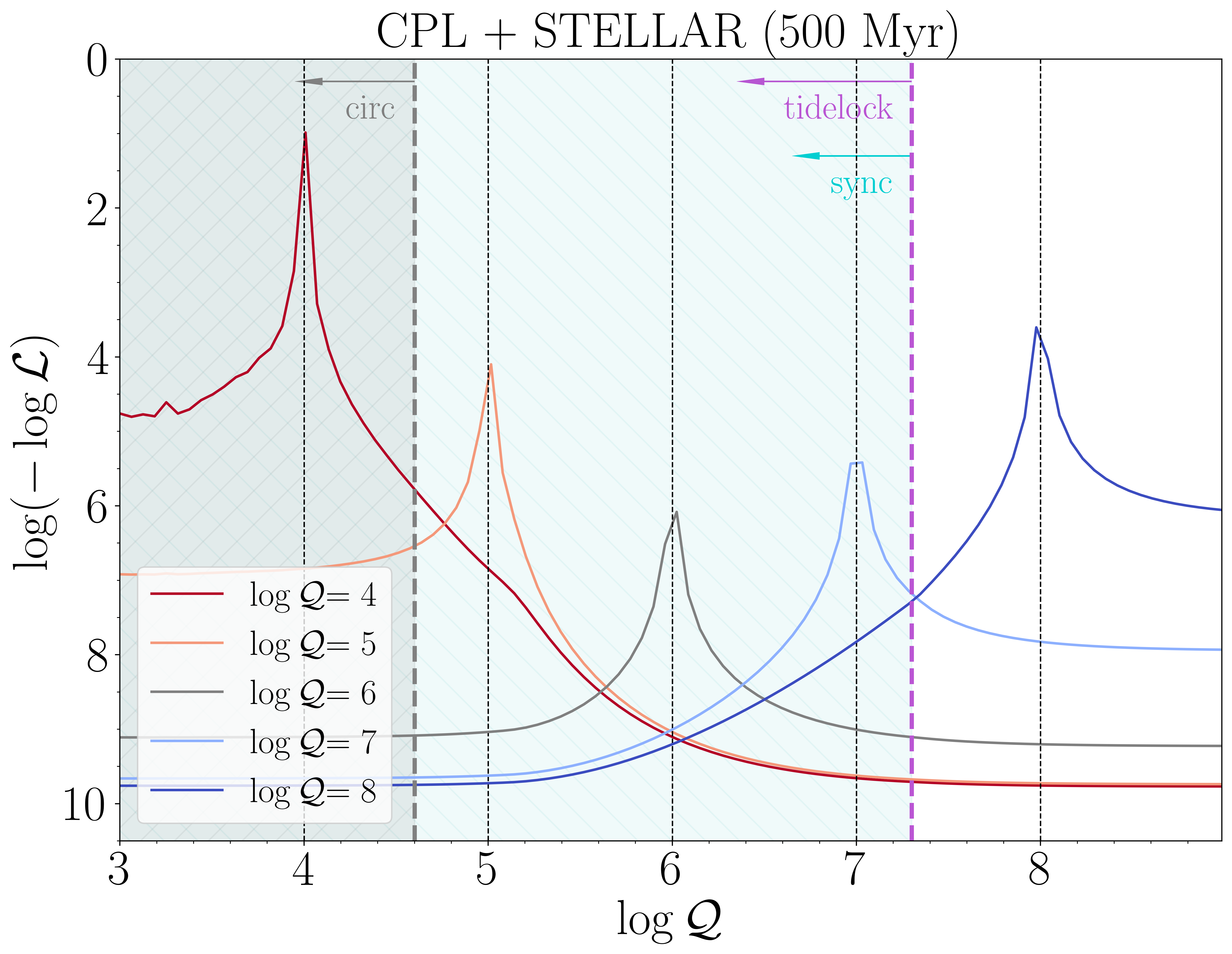} 
    \vspace{0.5em}
    \caption{\normalsize Same as Figure \ref{fig:degeneracy1d}, but for ages 500 Myr.} 
    \label{fig:degeneracy1d_2}
\end{center}
\end{figure}

\begin{figure}[ht!]
\begin{center}
    \includegraphics[width=.49\linewidth]{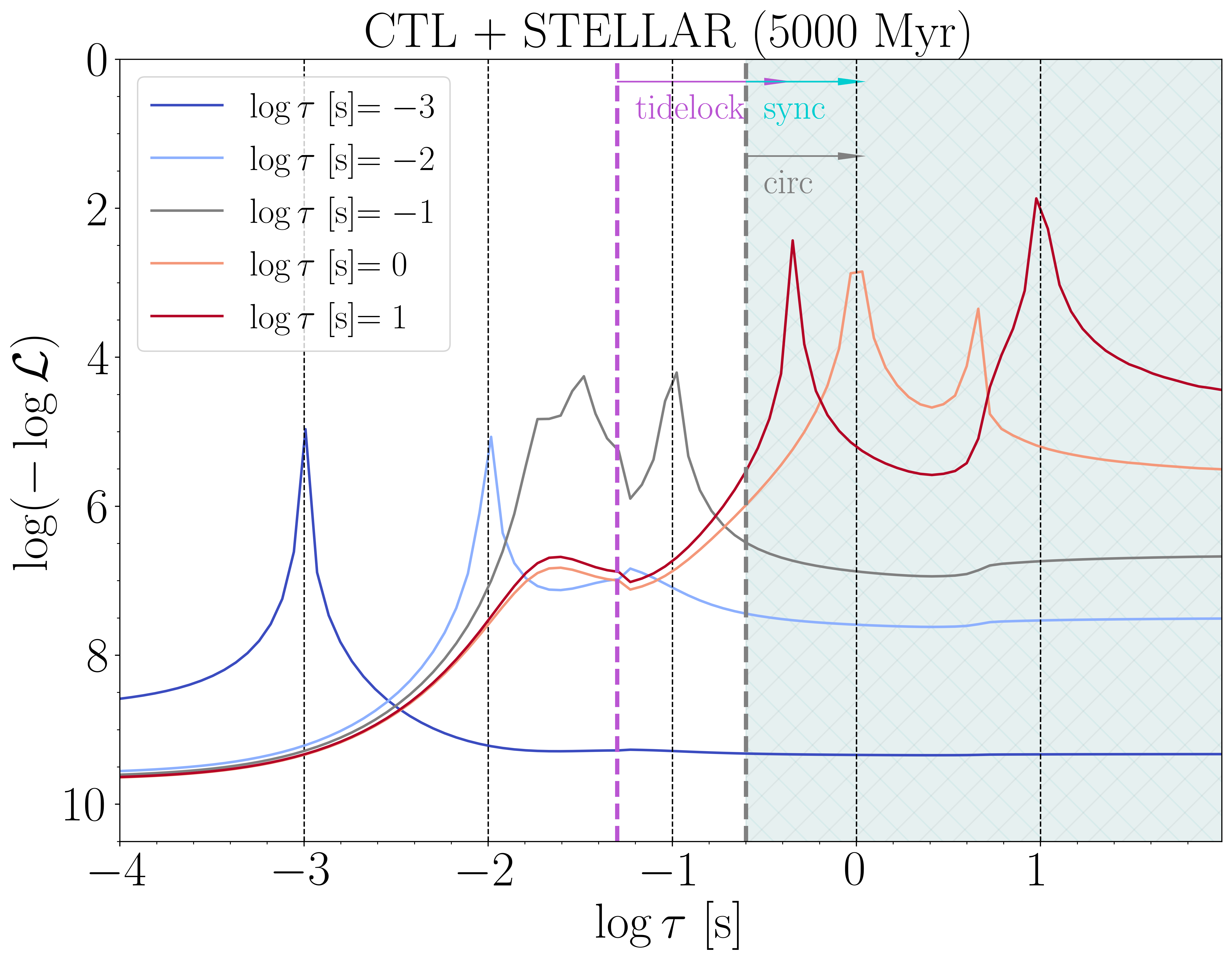} 
    \includegraphics[width=.49\linewidth]{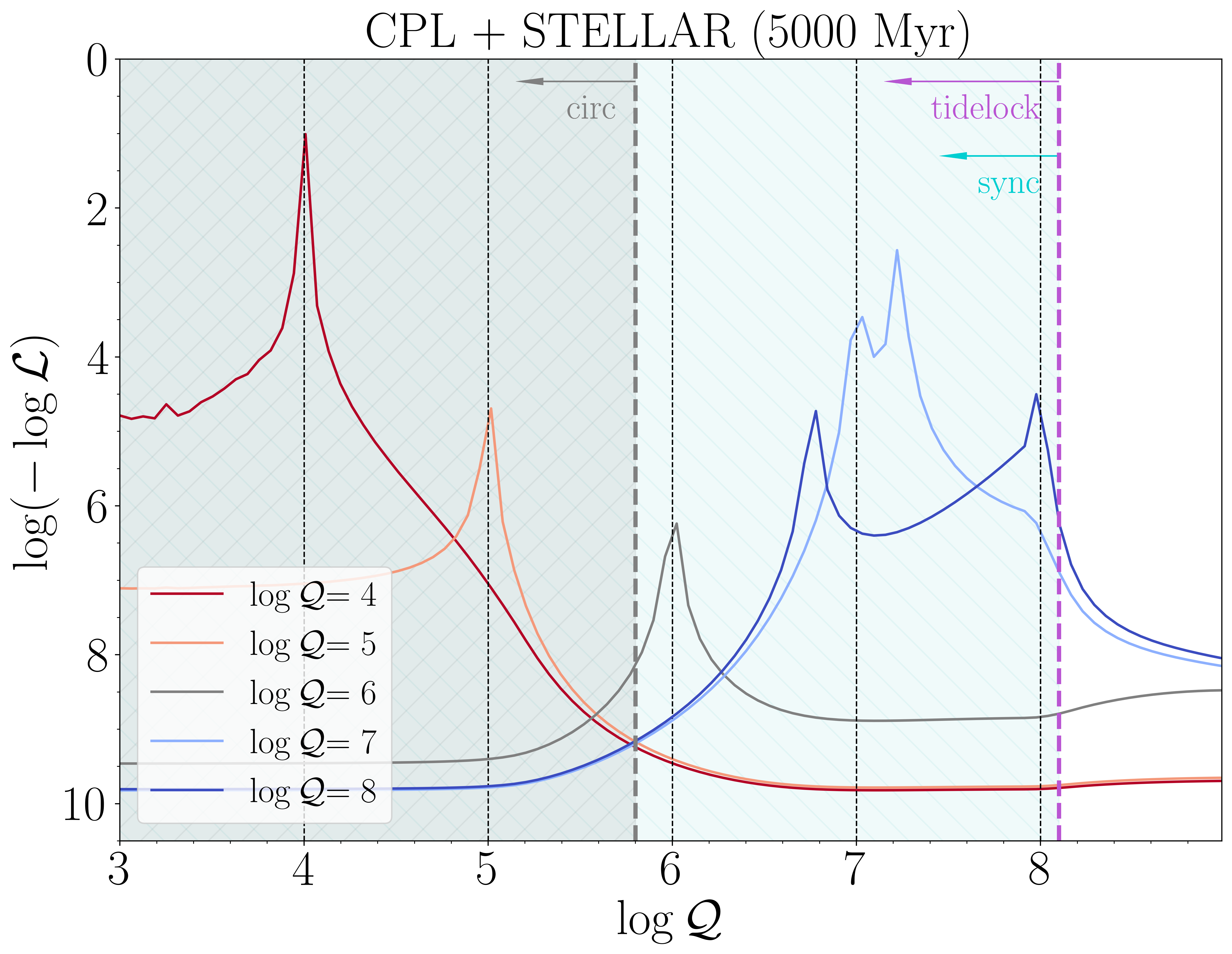} 
    \vspace{0.5em}
    \caption{\normalsize Same as Figure \ref{fig:degeneracy1d}, but for 5 Gyr.} 
    \label{fig:degeneracy1d_3}
\end{center}
\end{figure}

In Figures \ref{fig:degeneracy1d}--\ref{fig:degeneracy1d_3} we show the simulated 1-dimensional likelihood for the CTL and CPL models at three different ages: 50 Myr, 500 Myr, and 5 Gyr. The CTL model is shown on the left panels, and the CPL model is shown on the right panels. The true values of $\log\tau$ ($-3, -2, -1, 0, +1$) and $\log\mathcal{Q}$ ($4, 5, 6, 7, 8$) are indicated with dashed vertical lines. The likelihood values as a function of $\log\tau$ or $\log\mathcal{Q}$ are represented by distinct colored lines. For both CTL and CPL, the blue lines represent weaker tides and the red lines represent stronger tides (on the same color scale as Figures \ref{fig:evolution_ctl} and \ref{fig:evolution_cpl}). 


A single peak in a given curve represents a unimodal solution in which the true value of \ttau\ or \tq\ could theoretically be recovered if \ttau\ or \tq\ were the only free variable. However, degeneracies do occur for both the CTL and CPL models.
The origin of these double peaks can be understood when we examine the evolution trajectories. In particular (as we noted in Section \ref{subsec:sensitivity_results}), the constraint on orbital period dominates the likelihood. In Figure \ref{fig:evolution_ctl} (showing the trajectories for varying \ttau) we see that there is a region in orbital period and eccentricity where solutions with different \ttau\ overlap. Similarly (but to a lesser extent), we also see in Figure \ref{fig:evolution_cpl} that there's a region in the orbital period evolution of the CPL model where solutions with different \tq\ overlap.

\begin{figure}[ht!]
\begin{center}
    \includegraphics[width=\linewidth]{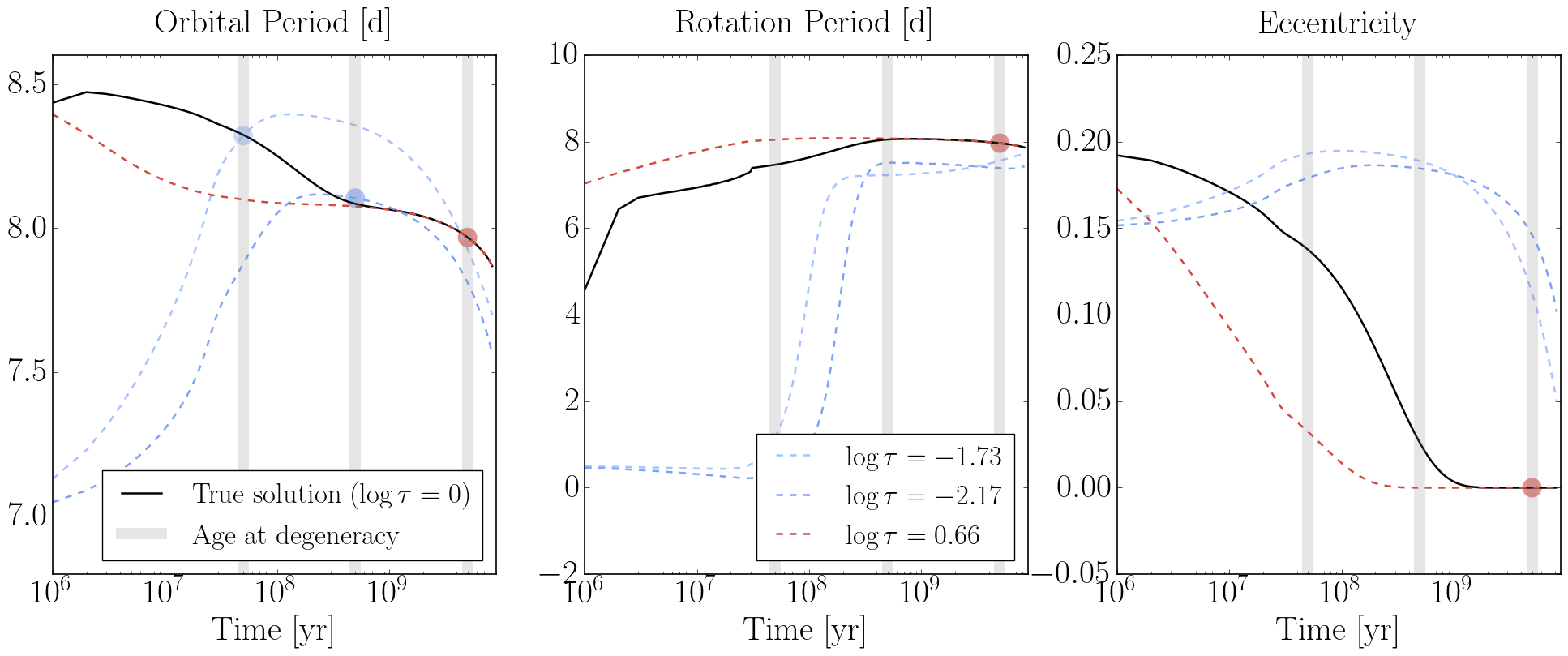}
\vspace{0.5em}
\caption{\normalsize Examples showing the how solutions with different $\log\tau$ are degenerate at different ages. The black line highlights the solution for $\log\tau=0$. The thick grey vertical lines mark the ages 50 Myr, 500 Myr, and 5 Gyr as used in Figures \ref{fig:degeneracy1d}--\ref{fig:degeneracy1d_3}. The colored dashed lines show the solutions for the secondary likelihood peak at each age: at 50 Myr the likelihood has peaks at $\log\tau=0$ and $-1.73$; at 500 Myr the likelihood has peaks at $\log\tau=0$ and $-2.17$; and at 5 Gyr the likelihood has peaks at $\log\tau=0$ and $0.66$. The color scale is consistent with Figures \ref{fig:evolution_ctl}--\ref{fig:evolution_cpl} and \ref{fig:degeneracy1d}--\ref{fig:degeneracy1d_3}, where darker red represents stronger tides and darker blue represents weaker tides. The colored dots highlight where the evolutions intersect.}
\label{fig:1d_degeneracy_evolution_tau_n1}
\end{center}
\end{figure}

Figure \ref{fig:1d_degeneracy_evolution_tau_n1} illustrates the degenerate solutions for $\log\tau=0$:
at 50 Myr the likelihood has peaks at $\log\tau=0$ and $-1.73$; at age 500 Myr the likelihood has peaks at $\log\tau=0$ and $-2.17$; and at age 5 Gyr the likelihood has peaks at $\log\tau=0$ and $0.66$. The black lines in Figure \ref{fig:1d_degeneracy_evolution_tau_n1} show the solution for $\log\tau=0$, and the dashed lines show the solutions of the secondary peaks. Points where the solutions overlap are highlighted with dots.
In the left panel of Figure \ref{fig:1d_degeneracy_evolution_tau_n1} (orbital period), there are intersections with $\log\tau=0$ at each of the ages, but in the other two panels (rotation period and eccentricity), there is only an intersection with $\log\tau=0$ at 5 Gyr. Thus, we see that the orbital period dominates the likelihood constraint, and the double-peaked degeneracies seen in the CTL + STELLAR likelihood (Figures \ref{fig:degeneracy1d}--\ref{fig:degeneracy1d_3}) occur when there are degeneracies in orbital period for different $\log\tau$ values.

We can gain further insight into the structure of the likelihood plots by examining the \ttau\ values where tidal locking, synchronization, and circularization happen. Tidal locking refers to the point in the evolution at which there is no net tidal torque acting on the star. Synchronization means that the rotation period and orbital period are identical. Circularization occurs when the eccentricity reaches a value of 0. For the CTL model, tidal locking occurs first, followed by synchronization and circularization, which occur at the same time. For the CPL model, however, tidal locking and synchronization occur first at the same time, and circularization follows later.

Table \ref{tab:circ_sync_tidelock} gives the values of \ttau\ or \tq\ for tidal locking, synchronization, and circularization for each age in the likelihood plots (Figures \ref{fig:degeneracy1d}--\ref{fig:degeneracy1d_3}). We also see in Figures \ref{fig:degeneracy1d}--\ref{fig:degeneracy1d_3} that there tends to be a symmetry to the degenerate peaks, where one solution occurs when the orbital period is rising, while the other solution occurs when the orbital period is decaying. This symmetry point is roughly the $\log\tau$ at which the stars become tidally locked. This ``turn-over'' in orbital period (as discussed in \citealt{fleming_lack_2018}) happens when a system evolves from an initial state where the rotation period is faster than the orbital period. Assuming that the initial rotation periods of stars in a binary system is around the same distribution as the rotation periods of young single stars, it is reasonable to expect fast initial rotation periods of $P_{\rm rot} \lesssim 1$ day \citep{stassun_rotation_1999,rebull_correlation_2006,marilli_rotational_2007}.

In the case of initial $P_{\rm rot} < P_{\rm orb}$, tidal locking would cause angular momentum to be transferred from the rotation period to the orbit, until the two approach synchronization. At the point of tidal locking, the magnetic braking acting to slow the rotation of the star is balanced by tidal forces that speed up the rotation of the star (Section \ref{sec:stellar}). The angular momentum lost due to magnetic braking must come from the angular momentum of the orbit, causing orbital decay. Thus, we find that coupling magnetic braking with the constant time lag model results in a ``turn-over'' in orbital period evolution (Figure \ref{fig:1d_degeneracy_evolution_tau_n1}), which results in degenerate solutions (Figures \ref{fig:degeneracy1d}--\ref{fig:degeneracy1d_3}).

\begin{table}
    \centering
    \begin{tabular}{|l|l|l|l|}
    \hline
    \multicolumn{4}{|l|}{CTL+STELLAR} \\
    \hline
    Age & tidelock & sync & circ \\
    \hline
    50 Myr & $\tau>-1.1$ & $\tau>1.2$ & $\tau>1.2$ \\
    500 Myr & $\tau>-1.2$ & $\tau>0.5$ & $\tau>0.5$ \\
    5000 Myr & $\tau>-1.3$ & $\tau>-0.6$ & $\tau>-0.6$ \\
    \hline
    \hline
    \multicolumn{4}{|l|}{CPL+STELLAR} \\
    \hline
    Age & tidelock & sync & circ \\
    \hline
    50 Myr & $\mathcal{Q}<5.2$ & $\mathcal{Q}<5.2$ & $\mathcal{Q}<3.8$ \\
    500 Myr & $\mathcal{Q}<7.3$ & $\mathcal{Q}<7.3$ & $\mathcal{Q}<4.6$ \\
    5000 Myr & $\mathcal{Q}<8.1$ & $\mathcal{Q}<8.1$ & $\mathcal{Q}<5.8$ \\
    \hline
    \end{tabular}
    \vspace{1em}
    \caption{\normalsize Range of tidal values in which tides are strong enough to cause tidal locking, synchronization, or circularization at a given age. In Figures \ref{fig:degeneracy1d}--\ref{fig:degeneracy1d_3}, the tidelock range is shown in magenta, the synchronization range is shown in cyan, and the circularization range is shown in gray.}
    \label{tab:circ_sync_tidelock}
\end{table}


\begin{figure}[ht!]
\begin{center}
    \includegraphics[width=\linewidth]{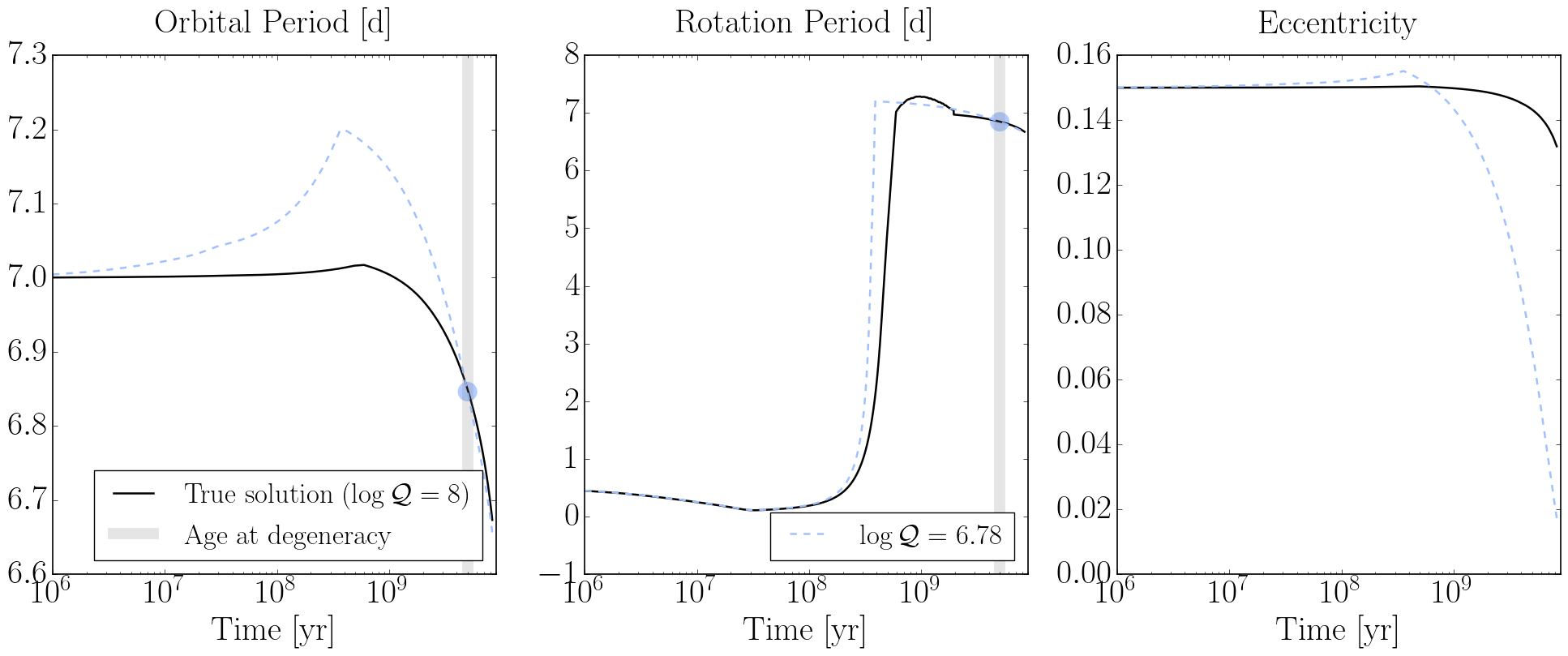}
\vspace{0.5em}
\caption{\normalsize Same as Figure \ref{fig:1d_degeneracy_evolution_tau_n1}, but for the CPL model. In Figure \ref{fig:degeneracy1d_3}, the solution for $\log\mathcal{Q}=8$ shows a double peak in likelihood. The black line highlights the solution for $\log\mathcal{Q}=8$. The blue dashed line shows the solution for the secondary likelihood peak: at 5 Gyr the likelihood has peaks at $\log\mathcal{Q}=8$ and $\log\mathcal{Q}=6.78$.}
\label{fig:1d_degeneracy_evolution_q_8}
\end{center}
\end{figure}

\clearpage

\section{Discussion} \label{sec:discussion}

The previous sections revealed that the path to constraining tidal evolution due to equilibrium tides faces fundamental challenges:
\begin{enumerate}
    \item  The evolution is governed by first-order ordinary differential equations that are high dimensional (18-dimensional phase space) and nonlinear. Thus, it is important to systematically understand how inference is sensitive to each model unknown and uncertainty.
    
    \item We cannot observationally constrain the full phase space of a system: we can measure a system's present-day configuration (\ie\ final $P_{\rm orb}$, $P_{\rm rot}$, $e$ $R$), but the time derivatives (\ie\ \dporb, \dprot, \decc, $dR/dt$) are significantly more difficult to constrain. 
    
    \item There are strong degeneracies between the input parameters, particularly between \tq\ and initial orbital conditions ($P_{\rm orb}$, $P_{\rm rot}$, $e$).

    \item The observable constraints (final $\porb$, $\prot$, $e$, etc.) are more sensitive to the initial state conditions than tidal \tq.

    \item For old systems (older than 5 Gyrs), the ratio $\rm P_{orb}/P_{rot}$ is more sensitive to tidal \tq\ than initial conditions. We find that populations of old systems may be useful for constraining upper or lower limits on tidal \tq. 

    \item Our analysis shows that for fixed tidal \tq\ for all stars that tidal \tq\ cannot be inferred to order-of-magnitude precision for individual systems by Bayesian methods, even when considering idealized uncertainties on present day orbital states and perfect priors (fixed at true values) for system masses and age.
\end{enumerate}

In addition, when models are applied to real data (without inspection on simulated data), there is the confounding factor that the model may not accurately represent the true physics of the system. Although the ultimate goal is to make meaningful inferences on real data, having a poor understanding of the systematic biases of the model itself (as stated above) makes it difficult to unambiguously interpret inferences and disentangle which factors can be improved. Thus, in this paper we have investigated such systematic model biases in the context of statistical inference.
Given these systematic challenges when it comes to inference of tidal \tq, we caution the the community to consider model biases when interpreting inferred values of tidal \tq, particularly based on inference of individual binary systems.

A recent study by \cite{patel_constraining_2022} reports a modified tidal $\mathcal{Q}'$ values for 41 low-mass ($0.4 < M_* < 1.2M_\odot$) eclipsing binary systems from \kepler. 
That study does not publish the full posteriors or summary statistics for their individual objects, but reports an overall constraint of $\log\,\mathcal{Q}' = 7.818 \pm 0.035$ based on the joint posterior of the 41 combined objects.
However, it is unclear in the methodology of \cite{patel_constraining_2022} (and similar studies, including \citealt{penev_2018}) how they handle model degeneracies in their inference. 
Their paper states that constraints on the initial orbital state come from finding ``the initial orbital period and initial eccentricity, which when evolved to the sampled age of the system matches the orbital period and eccentricity sampled'' \citep[Section 4.1;][]{patel_constraining_2022}. 
This assumption certainly contradicts our findings that the initial orbital period and eccentricity are not unique to a given fit of final values (Section \ref{subsec:posterior} and Figures \ref{fig:posterior_ctl_5gyr_zoom}--\ref{fig:posterior_cpl_5gyr_zoom}).
Their paper also states that constraints on mass and age come from fitting stellar isochrones \citep{paxton_modules_2010} to the estimated effective temperatures, surface gravity, and metallicity from the \cite{mathur_revised_2017} catalog of combined photometric and spectroscopic stellar parameters of \kepler targets. 
The study does not report uncertainties on their mass or age estimates, although isochrone fitting is known to be an unreliable age metric, particularly for low-mass stars, which may be on the main sequence \citep{soderblom_ages_2010}. It is furthermore unclear whether the catalog estimates from \cite{mathur_revised_2017} represent only the primary, or are biased by the blended photometry or spectroscopy of the secondary (\citealt{mathur_revised_2017} notes in Section 5.3 that the catalog is intended for single stars, and is likely biased for multistar systems).
We argue that more work is necessary in this field to identify a reliable constraint on tidal \tq\ than what is present in the existing literature. 

\subsection{Physical Origin of Short-period Binaries}

In addition to the statistical challenge of constraining the physical parameters of equilibrium tides, there is also the fundamental question of what the constraints on the initial orbital parameters tell us about the origin of short-period binaries, which remains an open question in the field. The formation of stars is expected to result from a hierarchical collapse of gas within molecular clouds \citep{vazquez-semadeni_global_2019}. However, the amount of thermal pressure expected during the initial collapse precudes the formation of systems at very short separations (with $a \lesssim 0.1\,$AU or $P_{\rm orb} \lesssim 10\,$days), implying that additional mechanisms of dissipation must take place to explain the existence of short-period binaries \citep{sterzik_how_2003,tokovinin_architecture_2021}. 

Results from the simulated posteriors (Figures \ref{fig:posterior_ctl_5myr}--\ref{fig:posterior_cpl_5gyr}) suggest that present-day short period systems (with orbital periods $P_{\rm orb} < 10\,$days) evolved from an initial configuration that started with a short orbital period, as seen by the high density of points with orbital periods between $\sim5$ and $10\,$days. In other words, equilibrium tides (even for $\log \tau$ as high as 1.0, or $\log\mathcal{Q}$ as low as 4.0) are not an efficient enough dissipation mechanism to drive significant orbital migration. 

A number of explanations have been proposed and investigated, including evolution due to a tertiary companion \citep{fabrycky_shrinking_2007}, or dissipation after collapse due to interactions with primordial gas \citep{moe_dynamical_2018}.
From a formation standpoint, it is worth mentioning that the plausibility of this conclusion (that short-period systems initially start out with a short orbital period) relies on the assumption that orbits undergo additional dissipation, causing inward migration to orbital periods of $<10\,$days within the first $\sim5\,$Myrs of their formation. For now it is understood that young, short-period systems may be plausible, but more work is needed to understand the dominant formation mechanisms of short-period binaries \citep{moe_dynamical_2018,tokovinin_formation_2020}, as well as the role of dissipation due to dynamical tides, which has been shown to be more efficient than equilibrium tides during pre main-sequence evolution \citep{zanazzi_tidal_2021}.

\subsection{Outlook} 

Our analysis (as well as recent work by \citealt{mirouh_detailed_2023}) suggests that constraining tidal dissipation, even using the best possible observational constraints \citep[eclipsing binaries in open clusters with precise ages, orbital prameters, and rotation periods; \eg,][]{southworth_eclipsing_2006,david_k2_2015,david_new_2016,gillen_new_2017,torres_eclipsing_2018}, is difficult to near-impossible. While the prospects of constraining tides based on existing metrics (\eg, measuring the tidal circularization period, or performing Bayesian inference of \tq\ based on present-day orbital state) may not be promising, there may be alternative methods worth exploring. 

\subsubsection{Constraints on Orbital Period Decay} \label{subsec:porb_decay}

In particular, it is worth exploring alternative methods to constrain the derivative states of the evolutions. There are a few cases in which orbital period decay has been measured, including the planetary systems WASP-12b \citep{patra_apparently_2017,yee_orbit_2019,patra_continuing_2020} and Kepler-1658b \citep{chontos_curious_2019,vissapragada_possible_2022} with implications for stellar tidal \tq's, though orbital decay has not been widely measured for many stellar systems due to the long baseline of observation time required. Moreover, in general, we expect that it would be difficult to disambiguate how tidal \tq\ influences orbital decay, apart from other factors (\eg, magnetic braking, presence of unobserved companion) that could influence the angular momentum of the system. This limitation is significant when it comes to inferring timescales related to parameters such as tidal \tq\ or tidal \ttau, which characterize tidal energy dissipation.

\subsubsection{Understanding the Attractor Space} \label{subsec:attractor}

An alternative approach would seek to gain constraints based on a population of systems. The full dynamical phase space of this problem is 18-dimensional, including the nine state variables and their derivatives (Table \ref{tab:parameters}).
However, as predicted from tidal models, the evolution of a binary system is dissipative (\ie\ flows contract in volume in phase space), due to the energy being lost to the interior of the stars in the form of heat due to tides.


\begin{figure}[!htp]
    \includegraphics[width=\textwidth]{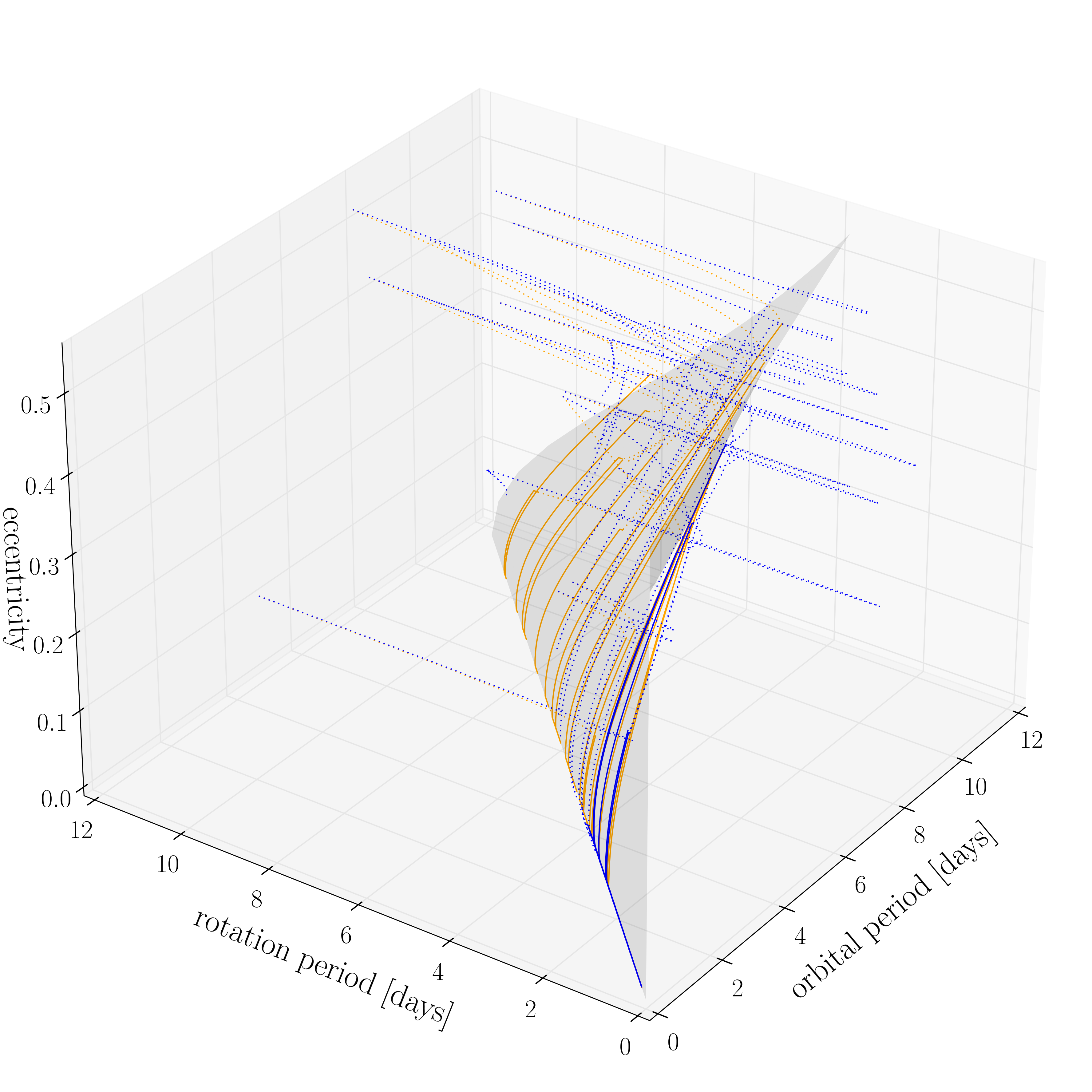} 
    \caption{\normalsize Orbital and rotational evolution for randomized initial conditions with strong tides (orange, $\log\tau=0$) and weak tides (blue, $\log\tau=-2$). Axes plot the rotation period (days), orbital period (days), and eccentricity. The dashed portion of the lines show the trajectory before rotational torque balance, and the solid portion shows after. At rotational torque equilibrium (\ie\ when tidal torque balances with magnetic braking), evolution trajectories converge to the shaded manifold, in which all ($\porb$, $\prot$, $e$) solutions are constrained to a very precise 2D subspace of 3D.} 
    \label{fig:equilibrium}
\end{figure}

For the CTL model the predicted equilibrium rotation period occurs for $d\omega/dt = 0$, or when
\begin{equation} \label{eqn:psync}
    \frac{\omega_i}{n} = \frac{2 \cos\psi}{1 + \cos^2\psi} \frac{N(e)}{\Omega(e)}
    \approx \frac{N(e)}{\Omega(e)}.
\end{equation}
For obliquities $\psi < 30$ deg, the factor of (Eq. \ref{eqn:psync}) containing $\psi$ terms is insignificant to the equilibrium period by less than $1\%$. 
Since the CPL model we used is only applicable to low eccentricities \citep{greenberg_frequency_2009}, and can only predict pseudosynchronous ratios of 1:1 and 3:2 \citep{fleming_rotation_2019}, for now we apply only the equilibrium state analysis to the CTL model. 

Next we consider the limiting behaviors when tides are combined with stellar evolution and magnetic braking. Synchronization occurs when the net torque acting on the rotation of a star becomes zero and the rotation period locks with the orbital period. This happens when the torque due to magnetic braking balances with the torque due to tides ($\frac{dJ}{dt}|_{tides} +  \frac{dJ}{dt}|_{MB} = 0$). Finally, circularization occurs when ($de/dt=0$ and $e=0$). 


While stellar evolution plays a large role in early dynamical evolution (as seen by trajectories in Figure \ref{fig:equilibrium} that initially ``blow past'' torque equilibrium), due to small variations in stellar radius and moment of inertia during the main sequence, the limiting behavior is driven by the equilibrium tide, which reduces the potential parameter space of our analysis. For old systems (on the main sequence), this implies that the model-observation comparison of the dominant variable space reduces to ($\porb$, $P_{\rm rot,i}$, $e$), or equivalently ($n$, $\omega_i$, $e$). That means the predicted equilibrium states lie along a very precise, low-dimensional manifold of ($n$, $\omega_i$, $e$) space as shown in Fig.~\ref{fig:equilibrium}. Hence more work is needed to precisely constrain the ($n$, $\omega_i$, $e$) space of observed binaries. 
In particular (as discovered in observational work by \citealt{lurie_tidal_2017,hobson_ritz_2025}, and suggested by simulation work by \citealt{mirouh_detailed_2023}), more work is needed to characterize subsynchronously rotating binaries (systems near synchronization, but with spin-orbit ratios less than 1:1), which may provide key evidence of the precise balance between tidal forces and magnetic braking.

\clearpage

\subsubsection{Model Reparameterization} \label{subsec:reparam}

We can also consider the possibility of reparameterizing the model as a way to reduce the dimensionality of the problem and degenerate parameters. Considering only the effect of tides, it is possible to reduce the number of orbital states from ($n$, $\omega_i$, $e$) to just ($\omega_i$, $e$). This is because the total angular momentum of the system is given by
\begin{align}
	J_{\rm tot} &= J_{\rm orb} + J_{\rm rot,1} + J_{\rm rot,2} \\
		&= \frac{\alpha}{n} \sqrt{1 - e^2} + \mathcal{I}_1\omega_1 + \mathcal{I}_2 \omega_2, 
\end{align}
so the mean motion can be written in terms of the rotation frequencies of each star and the eccentricity:
\begin{equation}
	n = \left(\frac{\alpha \sqrt{1 - e^2}}{J_{\rm tot} - \mathcal{I}_1\omega_1 + \mathcal{I}_2 \omega_2} \right)
\end{equation}
where $\alpha = M_1 M_2 G^{2/3} / (M_1 + M_2)^{1/3}$. 

\begin{figure}
    \includegraphics[width=\textwidth]{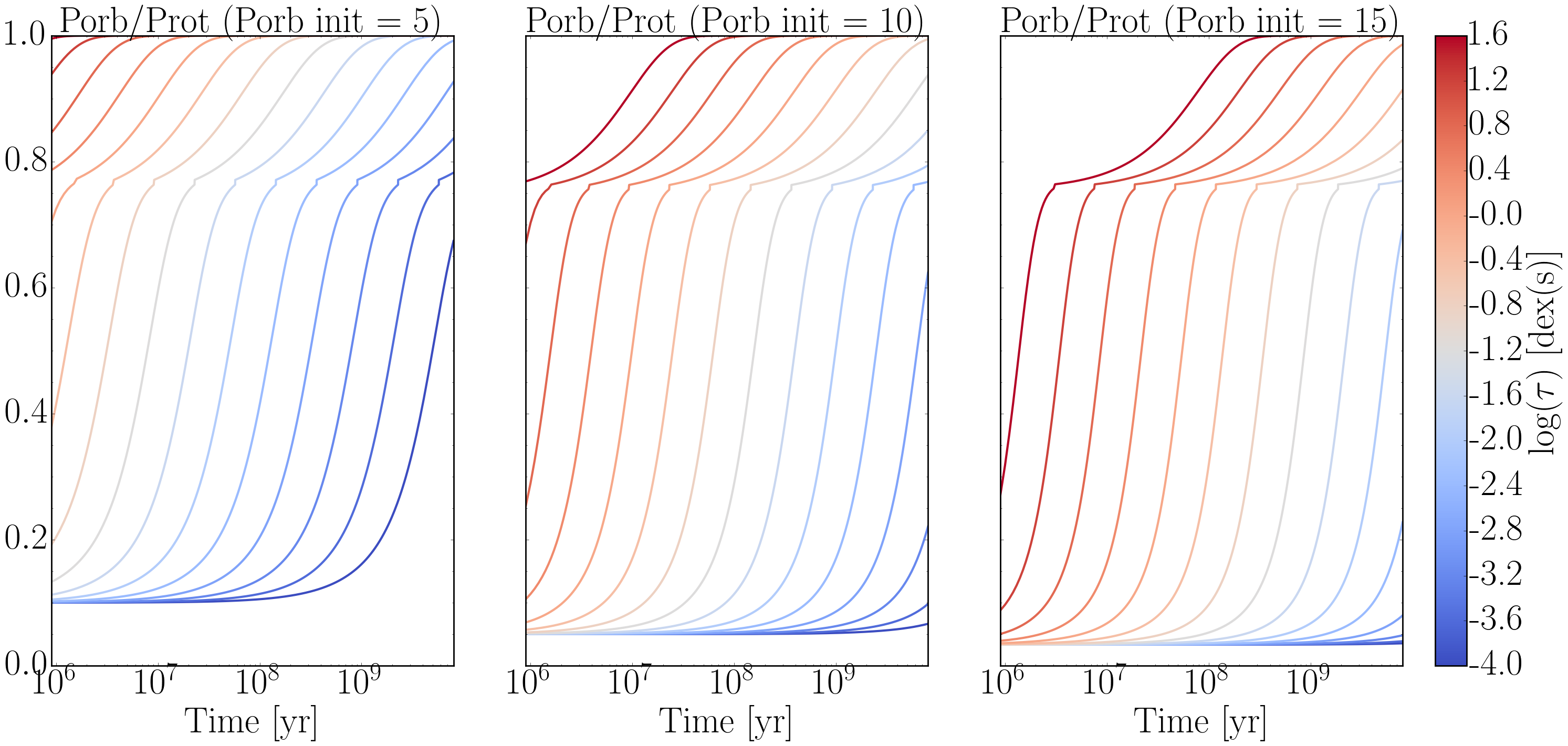} 
    \caption{\normalsize Evolution of the $P_{\rm orb} / P_{\rm rot}$ ratio for the primary star in a $1 M_{\odot} - 1 M_{\odot}$ binary system assuming only CTL tides. The system is evolved with an initial eccentricity of $e=0.2$ and initial rotation periods of $P_{\rm rot,1} = P_{\rm rot,2} = 0.5\,$d for different initial orbital periods: 5 days (left), 10 days (middle) and 15 days (right).} 
    \label{fig:ctl_ratio_eqtide}
\end{figure}

\begin{figure}
    \includegraphics[width=\textwidth]{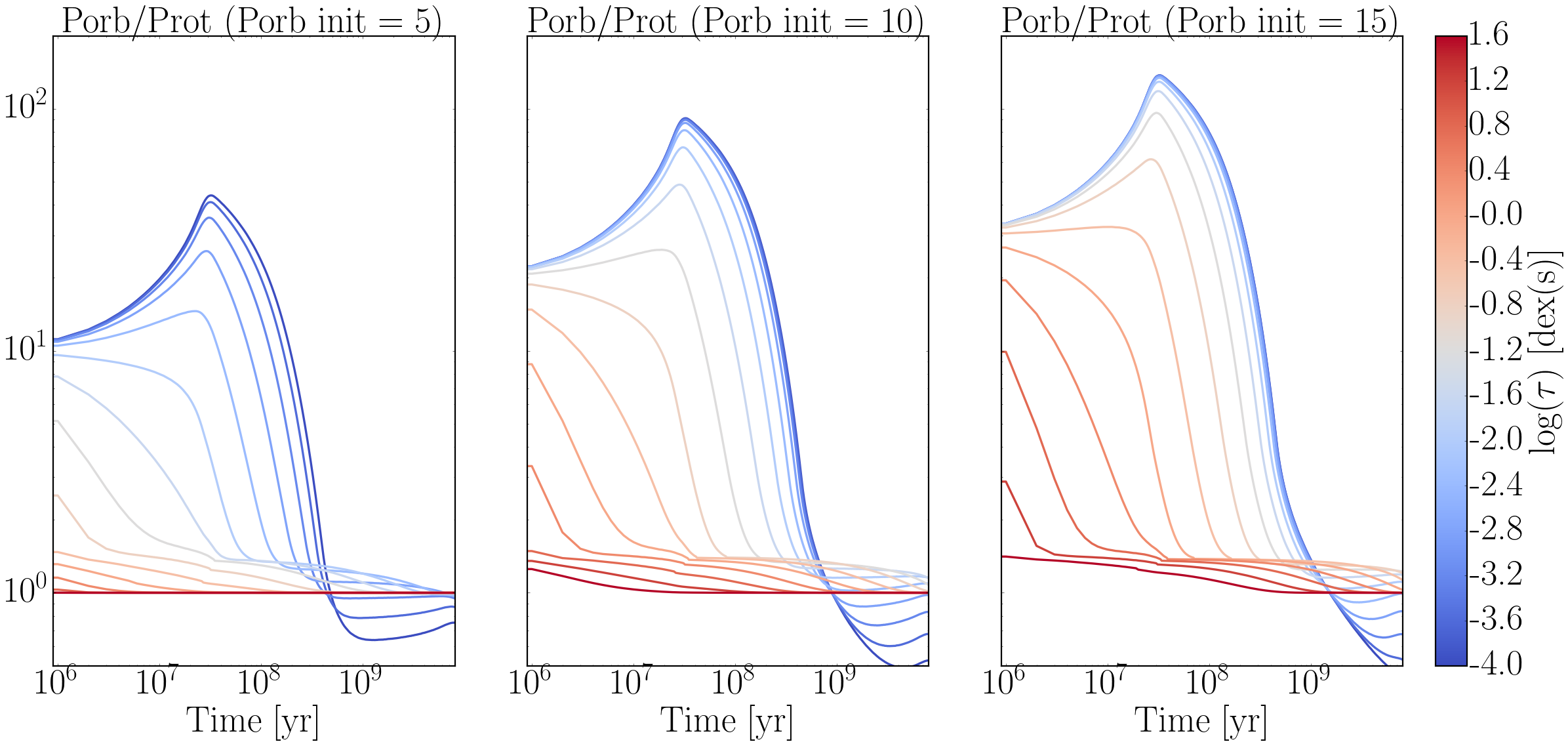} 
    \caption{\normalsize Same system as Figure \ref{fig:ctl_ratio_eqtide}, but modeled with CTL tides coupled with stellar evolution.} 
    \label{fig:ctl_ratio_stellar}
\end{figure} 

Figure \ref{fig:ctl_ratio_eqtide} shows the evolution of the ratio $\omega_i/n$ (or $P_{\rm orb} / P_{\rm rot}$) for the the primary star in a binary system evolved according to CTL tides. For systems with initial conditions of $P_{\rm rot} < P_{\rm orb}$, the rate at which the system synchronizes depends linearly on $\log\tau$, without crossing degeneracies.
However, this model reparameterization can only be done if the system conservatives angular momentum. When magnetic braking is coupled with tides (Section \ref{sec:stellar}), angular momentum is taken away from the system from stellar winds, introducing nonlinearities and crossing trajectories to the evolution of $P_{\rm orb} / P_{\rm rot}$ (Figure \ref{fig:ctl_ratio_stellar}). In this case an additional parameter $\Delta J_{\rm tot}$ (representing the amount of angular momentum lost from the system) would need to be sampled. This would not reduce the total number of parameters needed to be sampled, but would allow for the initial eccentricity and orbital period to be fit simultaneously and consistently, rather than independently. Further could test whether different parameterizations would be more efficient at sampling the parameter space.

\subsection{Limitations} \label{subsec:limitations}

The scope of this paper focuses on the limitations of equilibrium tidal dissipation theory (which is currently ill-constrained to orders of magnitude), not limitations of stellar evolution or magnetic braking (which we assume are correct by factors within an order of magnitude). 
In this study we only consider equilibrium tides, which only account for tidal dissipation when a hydrostatic bulge is raised on a star. We also assume a simplified tidal response, in which the mis-alignment of the bulge components are parameterized by either a time lag or phase lag that is constant in time. The CTL and CPL mechanisms do not account for all of the complexities of the stellar interior structures that have been proposed, including turbulent viscosity acting on equilibrium tides in the convective envelope \citep{barker_tidal_2020} or the fluid motions of dynamical tide prescriptions \citep{zahn_tidal_1989}. 

We also assume that stellar evolution influences tides, but we make no assumptions about how tides may influence stellar evolution. Studies such as \cite{casey_tidal_2019} suggest that tides may influence stellar evolution, including lithium production in red-giant stars. However, modeling the effects on stellar evolution would be a significant endeavor requiring the use of a stellar evolution code such as MESA \citep[Modules for Experiments in Stellar Astrophysics;][]{paxton_modules_2010}, rather than grids of evolution models \citep{baraffe_new_2015}. 
Futhermore, we only model low-mass systems through their main-sequence evolution phase with separations larger than the Roche radius, hence we do not model or make assumptions about any mass transfer between stars in the system. 

We also make the assumption that orbital evolution occurs for an isolated binary system, in which there are no additional perturbers (such as planet or nearby star) to influence the angular momentum of the system. The presence of additional bodies could introduce a variety of different effects, such as Kozai-Lidov effects, resonances, or chaos \citep{toonen_evolution_2016,fabrycky_shrinking_2007,naoz_eccentric_2016}.
Accounting for the presence of triple (or other higher order systems) would be an important role in future studies when comparing binary evolution models to observations, as unobserved companions could account for differences between the observed and simulated distributions of orbits.  \\

\section{Conclusion} \label{sec:conclusion}

In this paper, we investigated the prospects and limitations for constraining equilibrium tides in low-mass binary stars based on ideal uncertainties. We used the package \code{VPLanet} to simulate the coupled effects of stellar evolution, magnetic braking, and equilibrium tides. We then applied statistical methods including sensitivity analysis and simulated Bayesian inference to systematically analyze how model degeneracies and observational uncertainties limit constraints on tidal dissipation. 

Sensitivity analysis allowed us to systematically search high-dimensional and non-linear parameter spaces to identify which types of individual systems are most promising for constraining tidal dissipation, thus informing observational follow-up efforts.  We used sensitivity analysis to assess which input parameters dominate the final orbital and rotational states of the system. We found to first order that the final orbital state (in particular the final orbital period) is dominated by the initial orbital period of the system. We further investigated whether the final eccentricity and rotation states would be sensitive enough to gain meaningful constraints on tidal dissipation. To perform this assessment, we simulated inference using the package \code{alabi} to trace out 5-dimensional posteriors in a computationally efficient manner.   

Our analysis suggests that even when employing implausibly precise orbital parameters from \kepler/\tess, constraining tidal \tq\ remains ill-posed because of strong degeneracies with initial conditions and because we cannot determine how quickly a system evolves (\ie we have no constraints on \dporb, \dprot, \decc, etc. for any given system). An alternate approach would take advantage of the constraints from a large population of binaries. However,  different binary systems can originate from a wide range of initial conditions, which are degenerate and dependent on complex (and very ill-constrained, uncertain) formation mechanisms. 

An alternative approach worth investigating would seek to characterize the attractor state dynamics of tides.
Stellar evolution and magnetic braking dominate in early evolution; however, during the main-sequence tidal dissipation dominates, even for weak tides. Considering the case of constant radius and radius of gyration (which is approximately the case when stars reach the main sequence), the attractor phase space under torque equilibrium reduces to 6 dimensions ($e$, $a$, $\omega$, $de/dt$, $da/dt$, $d\omega/dt$). Thus, overdensities in ($\porb$, $e$, $\prot$) of observed binaries may be a promising way to constrain and validate theoretical tidal models, regardless of how fast these systems are evolving. 

In summary this work critically examines our understanding of equilibrium tides in binary systems. While there is certainly still much work left to realistically model the tidal response of stars and to understand how stellar structure plays a role in the gravitational interactions of stars, the impact of work in this subfield has many implications for broader astrophysics. 
Stars play a fundamental role on a broad range of scales from the long-term evolution of planets and planetary systems, to the dynamics of stellar clusters and galactic dynamics.
Thus the dynamical outcomes on the orbits of binary stars due to tidal evolution has the potential to give insight to many fundamental physical processes involving gravitational interactions with stars.

\clearpage

\appendix \label{sec:appendix} 
In this appendix, we present the equilibrium tide models used in this study. \\

\begin{center}
	\textit{A.1 Governing Equations}
\end{center}

The average energy dissipated $\langle \dot{E}_{\rm orb} \rangle_i$ and torque on each star in the system $\langle \dot{J}_{\rm rot} \rangle_i$ can be derived from the potential and an assumption for the phase lag dependence (e.g. constant time, or constant phase).
For the CTL model, the tidal energy and torque terms are given by \citep[][Equation A.21 \& Equation 9]{leconte_is_2010}:
\begin{equation} \label{eqn:dEorb_ctl}
    \langle \dot{E}_{\rm orb} \rangle_i
        = 2 K_i \, \tau_i \left[ 
            N(e) \cos\psi_i \frac{\omega_i}{n} - N_a(e)
        \right]
\end{equation}

\begin{equation} \label{eqn:torque_ctl}
    \langle \dot{J}_{\rm rot} \rangle_i 
        = - \frac{K_i \, \tau_i}{n} \left[ 
            \left(1 + \cos^2\psi_i \right) \Omega(e) \frac{\omega_i}{n}
            - 2 \cos\psi_i N(e)
        \right]
\end{equation} 
where the eccentricity functions $N(e)$, $N_a(e)$, $\Omega(e)$ from \cite{leconte_is_2010} are given by:
\begingroup
\allowdisplaybreaks
\begin{align} 
    N(e) &= \frac{1 + \frac{15}{2} e^2 + \frac{45}{8} e^4 + \frac{5}{16} e^6 }{(1 - e^2)^{6}} \label{eqn:eccN} \\
    N_a(e) &= \frac{1 + \frac{31}{2} e^2 + \frac{255}{8} e^4 + \frac{185}{16} e^6 + \frac{25}{64} e^8}{(1 - e^2)^{15/2}} \label{eqn:eccNa} \\
    \Omega_e(e) &= \frac{ 1 + \frac{3}{2} e^2 + \frac{1}{8} e^4 }{(1 - e^2)^{5}} \label{eqn:eccOmega} \\
    N_e(e) &= \frac{1 + \frac{15}{4} e^2 + \frac{15}{8} e^4 + \frac{5}{64} e^6}{(1 - e^2)^{13/2}} \\
    \Omega(e) &= \frac{ 1+ 3e^2 + \frac{3}{8}e^4 }{(1 - e^2)^{9/2}} \label{eqn:eccOmegae}
\end{align} 
\endgroup
and the intermediate coefficient variable $K_i$ is given by:
\begin{equation} 
    K_i = \frac{3}{2} G k_{2}  n^2
        \left(\frac{M_j^2 R_i^5}{a^6} \right). 
\end{equation}
Here $k_2$ is known as the tidal Love number.
The Love number $k_l$ describes the degree to which a stellar body deforms due to the tidal forces in the radial direction, where $k_2$ is the Love number due to the quadrupole order potential. In this study we adopt a value of $k_2 = 0.5$, but note that uncertainty in $k_2$ is degenerate with that of \tq\ and can also be written as the effective quality factor $\mathcal{Q}' =  3\mathcal{Q} / 2 k_2$, where $k_2 = 3/2$ and $\mathcal{Q}' =  \mathcal{Q}$ for a homogeneous fluid sphere \citep{jackson_tidal_2008}.

For the alternative linear equilibrium tide model, CPL, the tidal energy dissipation and torque terms are given by \citep[][Equation 48--49]{ferraz-mello_tidal_2008}:
\begin{equation} \label{eqn:dEorb_cpl}
    \langle \dot{E}_{\rm orb} \rangle_i 
        = \frac{K_i}{4 n}
        \left[ 
            4 \varepsilon_0 + e^2 (-20 \varepsilon_0 + \frac{147}{2} \varepsilon_1 + \frac{1}{2} \varepsilon_2 - 3 \varepsilon_5) 
            -4 \sin^2\psi (\varepsilon_0 - \varepsilon_8)
        \right]
\end{equation}

\begin{equation} \label{eqn:torque_cpl}
    \langle \dot{J}_{\rm rot} \rangle_i
        = - \frac{K_i}{4 n^2}
        \left[ 
            4 \varepsilon_0 + e^2 (-20 \varepsilon_0 + 49 \varepsilon_1 + \varepsilon_2) 
            +2 \sin^2\psi (-2\varepsilon_0 + \varepsilon_8 + \varepsilon_9)
        \right]
\end{equation}
Under the constant phase lag assumption, the amplitude of the phase lags are proportional to a constant $\mathcal{Q} \approx 1/\varepsilon$. For quadrupole order potential ($l=2$), we consider the dominant forcing frequencies $\varepsilon = kn - m\omega$ up to $m = \pm 2$, where the amplitude of each phase lag term corresponds to the following the frequencies \citep[Table 1,][]{ferraz-mello_tidal_2008}:
\begingroup
\allowdisplaybreaks
\begin{align*} 
    \varepsilon_{0,i} & = 2 \omega_i - 2n \\
    \varepsilon_{1,i} & = 2 \omega_i - 3n \\
    \varepsilon_{2,i} & = 2 \omega_i - n \\
    \varepsilon_{5,i} & = n \\
    \varepsilon_{6,i} & = 2 n \\
    \varepsilon_{7,i} & = 2 n \\
    \varepsilon_{8,i} & = \omega_i - 2n \\
    \varepsilon_{9,i} & = \omega_i \\
\end{align*}
\endgroup

\begin{center}
	\textit{A.2 Secular Evolution of Keplerian Elements}
\end{center}

This section elaborates on the equations for the total evolution of the Keplerian components: semi-major axis, eccentricity, rotation rate, and obliquity for both the constant-time-lag and constant-phase-lag equilibrium tidal models.
For any tidal model, the orbit-averaged evolution of the Keplerian orbital elements due to each component is given by:
\begin{align}
	\text{semi-major axis: \quad}
	\avg{\ddt{a}}    &= - \frac{a}{\Eorb} \avg{\ddt{\Eorb}},  \label{eqn:dadt} \\
	\text{eccentricity: \quad}
	\avg{\ddt{e}}    &= \left(\frac{1}{a} \ddt{a} - \frac{2}{\Jorb} \ddt{\Jorb} \right) \cdot \frac{1 - e^2}{2e},           \label{eqn:dedt} \\
\end{align}
where $\mtot = M_1 M_2 / (M_1 + M_2)$ is the reduced mass, and $\mathcal{I}_i = M_i r_{g,i}^2 R^2$ is the moment of inertia for each body. The spin evolution for each body is given by:
\begin{align}
	\text{rotation frequency: \quad}
	\avg{\ddt{\omega_i}} 
            &= \frac{1}{\Imom_i} \left(\ddt{\Jtide} + \ddt{\Jmb} - \omega_i \ddt{\Imom_i} \right),    \label{eqn:dwdt}  
\end{align}
where $i=1$ is the primary, $i=2$ is the secondary. The terms bracketed by $\langle \cdot \rangle$ denote quantities that are orbit-averaged. 

\begin{center}
	\textit{A.3 Semi-major axis}
\end{center}

The equation for the semi-major axis evolution when coupled with stellar evolution and magnetic braking remains the same as when only considering equilibrium tide. The total semi-major axis derivative is the sum of the contributions from the primary ($i=1$) and secondary ($i=2$):
\begin{equation}
    \avg{\ddt{a}}_{\rm tot} = \sum\limits_{i = 1}^2 \avg{\ddt{a_i}}_{\rm tide}.
\end{equation}
For the constant time lag model, the equation for the orbit-averaged derivative of semi-major axis can be derived from substituting Equations (\ref{eqn:dadt}) and (\ref{eqn:dEorb_ctl}): 
\begin{equation}\label{eqn:ctl:a}
    \avg{\ddt{a_i}}_{\rm CTL} = \  
        \frac{4a^2}{G M_1 M_2} 
        K_i \, \tau_i
        \left[ 
            N(e) \cos\psi_i \frac{\omega_i}{n} - N_a(e)
        \right].
\end{equation} 
Similarly, the orbit-averaged derivative of semi-major axis for the constant phase lag assumption is found from substituting Equations (\ref{eqn:dadt}) and (\ref{eqn:dEorb_cpl}): 
\begin{equation}
    \avg{\ddt{a_i}}_{\rm CPL}  = \  
        \frac{a^2}{G M_1 M_2} 
        \frac{K_i}{2 n \mathcal{Q}_i} 
            \left[ 
                4 \hat{\varepsilon}_0 + e^2 (-20 \hat{\varepsilon}_0 + \frac{147}{2} \hat{\varepsilon}_1 + \frac{1}{2} \hat{\varepsilon}_2 - 3 \hat{\varepsilon}_5) 
                -4 \sin^2\psi (\hat{\varepsilon}_0 - \hat{\varepsilon}_8)
            \right].
\end{equation}
Applying the constant phase lag assumption ($\varepsilon = \hat{\varepsilon}/\mathcal{Q}$ where \tq\ is a constant) lets us factor out the magnitude of the phase lag \tq\ on the denominator, and each $\hat{\varepsilon}$ term corresponds to the sign of each phase lag term \citep{heller_tidal_2011}, $\hat{\varepsilon}_i = \mathrm{sign}(\varepsilon_i)$.

\begin{center}
	\textit{A.3 Eccentricity}
\end{center}

The eccentricity evolution can be obtained by differentiating the orbital angular momentum and rearranging to find:
\begin{equation} \label{eqn:desq}
    \avg{\ddt{e}} = -\frac{1}{\Eorb} \left[ 
        \ddt{\Eorb} - \frac{n}{\sqrt{1 - e^2}} \ \ddt{\Jorb}
    \right] \cdot \frac{1 - e^2}{2e}.
\end{equation}
From conservation of angular momentum, the change in orbital angular momentum is the negative of the magnitude of the change in rotational angular momentum, which is the sum of the torques due to tides and magnetic braking:
\begin{equation}
    \ddt{\Jorb} = - \sum_{i=1}^2 \ddt{J_{\rm rot, i}} 
        = - \sum_{i=1}^2 \left(\ddt{\Jtide} + \ddt{\Jmb} \right)_i.
    \label{eqn:dJorb}
\end{equation}
Equation (\ref{eqn:desq}) can thus be broken into components, $\avg{de/dt} = \avg{de/dt}_{\rm tide} + \avg{de/dt}_{\rm mb}$. Following the assumption of \cite{repetto_coupled_2014} and \cite{fleming_rotation_2019}, we assume $\avg{de/dt}_{\rm mb}$ is relatively small compared to $\avg{de/dt}_{\rm tide}$, and thus just consider the effect of tides on eccentricity evolution.

The contribution to eccentricity due to tides for the constant time lag model, can be obtained by substituting Eqns (\ref{eqn:dEorb_ctl}) and (\ref{eqn:torque_ctl}) into Eqn. (\ref{eqn:dedt}) to get: 
\begin{equation}
    \avg{\ddt{e_i}}_{\rm CTL} 
        = \frac{a e}{G M_1 M_2} K_i \tau_i 
        \left[11 \cos\psi_i \Omega_e(e)  \frac{\omega_i}{n} 
        - 18 N_e(e) \right],
\end{equation}
and for the constant phase lag model, can be obtained by substituting Eqns (\ref{eqn:dEorb_cpl}) and (\ref{eqn:torque_cpl}) into Eqn. (\ref{eqn:desq}): 
\begin{equation}
    \avg{\ddt{e_i}}_{\rm CPL} 
        = -\frac{a e}{4 G M_1 M_2} \frac{K_i}{n \mathcal{Q}} 
        \left[ 2 \hat{\varepsilon}_0 - \frac{49}{2} \hat{\varepsilon}_1 
        + \frac{1}{2} \hat{\varepsilon}_2 + 3 \hat{\varepsilon}_5 \right].
\end{equation}

\begin{center}
	\textit{A.4 Rotation}
\end{center}

The secular rotational evolution for each individual body $i$ in the system is solved from the sum of torques due to both tides and magnetic braking acting on the body:
\begin{equation}
	\avg{\ddt{\omega_i}} = \avg{\ddt{\omega_i}}_{\rm tide} + \avg{\ddt{\omega_i}}_{\rm mb}
		= \frac{1}{\Imom} \left(\Ttide + \Tmb \right),
\end{equation}
where the total rotational evolution due to CTL tides and stellar evolution is given by the sum of Equation (\ref{eqn:torque_ctl}) and (\ref{eqn:torque_mb}) divided by moment of inertia, $\Imom = M r_g^2 R^2$:
\begin{equation}\label{eqn:ctl:omega}
    \avg{\ddt{\omega_i}}_{\rm CTL} = \ 
        - \frac{K_i \, \tau_i}{n \, \mathcal{I}_i} \left[ 
            \left(1 + \cos^2\psi_i \right) \Omega(e) \frac{\omega_i}{n}
            - 2 \cos\psi_i N(e)
        \right].
\end{equation}
Similarly, the total rotational evolution due to CPL tides and stellar evolution is given by the sum of Equation (\ref{eqn:torque_cpl}) and (\ref{eqn:torque_mb}) divided by moment of inertia:
\begin{equation} \label{eqn:cpl:omega}
    \avg{\ddt{\omega_i}}_{\rm CPL} = \
        - \frac{K_i}{4 n^2 \, \mathcal{Q}_i\, \mathcal{I}_i}
            \left[ 
                4 \hat{\varepsilon}_0 + e^2 (-20 \hat{\varepsilon}_0 + 49 \hat{\varepsilon}_1 + \hat{\varepsilon}_2) 
                +2 \sin^2\psi (-2\hat{\varepsilon}_0 + \hat{\varepsilon}_8 + \hat{\varepsilon}_9)
            \right].
\end{equation}

\begin{center}
	\textit{A.5 Obliquity}
\end{center}

Finally, the secular evolution of the obliquity for each star in the system for the CTL model was derived in \cite{leconte_is_2010}:
\begin{equation} \label{eqn:dpsi_dt_ctl}
    \avg{\ddt{\psi_i}} = \frac{K_i \sin\psi_i}{\Imom_i n} 
        \left[(\cos\psi_i - \eta_i) \Omega(e) \frac{\omega_i}{n} 
            - 2 N(e) \right],
\end{equation}
where $\eta_i = J_{\rm rot,i}/\Jorb$ is the ratio of rotational over orbital angular momentum:
\begin{equation} \label{eqn:eta_ctl}
    \eta_i = \frac{M_1 + M_2}{M_1 M_2} \frac{\Imom_i \omega_i}{a^2 n \sqrt{1-e^2}}.
\end{equation}
For the CPL model, we use the obliquity evolution derived in \cite{ferraz-mello_tidal_2008}:
\begin{equation} \label{eqn:dpsi_dt_cpl}
    \avg{\ddt{\psi_i}} = \frac{K_i \sin\psi_i}{2 \omega_i \Imom_i n^2 \mathcal{Q}}
        \left[\left(1 - \xi_i \right) \hat\varepsilon_0 + 
          \left(1 + \xi_i \right) (\hat\varepsilon_8 - \hat\varepsilon_9)  \right],
\end{equation} 
where $\xi_i$ is defined as
\begin{equation} \label{eqn:xi_cpl}
    \xi_i = \frac{\omega_i \Imom_i a n}{G M_i M_j}. \\
\end{equation}


\clearpage
	The authors would like to acknowledge Brian Jackson (Boise State), Eric Agol (University of Washington; UW), Sean Matt (University of Oklahoma), Scott Anderson (UW), Victoria Meadows (UW), Matthew McQuinn (UW), \u{Z}eljko Iveci\'{c} (UW), David Fleming (formerly UW), Hoony Kang (University of Maryland), and Zoe (parrot) for various constructive discussions in the process of this project. 
	JB acknowledges funding support from NSF Graduate Research Fellowship grant number DGE-1762114 and a Scialog grant supported by the Heissing-Simmons Foundation.
	This work was facilitated through the use of advanced computational, storage, and networking infrastructure provided by the Hyak supercomputer system and funded by the STF at the University of Washington. The code for reproducing this study is available on Github: \href{https://github.com/jbirky/tidal_inference}{https://github.com/jbirky/tidal\_inference}. 

\software{
        \code{alabi} \citep{birky_alabi}, \,
	\code{Astropy} \citep{astropy_collaboration_astropy_2013,astropy_collaboration_astropy_2018}, \,
        \code{corner} \citep{foreman-mackey_cornerpy_2016}, \,
	\code{matplotlib} \citep{hunter_matplotlib_2007}, \,
	\code{numpy} \citep{van_der_walt_numpy_2011}, \,
	\code{scipy} \citep{jones_scipy_2001}, \,
	\code{SALib} \citep{usher_salibsalib_2016}, \,
	\code{VPLanet} \citep{Barnes20} \\
}


\bibliographystyle{aasjournal}
\bibliography{min.bib}

\begin{thebibliography}{}
\expandafter\ifx\csname natexlab\endcsname\relax\def\natexlab#1{#1}\fi
\providecommand{\url}[1]{\href{#1}{#1}}
\providecommand{\dodoi}[1]{doi:~\href{http://doi.org/#1}{\nolinkurl{#1}}}
\providecommand{\doeprint}[1]{\href{http://ascl.net/#1}{\nolinkurl{http://ascl.net/#1}}}
\providecommand{\doarXiv}[1]{\href{https://arxiv.org/abs/#1}{\nolinkurl{https://arxiv.org/abs/#1}}}

\bibitem[{Aigrain {et~al.}(2015)Aigrain, Llama, Ceillier, Chagas, Davenport,
  Garcia, Hay, Lanza, McQuillan, Mazeh, de~Medeiros, Nielsen, \&
  Reinhold}]{aigrain_testing_2015}
Aigrain, S., Llama, J., Ceillier, T., {et~al.} 2015, Monthly Notices of the
  Royal Astronomical Society, 450, 3211, \dodoi{10.1093/mnras/stv853}

\bibitem[{Albrecht {et~al.}(2007)Albrecht, Reffert, Snellen, Quirrenbach, \&
  Mitchell}]{albrecht_spin_2007}
Albrecht, S., Reffert, S., Snellen, I., Quirrenbach, A., \& Mitchell, D.~S.
  2007, Astronomy \& Astrophysics, 474, 565, \dodoi{10.1051/0004-6361:20077953}

\bibitem[{Albrecht {et~al.}(2022)Albrecht, Dawson, \&
  Winn}]{albrecht_stellar_2022}
Albrecht, S.~H., Dawson, R.~I., \& Winn, J.~N. 2022, arXiv:2203.05460
  [astro-ph].
\newblock \url{http://arxiv.org/abs/2203.05460}

\bibitem[{{Astropy Collaboration} {et~al.}(2013){Astropy Collaboration},
  Robitaille, Tollerud, Greenfield, Droettboom, Bray, Aldcroft, Davis,
  Ginsburg, Price-Whelan, Kerzendorf, Conley, Crighton, Barbary, Muna,
  Ferguson, Grollier, Parikh, Nair, Unther, Deil, Woillez, Conseil, Kramer,
  Turner, Singer, Fox, Weaver, Zabalza, Edwards, Azalee~Bostroem, Burke, Casey,
  Crawford, Dencheva, Ely, Jenness, Labrie, Lim, Pierfederici, Pontzen, Ptak,
  Refsdal, Servillat, \& Streicher}]{astropy_collaboration_astropy_2013}
{Astropy Collaboration}, Robitaille, T.~P., Tollerud, E.~J., {et~al.} 2013,
  {\textbackslash}aap, 558, A33, \dodoi{10.1051/0004-6361/201322068}

\bibitem[{{Astropy Collaboration} {et~al.}(2018){Astropy Collaboration},
  Price-Whelan, Sipőcz, Günther, Lim, Crawford, Conseil, Shupe, Craig,
  Dencheva, Ginsburg, VanderPlas, Bradley, Pérez-Suárez, de~Val-Borro,
  Aldcroft, Cruz, Robitaille, Tollerud, Ardelean, Babej, Bach, Bachetti,
  Bakanov, Bamford, Barentsen, Barmby, Baumbach, Berry, Biscani, Boquien,
  Bostroem, Bouma, Brammer, Bray, Breytenbach, Buddelmeijer, Burke, Calderone,
  Cano~Rodríguez, Cara, Cardoso, Cheedella, Copin, Corrales, Crichton,
  D'Avella, Deil, Depagne, Dietrich, Donath, Droettboom, Earl, Erben, Fabbro,
  Ferreira, Finethy, Fox, Garrison, Gibbons, Goldstein, Gommers, Greco,
  Greenfield, Groener, Grollier, Hagen, Hirst, Homeier, Horton, Hosseinzadeh,
  Hu, Hunkeler, Ivezić, Jain, Jenness, Kanarek, Kendrew, Kern, Kerzendorf,
  Khvalko, King, Kirkby, Kulkarni, Kumar, Lee, Lenz, Littlefair, Ma, Macleod,
  Mastropietro, McCully, Montagnac, Morris, Mueller, Mumford, Muna, Murphy,
  Nelson, Nguyen, Ninan, Nöthe, Ogaz, Oh, Parejko, Parley, Pascual, Patil,
  Patil, Plunkett, Prochaska, Rastogi, Reddy~Janga, Sabater, Sakurikar,
  Seifert, Sherbert, Sherwood-Taylor, Shih, Sick, Silbiger, Singanamalla,
  Singer, Sladen, Sooley, Sornarajah, Streicher, Teuben, Thomas, Tremblay,
  Turner, Terrón, van Kerkwijk, de~la Vega, Watkins, Weaver, Whitmore,
  Woillez, Zabalza, \& {Astropy
  Contributors}}]{astropy_collaboration_astropy_2018}
{Astropy Collaboration}, Price-Whelan, A.~M., Sipőcz, B.~M., {et~al.} 2018,
  {\textbackslash}aj, 156, 123, \dodoi{10.3847/1538-3881/aabc4f}

\bibitem[{Baraffe {et~al.}(2015)Baraffe, Homeier, Allard, \&
  Chabrier}]{baraffe_new_2015}
Baraffe, I., Homeier, D., Allard, F., \& Chabrier, G. 2015, Astronomy \&
  Astrophysics, 577, A42, \dodoi{10.1051/0004-6361/201425481}

\bibitem[{Barker(2020)}]{barker_tidal_2020}
Barker, A.~J. 2020, arXiv:2008.03262 [astro-ph].
\newblock \url{http://arxiv.org/abs/2008.03262}

\bibitem[{Barker(2022)}]{barker_2022}
---. 2022, The Astrophysical Journal Letters, 927, L36,
  \dodoi{10.3847/2041-8213/ac5b63}

\bibitem[{Barker \& Ogilvie(2009)}]{barker_tidal_2009}
Barker, A.~J., \& Ogilvie, G.~I. 2009, {\textbackslash}mnras, 395, 2268,
  \dodoi{10.1111/j.1365-2966.2009.14694.x}

\bibitem[{Barnes(2017)}]{barnes_tidal_2017}
Barnes, R. 2017, Tidal {Locking} of {Habitable} {Exoplanets},
  \dodoi{10.1007/s10569-017-9783-7}

\bibitem[{Barnes {et~al.}(2013)Barnes, Mullins, Goldblatt, Meadows, Kasting, \&
  Heller}]{barnes2013}
Barnes, R., Mullins, K., Goldblatt, C., {et~al.} 2013, Astrobiology, 13, 225,
  \dodoi{10.1089/ast.2012.0851}

\bibitem[{Barnes {et~al.}(2020)Barnes, Luger, Deitrick, Driscoll, Quinn,
  Fleming, Smotherman, McDonald, Wilhelm, Garcia, Barth, Guyer, Meadows, Bitz,
  Gupta, Domagal-Goldman, \& Armstrong}]{barnes_vplanet_2020}
Barnes, R., Luger, R., Deitrick, R., {et~al.} 2020, Publications of the
  Astronomical Society of the Pacific, 132, 024502,
  \dodoi{10.1088/1538-3873/ab3ce8}

\bibitem[{{Barnes} {et~al.}(2020){Barnes}, {Luger}, {Deitrick}, {Driscoll},
  {Quinn}, {Fleming}, {Smotherman}, {McDonald}, {Wilhelm}, {Garcia}, {Barth},
  {Guyer}, {Meadows}, {Bitz}, {Gupta}, {Domagal-Goldman}, \&
  {Armstrong}}]{Barnes20}
{Barnes}, R., {Luger}, R., {Deitrick}, R., {et~al.} 2020, \pasp, 132, 024502,
  \dodoi{10.1088/1538-3873/ab3ce8}

\bibitem[{Barnes(2003)}]{barnes_rotational_2003}
Barnes, S.~A. 2003, The Astrophysical Journal, 586, 464, \dodoi{10.1086/367639}

\bibitem[{Bashi {et~al.}(2023)Bashi, Mazeh, \& Faigler}]{bashi_features_2023}
Bashi, D., Mazeh, T., \& Faigler, S. 2023, Monthly Notices of the Royal
  Astronomical Society, stad999, \dodoi{10.1093/mnras/stad999}

\bibitem[{Birky {et~al.}(2021)Birky, Barnes, \& Fleming}]{birky_improved_2021}
Birky, J., Barnes, R., \& Fleming, D.~P. 2021, Research Notes of the AAS, 5,
  122, \dodoi{10.3847/2515-5172/ac034c}

\bibitem[{Birky {et~al.}(in prep.)Birky, Fleming, \& Barnes}]{birky_alabi}
Birky, J., Fleming, D.~P., \& Barnes, R.~K. in prep.

\bibitem[{Bolmont \& Mathis(2016)}]{bolmont_effect_2016}
Bolmont, E., \& Mathis, S. 2016, Celestial Mechanics and Dynamical Astronomy,
  126, 275, \dodoi{10.1007/s10569-016-9690-3}

\bibitem[{Borucki {et~al.}(2010)Borucki, Koch, Basri, Batalha, Brown, Caldwell,
  Caldwell, Christensen-Dalsgaard, Cochran, DeVore, Dunham, Dupree, Gautier,
  Geary, Gilliland, Gould, Howell, Jenkins, Kondo, Latham, Marcy, Meibom,
  Kjeldsen, Lissauer, Monet, Morrison, Sasselov, Tarter, Boss, Brownlee, Owen,
  Buzasi, Charbonneau, Doyle, Fortney, Ford, Holman, Seager, Steffen, Welsh,
  Rowe, Anderson, Buchhave, Ciardi, Walkowicz, Sherry, Horch, Isaacson,
  Everett, Fischer, Torres, Johnson, Endl, MacQueen, Bryson, Dotson, Haas,
  Kolodziejczak, Van~Cleve, Chandrasekaran, Twicken, Quintana, Clarke, Allen,
  Li, Wu, Tenenbaum, Verner, Bruhweiler, Barnes, \& Prsa}]{borucki_kepler_2010}
Borucki, W.~J., Koch, D., Basri, G., {et~al.} 2010, Science, 327, 977,
  \dodoi{10.1126/science.1185402}

\bibitem[{Breimann {et~al.}(2021)Breimann, Matt, \&
  Naylor}]{breimann_statistical_2021}
Breimann, A.~A., Matt, S.~P., \& Naylor, T. 2021, The Astrophysical Journal,
  913, 75, \dodoi{10.3847/1538-4357/abf0a3}

\bibitem[{Burkart {et~al.}(2014)Burkart, Quataert, \&
  Arras}]{burkart_dynamical_2014}
Burkart, J., Quataert, E., \& Arras, P. 2014, Monthly Notices of the Royal
  Astronomical Society, 443, 2957, \dodoi{10.1093/mnras/stu1366}

\bibitem[{Cameron \& Jardine(2018)}]{cameron_hierarchical_2018}
Cameron, A.~C., \& Jardine, M. 2018, Monthly Notices of the Royal Astronomical
  Society, 476, 2542, \dodoi{10.1093/mnras/sty292}

\bibitem[{Casey {et~al.}(2019)Casey, Ho, Ness, Rix, Angelou, Hekker, Tout,
  Lattanzio, Karakas, Woods, Price-Whelan, \& Schlaufman}]{casey_tidal_2019}
Casey, A.~R., Ho, A. Y.~Q., Ness, M., {et~al.} 2019, The Astrophysical Journal,
  880, 125, \dodoi{10.3847/1538-4357/ab27bf}

\bibitem[{Chontos {et~al.}(2019)Chontos, Huber, Latham, Bieryla, Eylen,
  Bedding, Berger, Buchhave, Campante, Chaplin, Colman, Coughlin, Davies,
  Hirano, Howard, \& Isaacson}]{chontos_curious_2019}
Chontos, A., Huber, D., Latham, D.~W., {et~al.} 2019, The Astronomical Journal,
  157, 192, \dodoi{10.3847/1538-3881/ab0e8e}

\bibitem[{Colombo \& Shapiro(1966)}]{colombo_rotation_1966}
Colombo, G., \& Shapiro, I.~I. 1966, The Astrophysical Journal, 145, 296,
  \dodoi{10.1086/148762}

\bibitem[{Correia \& Laskar(2004)}]{correia_mercurys_2004}
Correia, A. C.~M., \& Laskar, J. 2004, Nature, 429, 848,
  \dodoi{10.1038/nature02609}

\bibitem[{Counselman(1973)}]{counselman_outcomes_1973}
Counselman, III, C.~C. 1973, {\textbackslash}apj, 180, 307,
  \dodoi{10.1086/151964}

\bibitem[{Cranmer \& Saar(2011)}]{cranmer_testing_2011}
Cranmer, S.~R., \& Saar, S.~H. 2011, {\textbackslash}apj, 741, 54,
  \dodoi{10.1088/0004-637X/741/1/54}

\bibitem[{David {et~al.}(2015)David, Hillenbrand, Cody, Carpenter, \&
  Howard}]{david_k2_2015}
David, T.~J., Hillenbrand, L.~A., Cody, A.~M., Carpenter, J.~M., \& Howard,
  A.~W. 2015, The Astrophysical Journal, 816, 21,
  \dodoi{10.3847/0004-637X/816/1/21}

\bibitem[{David {et~al.}(2016)David, Conroy, Hillenbrand, Stassun, Stauffer,
  Rebull, Cody, Isaacson, Howard, \& Aigrain}]{david_new_2016}
David, T.~J., Conroy, K.~E., Hillenbrand, L.~A., {et~al.} 2016, The
  Astronomical Journal, 151, 112, \dodoi{10.3847/0004-6256/151/5/112}

\bibitem[{{Duquennoy} \& {Mayor}(1991)}]{DuquennoyMayor91}
{Duquennoy}, A., \& {Mayor}, M. 1991, \aap, 248, 485

\bibitem[{Efroimsky \& Williams(2009)}]{efroimsky_tidal_2009}
Efroimsky, M., \& Williams, J.~G. 2009, Celestial Mechanics and Dynamical
  Astronomy, 104, 257, \dodoi{10.1007/s10569-009-9204-7}

\bibitem[{El-Badry {et~al.}(2018)El-Badry, Ting, Rix, Quataert, Weisz, Cargile,
  Conroy, Hogg, Bergemann, \& Liu}]{el-badry_discovery_2018}
El-Badry, K., Ting, Y.-S., Rix, H.-W., {et~al.} 2018, {\textbackslash}mnras,
  476, 528, \dodoi{10.1093/mnras/sty240}

\bibitem[{Fabrycky \& Tremaine(2007)}]{fabrycky_shrinking_2007}
Fabrycky, D., \& Tremaine, S. 2007, The Astrophysical Journal, 669, 1298,
  \dodoi{10.1086/521702}

\bibitem[{Ferraz-Mello {et~al.}(2008)Ferraz-Mello, Rodríguez, \&
  Hussmann}]{ferraz-mello_tidal_2008}
Ferraz-Mello, S., Rodríguez, A., \& Hussmann, H. 2008, Celestial Mechanics and
  Dynamical Astronomy, 101, 171, \dodoi{10.1007/s10569-008-9133-x}

\bibitem[{Fleming {et~al.}(2019)Fleming, Barnes, Davenport, \&
  Luger}]{fleming_rotation_2019}
Fleming, D.~P., Barnes, R., Davenport, J. R.~A., \& Luger, R. 2019, The
  Astrophysical Journal, 881, 88, \dodoi{10.3847/1538-4357/ab2ed2}

\bibitem[{Fleming {et~al.}(2018)Fleming, Barnes, Graham, Luger, \&
  Quinn}]{fleming_lack_2018}
Fleming, D.~P., Barnes, R., Graham, D.~E., Luger, R., \& Quinn, T.~R. 2018, The
  Astrophysical Journal, 858, 86, \dodoi{10.3847/1538-4357/aabd38}

\bibitem[{Fleming {et~al.}(2020)Fleming, Barnes, Luger, \&
  VanderPlas}]{fleming_xuv_2020}
Fleming, D.~P., Barnes, R., Luger, R., \& VanderPlas, J.~T. 2020, The
  Astrophysical Journal, 891, 155, \dodoi{10.3847/1538-4357/ab77ad}

\bibitem[{Fleming \& VanderPlas(2018)}]{fleming_approxposterior_2018}
Fleming, D.~P., \& VanderPlas, J. 2018, Journal of Open Source Software, 3,
  781, \dodoi{10.21105/joss.00781}

\bibitem[{Foreman-Mackey(2016)}]{foreman-mackey_cornerpy_2016}
Foreman-Mackey, D. 2016, The Journal of Open Source Software, 1,
  \dodoi{10.21105/joss.00024}

\bibitem[{{Fuller} \& {Lai}(2012)}]{FullerLai12}
{Fuller}, J., \& {Lai}, D. 2012, \mnras, 421, 426,
  \dodoi{10.1111/j.1365-2966.2011.20320.x}

\bibitem[{Gillen {et~al.}(2017)Gillen, Hillenbrand, David, Aigrain, Rebull,
  Stauffer, Cody, \& Queloz}]{gillen_new_2017}
Gillen, E., Hillenbrand, L.~A., David, T.~J., {et~al.} 2017, The Astrophysical
  Journal, 849, 11, \dodoi{10.3847/1538-4357/aa84b3}

\bibitem[{Ginat \& Perets(2021)}]{ginat_analytical_2021}
Ginat, Y.~B., \& Perets, H.~B. 2021, Physical Review X, 11, 031020,
  \dodoi{10.1103/PhysRevX.11.031020}

\bibitem[{Goldreich \& Peale(1966)}]{goldreich_spin-orbit_1966}
Goldreich, P., \& Peale, S. 1966, The Astronomical Journal, 71, 425,
  \dodoi{10.1086/109947}

\bibitem[{Goldreich \& Soter(1966)}]{goldreich_q_1966}
Goldreich, P., \& Soter, S. 1966, Icarus, 5, 375,
  \dodoi{10.1016/0019-1035(66)90051-0}

\bibitem[{Greenberg(1974)}]{greenberg_outcomes_1974}
Greenberg, R. 1974, ıcarus, 23, 51, \dodoi{10.1016/0019-1035(74)90103-1}

\bibitem[{Greenberg(2009)}]{greenberg_frequency_2009}
---. 2009, The Astrophysical Journal, 698, L42,
  \dodoi{10.1088/0004-637X/698/1/L42}

\bibitem[{Hamer \& Schlaufman(2019)}]{hamer_hot_2019}
Hamer, J.~H., \& Schlaufman, K.~C. 2019, The Astronomical Journal, 158, 190,
  \dodoi{10.3847/1538-3881/ab3c56}

\bibitem[{Hamers {et~al.}(2021)Hamers, Rantala, Neunteufel, Preece, \&
  Vynatheya}]{hamers_multiple_2021}
Hamers, A.~S., Rantala, A., Neunteufel, P., Preece, H., \& Vynatheya, P. 2021,
  Monthly Notices of the Royal Astronomical Society, 502, 4479,
  \dodoi{10.1093/mnras/stab287}

\bibitem[{Hansen(2010)}]{hansen_calibration_2010}
Hansen, B. M.~S. 2010, The Astrophysical Journal, 723, 285,
  \dodoi{10.1088/0004-637X/723/1/285}

\bibitem[{Hatzes(2019)}]{Hatzes2019}
Hatzes, A.~P. 2019, in The Doppler Method for the Detection of Exoplanets,
  2514-3433 (IOP Publishing), 13--1 to 13--14,
  \dodoi{10.1088/2514-3433/ab46a3ch13}

\bibitem[{Heller {et~al.}(2011)Heller, Leconte, \& Barnes}]{heller_tidal_2011}
Heller, R., Leconte, J., \& Barnes, R. 2011, {\textbackslash}aap, 528, A27,
  \dodoi{10.1051/0004-6361/201015809}

\bibitem[{Herman \& Usher(2017)}]{herman_salib_2017}
Herman, J., \& Usher, W. 2017, Journal of Open Source Software, 2, 97,
  \dodoi{10.21105/joss.00097}

\bibitem[{{Hobson-Ritz} {et~al.}(2025){Hobson-Ritz}, {Birky}, {Peterson},
  {Gwartney}, {Wong}, {Delker}, {Gordon}, {Gilbert}, {Davenport}, \&
  {Barnes}}]{hobson_ritz_2025}
{Hobson-Ritz}, M., {Birky}, J., {Peterson}, L., {et~al.} 2025, arXiv e-prints,
  arXiv:2501.04082, \dodoi{10.48550/arXiv.2501.04082}

\bibitem[{Howell {et~al.}(2014)Howell, Sobeck, Haas, Still, Barclay, Mullally,
  Troeltzsch, Aigrain, Bryson, Caldwell, Chaplin, Cochran, Huber, Marcy,
  Miglio, Najita, Smith, Twicken, \& Fortney}]{howell_k2_2014}
Howell, S.~B., Sobeck, C., Haas, M., {et~al.} 2014, {\textbackslash}pasp, 126,
  398, \dodoi{10.1086/676406}

\bibitem[{Hunter(2007)}]{hunter_matplotlib_2007}
Hunter, J.~D. 2007, Computing in Science and Engineering, 9, 90,
  \dodoi{10.1109/MCSE.2007.55}

\bibitem[{Hut(1980)}]{hut_stability_1980}
Hut, P. 1980, Astronomy and Astrophysics, 92, 167.
\newblock \url{https://ui.adsabs.harvard.edu/abs/1980A&A....92..167H/abstract}

\bibitem[{Hut(1981)}]{hut_tidal_1981}
---. 1981, Astronomy and Astrophysics, 99, 126.
\newblock \url{http://adsabs.harvard.edu/abs/1981A%26A....99..126H}

\bibitem[{Jackson {et~al.}(2009)Jackson, Barnes, \&
  Greenberg}]{jackson_observational_2009}
Jackson, B., Barnes, R., \& Greenberg, R. 2009, {\textbackslash}apj, 698, 1357,
  \dodoi{10.1088/0004-637X/698/2/1357}

\bibitem[{Jackson {et~al.}(2008)Jackson, Greenberg, \&
  Barnes}]{jackson_tidal_2008}
Jackson, B., Greenberg, R., \& Barnes, R. 2008, {\textbackslash}apj, 678, 1396,
  \dodoi{10.1086/529187}

\bibitem[{Jermyn {et~al.}(2020)Jermyn, Tayar, \&
  Fuller}]{jermyn_differential_2020}
Jermyn, A.~S., Tayar, J., \& Fuller, J. 2020, Monthly Notices of the Royal
  Astronomical Society, 491, 690, \dodoi{10.1093/mnras/stz2983}

\bibitem[{Jones {et~al.}(2001)Jones, Oliphant, Peterson, \&
  {others}}]{jones_scipy_2001}
Jones, E., Oliphant, T., Peterson, P., \& {others}. 2001, {SciPy}: {Open}
  source scientific tools for {Python}.
\newblock \url{http://www.scipy.org/}

\bibitem[{Kandasamy {et~al.}(2017)Kandasamy, Schneider, \&
  Póczos}]{kandasamy_query_2017}
Kandasamy, K., Schneider, J., \& Póczos, B. 2017, Artificial Intelligence,
  243, 45 , \dodoi{https://doi.org/10.1016/j.artint.2016.11.002}

\bibitem[{Kaula(1966)}]{kaula_theory_1966}
Kaula, W.~M. 1966, Theory of satellite geodesy. {Applications} of satellites to
  geodesy

\bibitem[{Kounkel {et~al.}(2021)Kounkel, Covey, Stassun, Price-Whelan,
  Holtzman, Chojnowski, Longa-Peña, Román-Zúñiga, Hernandez, Serna,
  Badenes, De~Lee, Majewski, Stringfellow, Kratter, Moe, Frinchaboy, Beaton,
  Fernández-Trincado, Mahadevan, Minniti, Beers, Schneider, Barbá,
  Brownstein, García-Hernández, Pan, \& Bizyaev}]{kounkel_double-lined_2021}
Kounkel, M., Covey, K.~R., Stassun, K.~G., {et~al.} 2021, The Astronomical
  Journal, 162, 184, \dodoi{10.3847/1538-3881/ac1798}

\bibitem[{Lai(1997)}]{lai_dynamical_1997}
Lai, D. 1997, The Astrophysical Journal, 490, 847, \dodoi{10.1086/304899}

\bibitem[{Leconte {et~al.}(2010)Leconte, Chabrier, Baraffe, \&
  Levrard}]{leconte_is_2010}
Leconte, J., Chabrier, G., Baraffe, I., \& Levrard, B. 2010, Astronomy and
  Astrophysics, 516, A64, \dodoi{10.1051/0004-6361/201014337}

\bibitem[{Levrard {et~al.}(2007)Levrard, Correia, Chabrier, Baraffe, Selsis, \&
  Laskar}]{levrard_tidal_2007}
Levrard, B., Correia, A. C.~M., Chabrier, G., {et~al.} 2007, Astronomy \&
  Astrophysics, 462, L5, \dodoi{10.1051/0004-6361:20066487}

\bibitem[{Lurie {et~al.}(2017)Lurie, Vyhmeister, Hawley, Adilia, Chen,
  Davenport, Jurić, Puig-Holzman, \& Weisenburger}]{lurie_tidal_2017}
Lurie, J.~C., Vyhmeister, K., Hawley, S.~L., {et~al.} 2017, The Astronomical
  Journal, 154, 250, \dodoi{10.3847/1538-3881/aa974d}

\bibitem[{MacDonald(1964)}]{macdonald_tidal_1964}
MacDonald, G. J.~F. 1964, Reviews of Geophysics, 2, 467,
  \dodoi{10.1029/RG002i003p00467}

\bibitem[{Majewski {et~al.}(2015)Majewski, Schiavon, Frinchaboy,
  Allende~Prieto, Barkhouser, Bizyaev, Blank, Brunner, Burton, Carrera,
  Chojnowski, Cunha, Epstein, Fitzgerald, García~Pérez, Hearty, Henderson,
  Holtzman, Johnson, Lam, Lawler, Maseman, Mészáros, Nelson, Nguyen, Nidever,
  Pinsonneault, Shetrone, Smee, Smith, Stolberg, Skrutskie, Walker, Wilson,
  Zasowski, Anders, Basu, Beland, Blanton, Bovy, Brownstein, Carlberg, Chaplin,
  Chiappini, Eisenstein, Elsworth, Feuillet, Fleming, Galbraith-Frew, García,
  García-Hernández, Gillespie, Girardi, Gunn, Hasselquist, Hayden, Hekker,
  Ivans, Kinemuchi, Klaene, Mahadevan, Mathur, Mosser, Muna, Munn, Nichol,
  O'Connell, Parejko, Robin, Rocha-Pinto, Schultheis, Serenelli, Shane,
  Silva~Aguirre, Sobeck, Thompson, Troup, Weinberg, \&
  Zamora}]{majewski_apache_2015}
Majewski, S.~R., Schiavon, R.~P., Frinchaboy, P.~M., {et~al.} 2015,
  {\textbackslash}aj, 154, 94, \dodoi{10.3847/1538-3881/aa784d}

\bibitem[{Mardling \& Aarseth(2001)}]{mardling_tidal_2001}
Mardling, R.~A., \& Aarseth, S.~J. 2001, Monthly Notices of the Royal
  Astronomical Society, 321, 398, \dodoi{10.1046/j.1365-8711.2001.03974.x}

\bibitem[{Marilli {et~al.}(2007)Marilli, Frasca, Covino, Alcalá, Catalano,
  Fernández, Arellano~Ferro, Rubio-Herrera, \&
  Spezzi}]{marilli_rotational_2007}
Marilli, E., Frasca, A., Covino, E., {et~al.} 2007, {\textbackslash}aap, 463,
  1081, \dodoi{10.1051/0004-6361:20066458}

\bibitem[{{Mathieu} \& {Mazeh}(1988)}]{Mathieu_mazeh_1988}
{Mathieu}, R.~D., \& {Mazeh}, T. 1988, \apj, 326, 256, \dodoi{10.1086/166087}

\bibitem[{Mathieu {et~al.}(2004)Mathieu, Meibom, \& Dolan}]{mathieu_tidal_2004}
Mathieu, R.~D., Meibom, S., \& Dolan, C.~J. 2004, The Astrophysical Journal,
  602, L121, \dodoi{10.1086/382686}

\bibitem[{Mathur {et~al.}(2017)Mathur, Huber, Batalha, Ciardi, Bastien,
  Bieryla, Buchhave, Cochran, Endl, Esquerdo, Furlan, Howard, Howell, Isaacson,
  Latham, MacQueen, \& Silva}]{mathur_revised_2017}
Mathur, S., Huber, D., Batalha, N.~M., {et~al.} 2017, The Astrophysical Journal
  Supplement Series, 229, 30, \dodoi{10.3847/1538-4365/229/2/30}

\bibitem[{Matson {et~al.}(2016)Matson, Gies, Guo, \&
  Orosz}]{matson_fundamental_2016}
Matson, R.~A., Gies, D.~R., Guo, Z., \& Orosz, J.~A. 2016, The Astronomical
  Journal, 151, 139, \dodoi{10.3847/0004-6256/151/6/139}

\bibitem[{Matt {et~al.}(2015)Matt, Brun, Baraffe, Bouvier, \&
  Chabrier}]{matt_mass-dependence_2015}
Matt, S.~P., Brun, A.~S., Baraffe, I., Bouvier, J., \& Chabrier, G. 2015,
  {\textbackslash}apjl, 799, L23, \dodoi{10.1088/2041-8205/799/2/L23}

\bibitem[{Matt {et~al.}(2019)Matt, Brun, Baraffe, Bouvier, \&
  Chabrier}]{matt_erratum_2019}
---. 2019, {\textbackslash}apjl, 870, L27, \dodoi{10.3847/2041-8213/aafa1b}

\bibitem[{Matt {et~al.}(2012)Matt, MacGregor, Pinsonneault, \&
  Greene}]{matt_magnetic_2012}
Matt, S.~P., MacGregor, K.~B., Pinsonneault, M.~H., \& Greene, T.~P. 2012, The
  Astrophysical Journal Letters, 754, L26, \dodoi{10.1088/2041-8205/754/2/L26}

\bibitem[{{Mayor} \& {Mermilliod}(1984)}]{MayorMermilliod84}
{Mayor}, M., \& {Mermilliod}, J.~C. 1984, in Observational Tests of the Stellar
  Evolution Theory, 411

\bibitem[{Mazeh(2008)}]{mazeh_observational_2008}
Mazeh, T. 2008, EAS Publications Series, 29, 1, \dodoi{10.1051/eas:0829001}

\bibitem[{Meibom {et~al.}(2015)Meibom, Barnes, Platais, Gilliland, Latham, \&
  Mathieu}]{meibom_spin-down_2015}
Meibom, S., Barnes, S.~A., Platais, I., {et~al.} 2015, {\textbackslash}nat,
  517, 589, \dodoi{10.1038/nature14118}

\bibitem[{Meibom \& Mathieu(2005)}]{meibom_robust_2005}
Meibom, S., \& Mathieu, R.~D. 2005, {\textbackslash}apj, 620, 970,
  \dodoi{10.1086/427082}

\bibitem[{Meibom {et~al.}(2006)Meibom, Mathieu, \&
  Stassun}]{meibom_observational_2006}
Meibom, S., Mathieu, R.~D., \& Stassun, K.~G. 2006, {\textbackslash}apj, 653,
  621, \dodoi{10.1086/508252}

\bibitem[{Mirouh {et~al.}(2023)Mirouh, Hendriks, Dykes, Moe, \&
  Izzard}]{mirouh_detailed_2023}
Mirouh, G.~M., Hendriks, D.~D., Dykes, S., Moe, M., \& Izzard, R.~G. 2023,
  Monthly Notices of the Royal Astronomical Society, 524, 3978,
  \dodoi{10.1093/mnras/stad2048}

\bibitem[{Moe \& Kratter(2018)}]{moe_dynamical_2018}
Moe, M., \& Kratter, K.~M. 2018, The Astrophysical Journal, 854, 44,
  \dodoi{10.3847/1538-4357/aaa6d2}

\bibitem[{Morton \& Winn(2014)}]{morton_obliquities_2014}
Morton, T.~D., \& Winn, J.~N. 2014, The Astrophysical Journal, 796, 47,
  \dodoi{10.1088/0004-637X/796/1/47}

\bibitem[{Naoz(2016)}]{naoz_eccentric_2016}
Naoz, S. 2016, Annual Review of Astronomy and Astrophysics, 54, 441,
  \dodoi{10.1146/annurev-astro-081915-023315}

\bibitem[{Ogilvie(2013)}]{ogilvie_tides_2013}
Ogilvie, G.~I. 2013, Monthly Notices of the Royal Astronomical Society, 429,
  613, \dodoi{10.1093/mnras/sts362}

\bibitem[{Oleskiewicz \& Baugh(2019)}]{oleskiewicz_sensitivity_2019}
Oleskiewicz, P., \& Baugh, C.~M. 2019, arXiv:1910.01745 [astro-ph],
  \dodoi{10.1093/mnras/stz3560}

\bibitem[{{Patel} \& {Penev}(2022)}]{PatelPenev22}
{Patel}, R., \& {Penev}, K. 2022, \mnras, 512, 3651,
  \dodoi{10.1093/mnras/stac203}

\bibitem[{Patel \& Penev(2022)}]{patel_constraining_2022}
Patel, R., \& Penev, K. 2022, Monthly Notices of the Royal Astronomical
  Society, stac203, \dodoi{10.1093/mnras/stac203}

\bibitem[{Patra {et~al.}(2017)Patra, Winn, Holman, Yu, Deming, \&
  Dai}]{patra_apparently_2017}
Patra, K.~C., Winn, J.~N., Holman, M.~J., {et~al.} 2017, {\textbackslash}aj,
  154, 4, \dodoi{10.3847/1538-3881/aa6d75}

\bibitem[{Patra {et~al.}(2020)Patra, Winn, Holman, Gillon, Burdanov, Jehin,
  Delrez, Pozuelos, Barkaoui, Benkhaldoun, Narita, Fukui, Kusakabe, Kawauchi,
  Terada, Bouma, Weinberg, \& Broome}]{patra_continuing_2020}
---. 2020, The Astronomical Journal, 159, 150, \dodoi{10.3847/1538-3881/ab7374}

\bibitem[{Paxton {et~al.}(2010)Paxton, Bildsten, Dotter, Herwig, Lesaffre, \&
  Timmes}]{paxton_modules_2010}
Paxton, B., Bildsten, L., Dotter, A., {et~al.} 2010, The Astrophysical Journal
  Supplement Series, 192, 3, \dodoi{10.1088/0067-0049/192/1/3}

\bibitem[{Penev {et~al.}(2018)Penev, Bouma, Winn, \& Hartman}]{penev_2018}
Penev, K., Bouma, L.~G., Winn, J.~N., \& Hartman, J.~D. 2018, The Astronomical
  Journal, 155, 165, \dodoi{10.3847/1538-3881/aaaf71}

\bibitem[{Penev {et~al.}(2014)Penev, Zhang, \& Jackson}]{penev_poet_2014}
Penev, K., Zhang, M., \& Jackson, B. 2014, Publications of the Astronomical
  Society of the Pacific, 126, 553, \dodoi{10.1086/677042}

\bibitem[{Penev \& Schussler(2022)}]{penev_comprehensive_2022}
Penev, K.~M., \& Schussler, J.~A. 2022, Monthly Notices of the Royal
  Astronomical Society, stac2618, \dodoi{10.1093/mnras/stac2618}

\bibitem[{Price-Whelan \& Goodman(2018)}]{price-whelan_binary_2018}
Price-Whelan, A.~M., \& Goodman, J. 2018, The Astrophysical Journal, 867, 5,
  \dodoi{10.3847/1538-4357/aae264}

\bibitem[{Price-Whelan {et~al.}(2020)Price-Whelan, Hogg, Rix, Beaton, Lewis,
  Nidever, Almeida, Badenes, Barba, Beers, Carlberg, Lee, Fernández-Trincado,
  Frinchaboy, García-Hernández, Green, Hasselquist, Longa-Peña, Majewski,
  Nitschelm, Sobeck, Stassun, Stringfellow, \& Troup}]{price-whelan_close_2020}
Price-Whelan, A.~M., Hogg, D.~W., Rix, H.-W., {et~al.} 2020, The Astrophysical
  Journal, 895, 2, \dodoi{10.3847/1538-4357/ab8acc}

\bibitem[{Rasmussen \& Williams(2006)}]{rasmussen_gaussian_2006}
Rasmussen, C.~E., \& Williams, C. K.~I. 2006, Gaussian {Processes} for
  {Machine} {Learning}

\bibitem[{Rebull {et~al.}(2006)Rebull, Stauffer, Megeath, Hora, \&
  Hartmann}]{rebull_correlation_2006}
Rebull, L.~M., Stauffer, J.~R., Megeath, S.~T., Hora, J.~L., \& Hartmann, L.
  2006, {\textbackslash}apj, 646, 297, \dodoi{10.1086/504865}

\bibitem[{Reiners \& Mohanty(2012)}]{reiners_radius-dependent_2012}
Reiners, A., \& Mohanty, S. 2012, {\textbackslash}apj, 746, 43,
  \dodoi{10.1088/0004-637X/746/1/43}

\bibitem[{Repetto \& Nelemans(2014)}]{repetto_coupled_2014}
Repetto, S., \& Nelemans, G. 2014, Monthly Notices of the Royal Astronomical
  Society, 444, 542, \dodoi{10.1093/mnras/stu1454}

\bibitem[{{Ricker} {et~al.}(2015){Ricker}, {Winn}, {Vanderspek}, {Latham},
  {Bakos}, {Bean}, {Berta-Thompson}, {Brown}, {Buchhave}, {Butler}, {Butler},
  {Chaplin}, {Charbonneau}, {Christensen-Dalsgaard}, {Clampin}, {Deming},
  {Doty}, {De Lee}, {Dressing}, {Dunham}, {Endl}, {Fressin}, {Ge}, {Henning},
  {Holman}, {Howard}, {Ida}, {Jenkins}, {Jernigan}, {Johnson}, {Kaltenegger},
  {Kawai}, {Kjeldsen}, {Laughlin}, {Levine}, {Lin}, {Lissauer}, {MacQueen},
  {Marcy}, {McCullough}, {Morton}, {Narita}, {Paegert}, {Palle}, {Pepe},
  {Pepper}, {Quirrenbach}, {Rinehart}, {Sasselov}, {Sato}, {Seager},
  {Sozzetti}, {Stassun}, {Sullivan}, {Szentgyorgyi}, {Torres}, {Udry}, \&
  {Villasenor}}]{Ricker15}
{Ricker}, G.~R., {Winn}, J.~N., {Vanderspek}, R., {et~al.} 2015, Journal of
  Astronomical Telescopes, Instruments, and Systems, 1, 014003,
  \dodoi{10.1117/1.JATIS.1.1.014003}

\bibitem[{Saltelli {et~al.}(2017)Saltelli, Aleksankina, Becker, Fennell,
  Ferretti, Holst, Li, \& Wu}]{saltelli_why_2017}
Saltelli, A., Aleksankina, K., Becker, W., {et~al.} 2017, Why {So} {Many}
  {Published} {Sensitivity} {Analyses} {Are} {False}. {A} {Systematic} {Review}
  of {Sensitivity} {Analysis} {Practices}, Tech. Rep. arXiv:1711.11359, arXiv,
  \dodoi{10.48550/arXiv.1711.11359}

\bibitem[{Saltelli {et~al.}(2010)Saltelli, Annoni, Azzini, Campolongo, Ratto,
  \& Tarantola}]{saltelli_variance_2010}
Saltelli, A., Annoni, P., Azzini, I., {et~al.} 2010, Computer Physics
  Communications, 181, 259, \dodoi{10.1016/j.cpc.2009.09.018}

\bibitem[{Skumanich(1972)}]{skumanich_time_1972}
Skumanich, A. 1972, {\textbackslash}apj, 171, 565, \dodoi{10.1086/151310}

\bibitem[{Sobol(2001)}]{sobol_global_2001}
Sobol, I.~M. 2001, Mathematics and Computers in Simulation, 55, 271,
  \dodoi{10.1016/S0378-4754(00)00270-6}

\bibitem[{Soderblom(2010)}]{soderblom_ages_2010}
Soderblom, D.~R. 2010, Annual Review of Astronomy and Astrophysics, 48, 581,
  \dodoi{10.1146/annurev-astro-081309-130806}

\bibitem[{Song {et~al.}(2018)Song, Meynet, Maeder, Ekstrom, Eggenberger,
  Georgy, Qin, Fragos, Soerensen, Barblan, \& Wade}]{song_close_2018}
Song, H.~F., Meynet, G., Maeder, A., {et~al.} 2018, Astronomy \& Astrophysics,
  609, A3, \dodoi{10.1051/0004-6361/201731073}

\bibitem[{Southworth \& Clausen(2006)}]{southworth_eclipsing_2006}
Southworth, J., \& Clausen, J.~V. 2006, arXiv:astro-ph/0608016.
\newblock \url{http://arxiv.org/abs/astro-ph/0608016}

\bibitem[{Stassun {et~al.}(1999)Stassun, Mathieu, Mazeh, \&
  Vrba}]{stassun_rotation_1999}
Stassun, K.~G., Mathieu, R.~D., Mazeh, T., \& Vrba, F.~J. 1999,
  {\textbackslash}aj, 117, 2941, \dodoi{10.1086/300881}

\bibitem[{Sterzik {et~al.}(2003)Sterzik, Durisen, \&
  Zinnecker}]{sterzik_how_2003}
Sterzik, M.~F., Durisen, R.~H., \& Zinnecker, H. 2003, Astronomy \&
  Astrophysics, 411, 91, \dodoi{10.1051/0004-6361:20034219}

\bibitem[{Terquem {et~al.}(1998)Terquem, Papaloizou, Nelson, \&
  Lin}]{terquem_tidal_1998}
Terquem, C., Papaloizou, J. C.~B., Nelson, R.~P., \& Lin, D. N.~C. 1998, The
  Astrophysical Journal, 502, 788, \dodoi{10.1086/305927}

\bibitem[{Tokovinin(2021)}]{tokovinin_architecture_2021}
Tokovinin, A. 2021, Universe, 7, 352, \dodoi{10.3390/universe7090352}

\bibitem[{Tokovinin \& Moe(2020)}]{tokovinin_formation_2020}
Tokovinin, A., \& Moe, M. 2020, Monthly Notices of the Royal Astronomical
  Society, 491, 5158, \dodoi{10.1093/mnras/stz3299}

\bibitem[{Toonen {et~al.}(2016)Toonen, Hamers, \&
  Portegies~Zwart}]{toonen_evolution_2016}
Toonen, S., Hamers, A., \& Portegies~Zwart, S. 2016, Computational Astrophysics
  and Cosmology, 3, 6, \dodoi{10.1186/s40668-016-0019-0}

\bibitem[{Torres {et~al.}(2018)Torres, Curtis, Vanderburg, Kraus, \&
  Rizzuto}]{torres_eclipsing_2018}
Torres, G., Curtis, J.~L., Vanderburg, A., Kraus, A.~L., \& Rizzuto, A. 2018,
  The Astrophysical Journal, 866, 67, \dodoi{10.3847/1538-4357/aadca8}

\bibitem[{Triaud(2018)}]{triaud_rossiter-mclaughlin_2018}
Triaud, A. H. M.~J. 2018, 1375--1401, \dodoi{10.1007/978-3-319-55333-7_2}

\bibitem[{Usher {et~al.}(2016)Usher, Herman, Whealton, Hadka, xantares, Rios,
  bernardoct, Mutel, \& Engelen}]{usher_salibsalib_2016}
Usher, W., Herman, J., Whealton, C., {et~al.} 2016, {SALib}/{SALib}: {Launch}!,
   Zenodo, \dodoi{10.5281/zenodo.160164}

\bibitem[{van~der Walt {et~al.}(2011)van~der Walt, Colbert, \&
  Varoquaux}]{van_der_walt_numpy_2011}
van~der Walt, S., Colbert, S.~C., \& Varoquaux, G. 2011, Computing in Science
  and Engineering, 13, 22, \dodoi{10.1109/MCSE.2011.37}

\bibitem[{Vidal \& Barker(2020)}]{vidal_turbulent_2020}
Vidal, J., \& Barker, A.~J. 2020, The Astrophysical Journal, 888, L31,
  \dodoi{10.3847/2041-8213/ab6219}

\bibitem[{Vissapragada {et~al.}(2022)Vissapragada, Chontos, Greklek-McKeon,
  Knutson, Dai, González, Grunblatt, Huber, \&
  Saunders}]{vissapragada_possible_2022}
Vissapragada, S., Chontos, A., Greklek-McKeon, M., {et~al.} 2022, The
  Astrophysical Journal Letters, 941, L31, \dodoi{10.3847/2041-8213/aca47e}

\bibitem[{Vázquez-Semadeni {et~al.}(2019)Vázquez-Semadeni, Palau,
  Ballesteros-Paredes, Gómez, \&
  Zamora-Avilés}]{vazquez-semadeni_global_2019}
Vázquez-Semadeni, E., Palau, A., Ballesteros-Paredes, J., Gómez, G.~C., \&
  Zamora-Avilés, M. 2019, Monthly Notices of the Royal Astronomical Society,
  490, 3061, \dodoi{10.1093/mnras/stz2736}

\bibitem[{Willems(2000)}]{willems_modelling_2000}
Willems, J.~C. 2000, Mathematics and Computers in Simulation, 53, 227,
  \dodoi{10.1016/S0378-4754(00)00209-3}

\bibitem[{Windemuth {et~al.}(2019)Windemuth, Agol, Ali, \&
  Kiefer}]{windemuth_modeling_2019}
Windemuth, D., Agol, E., Ali, A., \& Kiefer, F. 2019, Monthly Notices of the
  Royal Astronomical Society, 489, 1644, \dodoi{10.1093/mnras/stz2137}

\bibitem[{{Witte} \& {Savonije}(1999)}]{WitteSavonije99}
{Witte}, M.~G., \& {Savonije}, G.~J. 1999, \aap, 350, 129.
\newblock \doarXiv{astro-ph/9909073}

\bibitem[{Witte \& Savonije(2002)}]{witte_orbital_2002}
Witte, M.~G., \& Savonije, G.~J. 2002, Astronomy \& Astrophysics, 386, 222,
  \dodoi{10.1051/0004-6361:20020155}

\bibitem[{Yee {et~al.}(2019)Yee, Winn, Knutson, Patra, Vissapragada, Zhang,
  Holman, Shporer, \& Wright}]{yee_orbit_2019}
Yee, S.~W., Winn, J.~N., Knutson, H.~A., {et~al.} 2019, The Astrophysical
  Journal Letters, 888, L5, \dodoi{10.3847/2041-8213/ab5c16}

\bibitem[{Zahn(1975)}]{zahn_dynamical_1975}
Zahn, J.-P. 1975, {\textbackslash}aap, 41, 329

\bibitem[{Zahn(2008)}]{zahn_tidal_2008}
---. 2008, EAS Publications Series, 29, 67, \dodoi{10.1051/eas:0829002}

\bibitem[{Zahn \& Bouchet(1989)}]{zahn_tidal_1989}
Zahn, J.-P., \& Bouchet, L. 1989, {\textbackslash}aap, 223, 112

\bibitem[{Zanazzi \& Wu(2021)}]{zanazzi_tidal_2021}
Zanazzi, J.~J., \& Wu, Y. 2021, The Astronomical Journal, 161, 263,
  \dodoi{10.3847/1538-3881/abf097}

\end{thebibliography}

\end{document}